\documentclass{article}
\usepackage[final]{neurips_2025}
\usepackage[utf8]{inputenc} % allow utf-8 input
\usepackage[T1]{fontenc}    % use 8-bit T1 fonts 
\usepackage[hidelinks]{hyperref} % optional, nice PDF links
\usepackage{url}            % simple URL typesetting
\usepackage{booktabs}       % professional-quality tables
\usepackage{amsfonts}       % blackboard math symbols
\usepackage{nicefrac}       % compact symbols for 1/2, etc.
\usepackage{microtype}      % microtypography
\usepackage{xcolor}         % colors
\usepackage{csquotes}
\usepackage{graphicx}
\usepackage{dblfloatfix}
\usepackage{enumitem}
\usepackage{multirow}
\usepackage{wrapfig}

\usepackage{caption} % allowed by NeurIPS
\usepackage{subcaption}
\makeatletter
\newcommand{\listofappendices}{%
  \section*{Technical Appendices}%
  \addcontentsline{toc}{section}{Technical Appendices}%
  \@starttoc{atoc}%
}

\newcommand{\beginappendixtoc}{%
  \let\Saved@addcontentsline\addcontentsline
  \renewcommand{\addcontentsline}[3]{%
    \def\@tempa{##2}%
    \def\@tempb{section}%
    \ifx\@tempa\@tempb
      \Saved@addcontentsline{atoc}{##2}{##3}%
    \fi
  }%
}
 
\makeatother
\title{STARC-9: A Large-scale Dataset for Multi-Class Tissue Classification for CRC Histopathology}

\author{%
  Barathi~Subramanian\textsuperscript{1,*} ~~ Rathinaraja~Jeyaraj\textsuperscript{1,*} ~~ Mitchell~Nevin~Peterson\textsuperscript{2} \\
  \textbf{Terry~Guo}\textsuperscript{\textbf{1}} ~~ \textbf{Nigam~Shah}\textsuperscript{\textbf{3}} ~~ \textbf{Curtis~Langlotz}\textsuperscript{\textbf{4}} ~~ \textbf{Andrew~Y.~Ng}\textsuperscript{\textbf{5,6}} ~~ \textbf{Jeanne~Shen}\textsuperscript{\textbf{1}} \\
  \{\textsuperscript{1}Department of Pathology, \textsuperscript{2}Department of Electrical Engineering, \textsuperscript{3}Department of Medicine, \\ \textsuperscript{4}Department of Radiology, \textsuperscript{5}Department of Computer Science\}, Stanford University, USA  \\ \textsuperscript{6}DeepLearning.AI, USA  
}

\begin{document}

\maketitle

\begingroup
\renewcommand\thefootnote{*}
\footnotetext{Equal contribution.}
\endgroup

\begin{abstract}
Multi-class tissue-type classification of colorectal cancer (CRC) histopathologic images is a significant step in the development of downstream machine learning models for diagnosis and treatment planning. However, publicly available CRC datasets used to build tissue classifiers often suffer from insufficient morphologic diversity, class imbalance, and low-quality image tiles, limiting downstream model performance and generalizability. To address this research gap, we introduce STARC-9 (STAnford coloRectal Cancer), a large-scale dataset for multi-class tissue classification. STARC-9 comprises 630,000 histopathologic image tiles uniformly sampled across nine clinically relevant tissue classes (each represented by 70,000 tiles), systematically extracted from hematoxylin \& eosin-stained whole-slide images (WSI) from 200 CRC patients at the Stanford University School of Medicine. To construct STARC-9, we propose a novel framework, DeepCluster++, consisting of two primary steps to ensure diversity within each tissue class, followed by pathologist verification. First, an encoder from an autoencoder trained specifically on histopathologic images is used to extract feature vectors from all tiles within a given input WSI. Next, K-means clustering groups morphologically similar tiles, followed by an equal-frequency binning method to sample diverse patterns within each tissue class. Finally, the selected tiles are verified by expert gastrointestinal pathologists to ensure classification accuracy. This semi-automated approach significantly reduces the manual effort required for dataset curation while producing high-quality training examples. To validate the utility of STARC-9, we benchmarked baseline convolutional neural networks, transformers, and pathology-specific foundation models on downstream multi-class CRC tissue classification and segmentation tasks when trained on STARC-9 versus publicly available datasets, demonstrating superior generalizability of models trained on STARC-9. Although we demonstrate the utility of DeepCluster++ on CRC as a pilot use-case, it is a flexible framework that can be used for constructing high-quality datasets from large WSI repositories across a wide range of cancer and non-cancer applications. \url{https://huggingface.co/datasets/Path2AI/STARC-9/tree/main}
\url{https://github.com/Path2AI/STARC-9/} 
\end{abstract}

\section{Introduction}
Colorectal cancer (CRC) is the 3rd most common cancer and the 2nd leading cause of cancer-related death worldwide \cite{Morgan338}. Histologic evaluation of CRC is essential for diagnosis, prognostication, and therapeutic decision-making. With the growing adoption of digital pathology, computational approaches, particularly those leveraging deep learning, will play an increasingly important role in automating and augmenting pathology workflows. Deep learning-based multi-class tissue classification represents one such foundational task in pathology \cite{QIMS135171}, \cite{PAN2025106933}, enabling models to distinguish between diverse tissue types such as tumor, normal epithelium, muscle, and necrotic regions, and supporting downstream applications such as tissue segmentation \cite{Bokhorst2023DeepLearning}, tissue composition analysis \cite{JIAO2021106047}, biomarker status prediction \cite{YAMASHITA2021132}, and survival analysis \cite{Kather2019}. Providing pathologists with visually intuitive tissue maps also reduces the mental burden of diagnosis, while enhancing the interpretability of AI-driven insights. However, publicly available CRC datasets for machine learning are limited. The NCT-CRC-HE-100K dataset from Kather et al. \cite{Kather2019} represented a significant contribution for multi-class CRC tissue classification. Recently, another dataset (HMU-GC-HE-30K) \cite{Lou2025GastricCancer} was made public with different tissue types for developing tissue classifiers. Although this dataset contains images obtained from gastric cancer specimens, many of the tissue classes overlap with those found in CRC. Additional publicly available CRC datasets include the TCGA COAD and READ (The Cancer Genome Atlas - Colorectal Adenocarcinoma and Rectal Adenocarcinoma) \cite{TCGANetwork2012ColonRectal} whole-slide image (WSI)-level datasets. Despite representing important contributions to the field, these and other currently available histopathologic datasets have been insufficient for building robust, generalizable models for tissue-type classification and other downstream applications for several reasons, including: (1) lack of morphologic diversity, with insufficient representation of the broad range of appearances of different tissue classes \cite{Tani2025}, (2) class imbalance, where samples from dominant tissue types (e.g., tumor epithelium) far outnumber other clinically significant classes (e.g., mucin or necrosis) \cite{Ignatov2025}, and (3) inclusion of non-representative (incorrectly classified) tiles and low-quality (artifact-containing) tiles \cite{Kheiri2025BiasFactors}, which hinder model interpretability and degrade downstream task-specific performance. Furthermore, the construction of new pathology datasets is labor intensive, with no standardized framework for capturing sufficient tissue diversity and morphologic variation. All of these represent significant barriers to the development of robust, generalizable tissue classification models.

\begin{figure}[!t]
	\centering
	\includegraphics[width=0.8\textwidth]{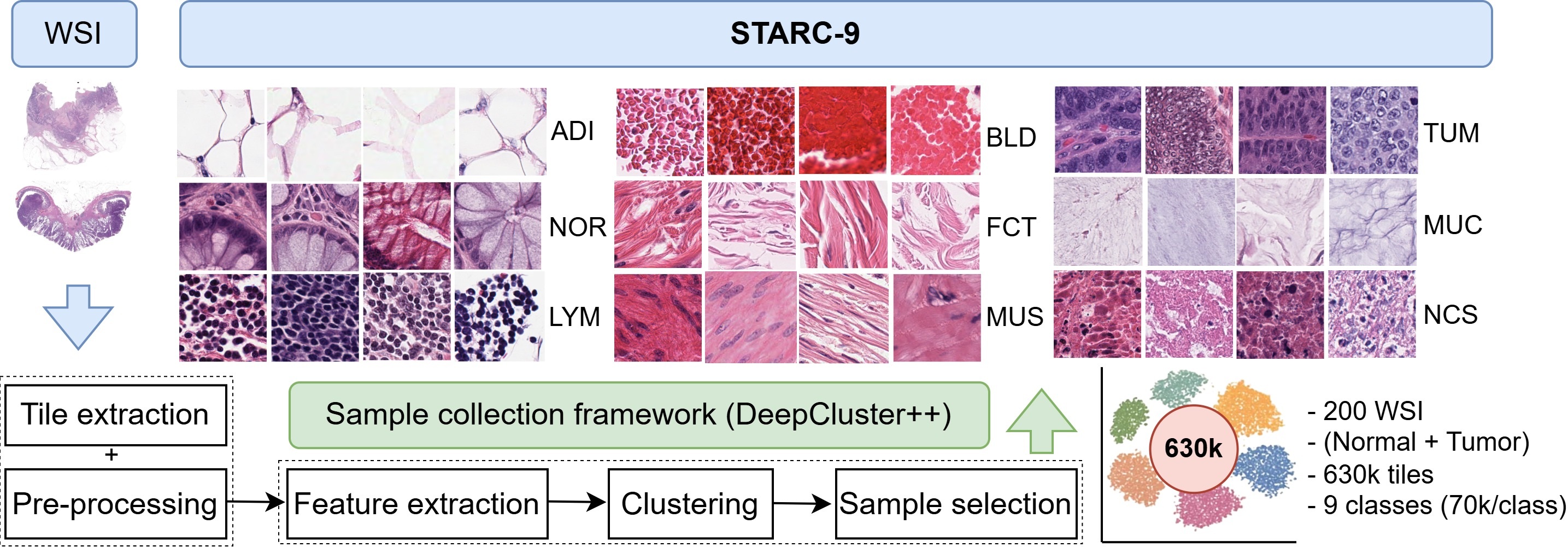}
	\caption{Overview of STARC-9 large-scale dataset generation.}
	\label{fig:1}
\end{figure}

To address these limitations, we introduce STARC-9 (STAnford coloRectal Cancer), a large-scale, high-quality dataset specifically designed for multi-class tissue classification for CRC histopathology, as well as a novel framework, DeepCluster++, used to construct STARC-9, which can readily be applied to other types of histopathologic WSI. STARC-9 comprises 630,000 non-overlapping tiles (256x256 pixels) systematically extracted using DeepCluster++ from hematoxylin \& eosin (H\&E)-stained WSI from approximately 200 CRC patients who underwent surgical resection of their CRC at the Stanford University School of Medicine. The following nine clinically relevant tissue classes are uniformly represented in the dataset: adipose tissue (ADI), lymphoid tissue (LYM), muscle (MUS), fibroconnective tissue (FCT), mucin (MUC), necrosis (NCS), blood (BLD), tumor (TUM), and normal mucosa (NOR), with each class containing 70,000 tiles. The overall DeepCluster++ workflow for constructing STARC-9 is illustrated in Figure \ref{fig:1}. Initially, tiles were extracted from the WSI and preprocessed to remove background and artifact tiles. Then, feature vectors for the remaining informative tiles were extracted using a CRC-specific autoencoder, pretrained on 100,000 histopathology images. K- means clustering was then employed to group the tiles based on morphologic similarity, wherein each cluster might represent a particular tissue type. To avoid oversampling in dense centroid regions, we partitioned each cluster into equal-frequency distance bins. Sampling across these bins ensured balanced intra-cluster diversity for robust classifier training. Repeating this pipeline across all WSI yielded 70,000 tiles per tissue type, resulting in 630,000 total high-quality tiles. Experienced pathologists then reviewed these samples to verify tissue-type classification accuracy, resulting in a robust, clinically relevant dataset. 

The DeepCluster++ framework for dataset construction significantly reduces the time and effort required for tile selection, compared to manual annotation using open-source tools such as QuPath \cite{Bankhead2017QuPath}. Traditional region-based sampling often leads to limited morphologic diversity, as pathologists tend to focus on visually similar regions within WSI, resulting in class imbalance and reduced generalizability of downstream models. Manual annotation is also subjective and inconsistent, relying heavily on the pathologist’s impression of whether a new region is sufficiently different from a previously annotated region to warrant inclusion in the dataset, making it difficult to ensure comprehensive representation of the entire morphologic spectrum within a WSI. In contrast, DeepCluster++ employs unsupervised clustering to group structurally similar tiles into coherent clusters, regardless of their location within a WSI. As a result, each cluster contains tissue tiles of a similar appearance sampled from diverse regions within the WSI. Sampling tiles from clusters corresponding to the same tissue type in this way enhances intra-class heterogeneity and tissue morphologic diversity (for instance, NOR, TUM, NCS in Figure \ref{fig:1}). This enhances dataset quality and increases the generalizability of models trained on the dataset by exposing them to a broad range of tissue morphologies important for downstream clinical applications. Although minimal manual review is still required, this method streamlines the overall dataset collection process, producing high-quality training examples for robust model development. Furthermore, DeepCluster++ is a flexible framework for constructing high-quality datasets which can be applied to both cancer and non-cancer WSI.

In a comprehensive evaluation, we trained both baseline and advanced multi-class classification models on the STARC-9, HMU-GC-HE-30K \cite{Lou2025GastricCancer}, and NCT-CRC-HE-100K \cite{Kather2019} datasets and evaluated their performance on independent Stanford and TCGA-CRC datasets using standard evaluation metrics, including precision, recall, F1-score, and accuracy. The baseline models included ResNet-50, EfficientNet-B7, KimiaNet, and ViT-base, as well as state-of-the-art (SOTA) transformer models such as DeiT-B, Swin Transformer-base, and ConvNeXT-Base. Pathology-specific foundation models, including CTransPath, HiPT, Prov-Gigapath, Path-DINO, CONCH, UNI, Virchow, and VIM4Path, were also benchmarked to evaluate their generalizability after fine-tuning on our dataset. In addition, a custom convolutional neural network (CNN) and a Histo-ViT model (DeiT-B) trained from scratch were also included in the analysis. In addition to these quantitative evaluations, we validated the practical utility of STARC-9 for tile-level segmentation on an independent TCGA-CRC dataset. In summary, our manuscript describes the following contributions:   

\begin{itemize}[noitemsep,topsep=0pt]
    \item STARC-9 dataset with 630,000 high-quality tiles across nine tissue types for model training
    \item Stanford (independent from STARC-9) and TCGA-CRC tile-level validation datasets
    \item Domain-specific feature extractor based on a custom-trained autoencoder
    \item Code repository for DeepCluster++ for generating datasets from any WSI 
    \item All models trained on the STARC-9 dataset
\end{itemize}

\section{Related Works}
\textit{\textbf{Publicly available tile-level CRC Datasets}}: Despite the growing interest in computational pathology, there are relatively few publicly available H\&E-stained tile-level CRC datasets for multi-class tissue classification. A significant contribution was made in \cite{Kather2016TextureAnalysis} with a dataset containing 5,000 non-overlapping 150×150 pixel image tiles across eight tissue categories. This was later expanded to 100,000 tiles in the NCT-CRC-HE-100K dataset \cite{Kather2019}, with 224×224 pixel patches covering nine tissue types from 86 WSI and an additional 7,180 images from 50 WSI for the validation set. These datasets have been widely adopted for various downstream tasks such as tissue classification \cite{bioimaging24}, segmentation \cite{Bokhorst2023DeepLearning}, and MSI prediction \cite{WANG2025107608}. Recently, the HMU-GC-HE-30K dataset \cite{Lou2025GastricCancer} was released, containing 30,000 224×224 pixel patches of gastric cancer with detailed tumor microenvironment (TME) annotations. While a few additional CRC datasets exist online \cite{Stettler_HistopathologyDatasets}, many of them either lack direct public access or do not provide comprehensive annotations or tissue-level labels. In contrast, datasets like TCGA-COAD/READ \cite{TCGANetwork2012ColonRectal} provide only unannotated WSI, requiring manual tile extraction for machine learning applications.

\textit{\textbf{Existing methodologies for building histopathologic image datasets}}: Manual annotation (e.g., using QuPath \cite{Bankhead2017QuPath}) of regions of interest (ROI) is slow and subjective and tends to favor easy regions, making it difficult to capture rare morphologies and maintain class balance as WSI size and complexity increase \cite{Pataki2022HunCRC}, \cite{Koziarski2024DiagSet}. Random sampling is susceptible to sampling error, wherein rare but clinically important morphologies are frequently missed, yielding imbalanced representations of tissue heterogeneity across WSI. Similarly, deep clustering (e.g., k-means on transfer-learned features \cite{Caron_2018_ECCV}, \cite{FU2024105826}) automates cluster formation, but sampling near cluster centroids biases toward common morphologies and under-represents intra-class variability required for robust supervised learning. Furthermore, active learning \cite{Li2024ActiveLearning} improves diversity by targeting model-uncertain samples, but requires iterative labeling and a seed of pre-labeled data.  

\textit{\textbf{Research gaps identified}}: Multi-class tissue classification for histopathology requires balanced, morphologically diverse datasets free of non-representative and low-quality tiles. However, existing datasets like NCT-CRC-HE-100K suffer from JPEG compression artifacts \cite{Ignatov2025}, and HMU-GC-HE-30K includes non-representative (e.g., incorrectly classified) tiles, leading models to learn spurious features. Similarly, TCGA-derived datasets exhibit sampling disparities and staining batch effects that affect model accuracy \cite{Kheiri2025BiasFactors}. While techniques like cross-entropy uncertainty and probabilistic local outlier detection \cite{Tani2025} can improve label quality, no cohesive pipeline exists for large-scale, balanced dataset curation. This limitation impacts downstream task performance, as observed in our initial experiments where models trained on HMU-GC-HE-30K and NCT-CRC-HE-100K achieved less than 90\% accuracy, reducing classification effectiveness on the independent Stanford dataset. In addition, existing dataset construction methods are slow, biased, and fail to capture rare morphologies effectively.

\section{DeepCluster++ for STARC-9 Construction   }   
To address the limitations of existing CRC datasets, we developed a semi-automated framework, DeepCluster++ (Figure \ref{fig:2}), to construct the STARC-9 dataset with 630,000 tiles across nine tissue types (ADI, LYM, MUS, FCT, MUC, NCS, BLD, TUM, NOR) shown in Figure \ref{fig:1}, from over 200 WSI (patient demographic details are provided in Technical Appendices Section \ref{sec:demographic_details}) representing a diverse morphologic spectrum of CRC surgical resection specimens. This approach integrates unsupervised feature extraction, clustering to group similar tiles, equal-frequency binning for tissue diversity, and an expert verification phase, resulting in the creation of a high-quality dataset for downstream tasks such as classification, tumor segmentation, and prognostication. 

\subsection{Phase 1: Autoencoder Training }
The first phase of our framework involves training an autoencoder (AE\_CRC) to learn domain-specific feature representations from histopathologic tiles. Autoencoders are unsupervised learning models that encode input images into low-dimensional latent vectors through a convolutional encoder, then reconstruct them via a symmetrical decoder. This process forces the encoder to capture compact, informative features which preserve critical structural and visual details. While autoencoders have been used for small grayscale image collections \cite{Ismail2022}, they have not been used to maximize diversity during tissue sample selection. For training AE\_CRC, we sampled 100,000 tiles of size 256×256 pixels from 10 representative WSI (5 tumor and 5 normal) independent of the STARC-9 training and validation sets, covering all nine histologic tissue types. Tile extraction was performed using histogram-based thresholding at a 32 down-sample factor with a 25\% tissue threshold to retain sparse tissues like ADI and MUC. Tile preprocessing included artifact removal and blank tile exclusion to create a high-quality ground truth set. Data augmentation techniques such as random rotations, flips, affine transformations, color jittering, and Gaussian blur were used to increase morphologic variability. The encoder consists of six convolutional layers with batch normalization and Leaky ReLU activations, producing a 32,768-dimensional latent vector. The decoder mirrors this architecture, using deconvolutional layers with a sigmoid activation in the final layer to reconstruct the input image. The AE\_CRC was trained using a structural similarity index (SSIM) loss function (see Technical Appendices Section \ref{sec:msevsssim} for details), which captures structural features crucial for histopathology. The reconstruction quality of AE\_CRC (as shown in Figure \ref{fig:2}(a) for NCS, NOR, LYM) was checked to ensure that the autoencoder learned a representation of diverse histologic patterns. We chose a custom autoencoder because its domain-specific, reconstruction-driven features produce finer morphology-sensitive embeddings with lower compute requirements than broad foundation and pretrained models, yielding more coherent clusters and better prototypical and edge-case coverage (see Technical Appendices Section \ref{sec:advantagesof_auto_encoder} for details).

\subsection{Phase 2: Clustering and Sampling Tiles  }
\begin{figure}[!t]
	\centering
	\includegraphics[width=1\textwidth]{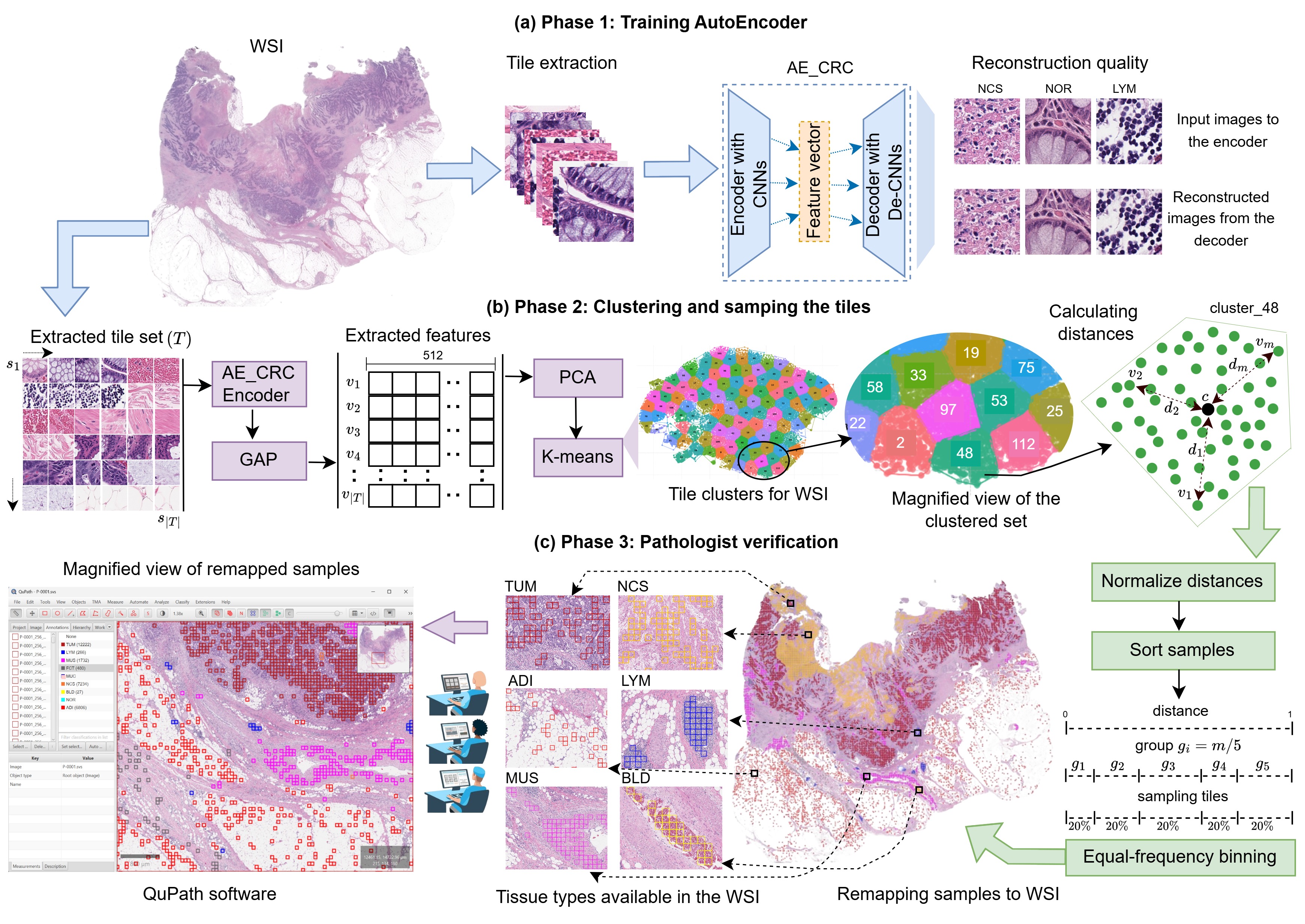}
	\caption{DeepCluster++ framework (Phases 1 and 2) followed by pathologist verification (Phase 3).  }
	\label{fig:2}
\end{figure}

We used only the frozen encoder of AE\_CRC to generate unsupervised embeddings for clustering candidate tiles in each WSI. These embeddings served solely to guide morphology‑aware tile sampling; they were not fed into any downstream classification or segmentation applications. Let the extracted tile set for a WSI be $T=\{s_1,s_2,…,s_{|T|}\}$, in which each tile $s_i$ was preprocessed as in Phase 1 and passed through the encoder to generate a 32,768-dimensional latent vector $v_i$. To reduce computational complexity and improve clustering performance, global average pooling (GAP) was applied to compress the feature vector to 512 dimensions, as shown in Figure \ref{fig:2}(b). We subsequently applied principal component analysis (PCA) to further reduce the latent dimensionality to 256, thereby decreasing computational complexity and improving efficiency. These feature vectors were then clustered using the K-means algorithm \cite{Caron_2018_ECCV}, as in \cite{RIASATIAN2021102032}, to group tiles with similar morphology. We set the number of samples per cluster ($m$) to 400 to balance tissue diversity and representation quality, based on our empirical evaluation (see Technical Appendices Section \ref{sec:ablation_study} for details), as higher values (e.g., 800) risk including mixed tissue types, while lower values (e.g., 100) may reduce morphologic variation. Additionally, K-means was preferred over methods like DBSCAN, which lacks consistent cluster sizing. This approach also preserved local morphologic coherence, as adjacent clusters often contained similar tissue types, facilitating efficient sampling of diverse tissue patterns. The next step involved sampling tiles from each cluster to preserve morphologic diversity. For each cluster (e.g., cluster\_48), we first computed the cluster centroid $c$ and calculated the Euclidean distance from centroid for each tile as $d_i=\left\| v_i -c \right\|$, as shown in Figure \ref{fig:2}(c). These distances were normalized to the range $[0,1]$ to ensure consistency across clusters of varying densities. Tiles were then sorted by distance, and equal-frequency binning was applied to divide the samples into five distance-based groups ($g=5$). This approach ensures that each group contains an equal number of tiles, preventing over-representation of dense regions near the cluster center and capturing a broad range of tissue patterns. Unlike equal-width binning, which often leads to imbalanced groups, this method maintains a uniform distribution of samples from near-centroid (homogeneous) to edge-of-cluster (diverse) tiles. Increasing (e.g., to 10) and decreasing (e.g., to 2) the number of bins respectively enhances and reduces variation, offering flexibility based on downstream requirements.

 We sampled 20\% of the tiles from each bin to ensure a comprehensive representation. The sampled tiles were then stored in separate class folders based on tissue type. Because these clusters did not carry semantic labels (e.g.,  \enquote{TUM} or \enquote{LYM}), we manually reviewed each output cluster to identify which tissue type it best reflected. To associate clusters with particular tissue types, once a cluster was labeled (for example, cluster\_48 in Figure \ref{fig:2}(b) was confirmed as \enquote{TUM}), we used the embedding space proximity to find adjacent clusters, such as 2, 97, 53, and 112, sharing similar feature representations implying similar histologic patterns. In our experiments, these neighboring clusters consistently contained the same tissue morphology, allowing us to extend the \enquote{TUM} label across additional clusters with minimal additional time and effort. This local continuity within the feature space enabled more efficient exploration and sampling of tissue diversity. By iterating through this process (labeling a seed cluster and propagating its label to nearby clusters), we efficiently mapped appropriate clusters to the nine target tissue classes with modest manual effort.    

\subsection{Phase 3: Pathologist Verification and Final Dataset Assembly}
The samples collected for all tissue types were mapped back to their original WSI location using QuPath \cite{Bankhead2017QuPath} (Figure \ref{fig:2}(c)) for pathologist verification of tissue-type classification. To support robust multi-tissue type classification, we fixed the number of tiles per tissue type at 70,000, resulting in a final STARC-9 dataset of 630,000 high-quality samples. While this decision might be debated, it was necessary for the consistent evaluation of 21,000 internal WSI. However, for other real-world applications, the number of tiles per class can be adjusted based on available data and downstream task requirements. All images were reviewed for classification accuracy by board-certified pathologists with subspecialty expertise in gastrointestinal (GI) pathology. In total, three board-certified GI pathologists participated in the review process: two pathologists with 13 and 41 years of experience, respectively, evaluated subsets of the dataset, while a third pathologist with 15 years of experience conducted a comprehensive final review of the entire dataset comprising 630,000 images.  

\section{Experiments and Results}
\textit{\textbf{Dataset description}}: STARC-9 is a comprehensive multi-tissue classification dataset consisting of 630,000 high-resolution non-overlapping 256×256 pixel tiles extracted from 40x magnification (0.25 micrometers/pixel) WSI. It includes nine clinically relevant tissue types: ADI, LYM, MUS, FCT, MUC, NCS, BLD, TUM, and NOR, capturing diverse, fine-grained tissue morphologies. To facilitate rigorous validation of models trained on STARC-9, two independent validation sets were prepared. (i)  STANFORD-CRC-HE-VAL-SMALL contains 18,000 tiles (2,000 per tissue type) obtained from 20 WSI separate from the cases used in STARC-9 and was used for preliminary model testing, yielding 79.59\% and 85.9\% accuracy for models trained on NCT-CRC-HE-100K (NCT) \cite{Kather2019} and HMU-GC-HE-30K (HMU) \cite{Lou2025GastricCancer}, respectively. This highlighted the need for larger, more diverse training sets, as these models struggled with mixed tissue-type tiles, achieving only 60\% overall per-WSI tissue mapping accuracy. (ii) The primary validation set, STANFORD-CRC-HE-VAL-LARGE, includes 54,000 tiles (6,000 per class) from 50 WSI independent from STARC-9 and STANFORD-CRC-HE-VAL-SMALL for performance evaluation. Training (STARC‑9) and validation (STANFORD-CRC-HE-VAL-SMALL and STANFORD-CRC-HE-VAL-LARGE) datasets were drawn from different patients, with no overlap between any of the datasets. Additionally, an external CURATED-TCGA-CRC-HE-VAL-20K dataset was prepared with 20,000 tiles extracted from 30 TCGA-CRC WSI to assess the robustness and generalizability of models trained on STARC-9. During benchmarking, models trained on STARC-9 (630,000 tiles, 9 classes), NCT (100,000 tiles, 9 classes), and HMU (30,000 tiles, 8 classes) were validated on seven overlapping tissue types (ADI, LYM, MUS, MUC, NCS, TUM, and NOR), ensuring fairness in performance evaluation. 
 
\textit{\textbf{Model description}}: To evaluate the utility of STARC-9, we conducted a series of benchmarking experiments using a diverse set of deep learning models, including baseline CNNs, SOTA transformer models, and pathology-specific foundation models. The objective was to assess classification performance, generalizability, and practical utility compared to models trained (fine-tuned) on publicly available datasets like NCT and HMU. Baseline models included ResNet-50, EfficientNet-B7, KimiaNet, and ViT-base, while SOTA models included DeiT-B, Swin Transformer-Base, and ConvNeXT-Base. Pathology-specific foundation models such as CTransPath, HiPT, Prov-Gigapath, Path-DINO, CONCH, UNI, Virchow, and VIM4Path were also tested to assess their generalizability on diverse tissue morphologies. Each model was trained on the STARC-9, NCT, and HMU datasets with Macenko normalization \cite{Macenko2009}. All models were fine-tuned with a batch size of 32, learning rate of 0.0001, weight decay of 1e-5, Adam optimization, and data augmentation (horizontal/vertical flips, random rotation, and color jittering) for 10 epochs. Maintaining identical training configurations, including batch size and optimizer settings, across datasets was important for a fair and unbiased model comparison. As our primary objective was to isolate the impact of the dataset on model performance, we kept the training configurations consistent to avoid possible confounding introduced by different hyperparameters being applied to the datasets. Model performance was evaluated using precision, recall, macro F1 score, accuracy, and the number of trainable parameters. STARC-9-trained models consistently outperformed models trained on other datasets, exhibiting better generalizability.   

\textit{\textbf{Resource description}}: All experiments were conducted on the following platforms: (i) a local server with 8x NVIDIA L40S 48GB GPUs, and the Stanford (ii) Carina \cite{Stanford_Carina} and (iii) Marlowe high-performance computing platforms \cite{kapfer_2025_14751899}. 

\begin{table} [b!]  
\large
\centering
\caption{Multi-class tissue classification performance of baseline, SOTA, pathology foundation, and custom models trained on HMU, NCT, and STARC-9 for seven common tissue types (ADI, LYM, MUS, MUC, NCS, TUM, NOR) evaluated on the STANFORD-CRC-HE-VAL-LARGE dataset. The highest accuracy models for each dataset are highlighted in bold. }
\label{table1}
\resizebox{\textwidth}{!}{%

\begin{tabular}{|lccccccccccccc|}
\hline
\multicolumn{1}{|c|}{\multirow{2}{*}{\textbf{Model}}} & \multicolumn{3}{c|}{\textbf{Precision}}                                                                       & \multicolumn{3}{c|}{\textbf{Recall}}                                                                          & \multicolumn{3}{c|}{\textbf{F1-macro}}                                                                        & \multicolumn{3}{c|}{\textbf{Accuracy}}                                                                            & \multirow{2}{*}{\textbf{\begin{tabular}[c]{@{}c@{}}No. of \\ params.\end{tabular}}} \\ \cline{2-13}
\multicolumn{1}{|c|}{}                                & \multicolumn{1}{c|}{\textbf{NCT}} & \multicolumn{1}{c|}{\textbf{HMU}} & \multicolumn{1}{c|}{\textbf{STARC-9}} & \multicolumn{1}{c|}{\textbf{NCT}} & \multicolumn{1}{c|}{\textbf{HMU}} & \multicolumn{1}{c|}{\textbf{STARC-9}} & \multicolumn{1}{c|}{\textbf{NCT}} & \multicolumn{1}{c|}{\textbf{HMU}} & \multicolumn{1}{c|}{\textbf{STARC-9}} & \multicolumn{1}{c|}{\textbf{NCT}}   & \multicolumn{1}{c|}{\textbf{HMU}}   & \multicolumn{1}{c|}{\textbf{STARC-9}} &                                                                                     \\ \hline
\multicolumn{14}{|c|}{\textbf{Baseline models}}                                                                                                                                                                                                                                                                                                                                                                                                                                                                                                                                                                 \\ \hline
\multicolumn{1}{|l|}{ResNet50 \cite{He2016DeepResidual}}                        & \multicolumn{1}{c|}{84.08}        & \multicolumn{1}{c|}{87.81}        & \multicolumn{1}{c|}{98.92}            & \multicolumn{1}{c|}{62.59}        & \multicolumn{1}{c|}{85.71}        & \multicolumn{1}{c|}{98.64}            & \multicolumn{1}{c|}{63.17}        & \multicolumn{1}{c|}{86.00}        & \multicolumn{1}{c|}{98.78}            & \multicolumn{1}{c|}{62.59}          & \multicolumn{1}{c|}{85.71}          & \multicolumn{1}{c|}{98.64}            & 24 M                                                                                \\ \hline
\multicolumn{1}{|l|}{EfficientNet-B7 \cite{Tan2019EfficientNet}}                  & \multicolumn{1}{c|}{89.99}        & \multicolumn{1}{c|}{88.65}        & \multicolumn{1}{c|}{99.11}            & \multicolumn{1}{c|}{82.47}        & \multicolumn{1}{c|}{87.45}        & \multicolumn{1}{c|}{98.80}            & \multicolumn{1}{c|}{84.55}        & \multicolumn{1}{c|}{87.87}        & \multicolumn{1}{c|}{98.95}            & \multicolumn{1}{c|}{82.47}          & \multicolumn{1}{c|}{84.45}          & \multicolumn{1}{c|}{98.80}            & 64 M                                                                                \\ \hline
\multicolumn{1}{|l|}{ViT-base \cite{Dosovitskiy2021ViT}}                        & \multicolumn{1}{c|}{92.71}        & \multicolumn{1}{c|}{91.57}        & \multicolumn{1}{c|}{98.49}            & \multicolumn{1}{c|}{84.25}        & \multicolumn{1}{c|}{90.29}        & \multicolumn{1}{c|}{98.09}            & \multicolumn{1}{c|}{87.30}        & \multicolumn{1}{c|}{90.87}        & \multicolumn{1}{c|}{98.28}            & \multicolumn{1}{c|}{\textbf{84.25}} & \multicolumn{1}{c|}{90.29}          & \multicolumn{1}{c|}{98.09}            & 86 M                                                                                \\ \hline
\multicolumn{14}{|c|}{\textbf{SOTA models}}                                                                                                                                                                                                                                                                                                                                                                                                                                                                                                                                                                     \\ \hline
\multicolumn{1}{|l|}{DeiT-B \cite{pmlr-v139-touvron21a}}                          & \multicolumn{1}{c|}{94.28}        & \multicolumn{1}{c|}{90.97}        & \multicolumn{1}{c|}{98.99}            & \multicolumn{1}{c|}{81.63}        & \multicolumn{1}{c|}{90.05}        & \multicolumn{1}{c|}{98.65}            & \multicolumn{1}{c|}{85.35}        & \multicolumn{1}{c|}{90.40}        & \multicolumn{1}{c|}{98.81}            & \multicolumn{1}{c|}{81.63}          & \multicolumn{1}{c|}{90.05}          & \multicolumn{1}{c|}{98.65}            & 86 M                                                                                \\ \hline
\multicolumn{1}{|l|}{Swin Trans-base \cite{Liu2022SwinTransformer}}                 & \multicolumn{1}{c|}{90.11}        & \multicolumn{1}{c|}{93.17}        & \multicolumn{1}{c|}{99.09}            & \multicolumn{1}{c|}{79.05}        & \multicolumn{1}{c|}{91.88}        & \multicolumn{1}{c|}{98.80}            & \multicolumn{1}{c|}{82.52}        & \multicolumn{1}{c|}{92.46}        & \multicolumn{1}{c|}{98.94}            & \multicolumn{1}{c|}{79.05}          & \multicolumn{1}{c|}{91.88}          & \multicolumn{1}{c|}{98.79}            & 87 M                                                                                \\ \hline
\multicolumn{1}{|l|}{KimiaNet \cite{RIASATIAN2021102032}}                        & \multicolumn{1}{c|}{87.25}        & \multicolumn{1}{c|}{88.60}        & \multicolumn{1}{c|}{99.03}            & \multicolumn{1}{c|}{71.53}        & \multicolumn{1}{c|}{86.67}        & \multicolumn{1}{c|}{98.72}            & \multicolumn{1}{c|}{71.53}        & \multicolumn{1}{c|}{87.04}        & \multicolumn{1}{c|}{98.87}            & \multicolumn{1}{c|}{68.69}          & \multicolumn{1}{c|}{86.67}          & \multicolumn{1}{c|}{98.72}            & 7M                                                                                  \\ \hline
\multicolumn{1}{|l|}{ConvNeXT-base \cite{Liu2022ConvNet2020s}}                   & \multicolumn{1}{c|}{91.95}        & \multicolumn{1}{c|}{92.09}        & \multicolumn{1}{c|}{99.01}            & \multicolumn{1}{c|}{82.82}        & \multicolumn{1}{c|}{91.07}        & \multicolumn{1}{c|}{98.36}            & \multicolumn{1}{c|}{85.56}        & \multicolumn{1}{c|}{91.50}        & \multicolumn{1}{c|}{98.68}            & \multicolumn{1}{c|}{82.82}          & \multicolumn{1}{c|}{91.07}          & \multicolumn{1}{c|}{98.36}            & 88 M                                                                                \\ \hline
\multicolumn{14}{|c|}{\textbf{Pathology foundation models}}                                                                                                                                                                                                                                                                                                                                                                                                                                                                                                                                                     \\ \hline
\multicolumn{1}{|l|}{CTransPath \cite{WANG2022102559}}                      & \multicolumn{1}{c|}{90.11}        & \multicolumn{1}{c|}{93.17}        & \multicolumn{1}{c|}{99.34}            & \multicolumn{1}{c|}{79.05}        & \multicolumn{1}{c|}{91.88}        & \multicolumn{1}{c|}{99.00}            & \multicolumn{1}{c|}{82.52}        & \multicolumn{1}{c|}{92.46}        & \multicolumn{1}{c|}{99.16}            & \multicolumn{1}{c|}{79.05}          & \multicolumn{1}{c|}{91.88}          & \multicolumn{1}{c|}{\textbf{99.00}}   & 87 M                                                                                \\ \hline
\multicolumn{1}{|l|}{HiPT \cite{Chen2022ScalingViT}}                            & \multicolumn{1}{c|}{90.92}        & \multicolumn{1}{c|}{93.21}        & \multicolumn{1}{c|}{98.64}            & \multicolumn{1}{c|}{74.51}        & \multicolumn{1}{c|}{91.99}        & \multicolumn{1}{c|}{98.32}            & \multicolumn{1}{c|}{77.41}        & \multicolumn{1}{c|}{92.54}        & \multicolumn{1}{c|}{98.47}            & \multicolumn{1}{c|}{74.51}          & \multicolumn{1}{c|}{\textbf{91.99}} & \multicolumn{1}{c|}{98.32}            & 86 M                                                                                \\ \hline
\multicolumn{1}{|l|}{ProvGigPath \cite{Xu2024WholeSlideFoundationModel}}                     & \multicolumn{1}{c|}{89.43}        & \multicolumn{1}{c|}{91.47}        & \multicolumn{1}{c|}{98.74}            & \multicolumn{1}{c|}{74.18}        & \multicolumn{1}{c|}{90.60}        & \multicolumn{1}{c|}{98.37}            & \multicolumn{1}{c|}{78.40}        & \multicolumn{1}{c|}{90.92}        & \multicolumn{1}{c|}{98.55}            & \multicolumn{1}{c|}{74.18}          & \multicolumn{1}{c|}{90.60}          & \multicolumn{1}{c|}{98.37}            & 305 M                                                                               \\ \hline
\multicolumn{1}{|l|}{PathDino \cite{Alfasly2024RotationAgnostic}}                        & \multicolumn{1}{c|}{92.93}        & \multicolumn{1}{c|}{91.19}        & \multicolumn{1}{c|}{98.67}            & \multicolumn{1}{c|}{77.35}        & \multicolumn{1}{c|}{89.64}        & \multicolumn{1}{c|}{98.37}            & \multicolumn{1}{c|}{81.71}        & \multicolumn{1}{c|}{90.22}        & \multicolumn{1}{c|}{98.51}            & \multicolumn{1}{c|}{77.35}          & \multicolumn{1}{c|}{89.64}          & \multicolumn{1}{c|}{98.37}            & 22 M                                                                                \\ \hline
\multicolumn{1}{|l|}{CONCH \cite{Lu2024VisualLanguageCP}}                           & \multicolumn{1}{c|}{91.53}        & \multicolumn{1}{c|}{91.41}        & \multicolumn{1}{c|}{98.56}            & \multicolumn{1}{c|}{75.69}        & \multicolumn{1}{c|}{90.02}        & \multicolumn{1}{c|}{98.19}            & \multicolumn{1}{c|}{78.08}        & \multicolumn{1}{c|}{90.52}        & \multicolumn{1}{c|}{98.37}            & \multicolumn{1}{c|}{75.69}          & \multicolumn{1}{c|}{90.02}          & \multicolumn{1}{c|}{98.19}            & 87 M                                                                                \\ \hline
\multicolumn{1}{|l|}{UNI \cite{Chen2024GeneralPurposeCP}}                             & \multicolumn{1}{c|}{94.55}        & \multicolumn{1}{c|}{93.03}        & \multicolumn{1}{c|}{98.67}            & \multicolumn{1}{c|}{80.43}        & \multicolumn{1}{c|}{91.80}        & \multicolumn{1}{c|}{98.25}            & \multicolumn{1}{c|}{84.42}        & \multicolumn{1}{c|}{92.36}        & \multicolumn{1}{c|}{98.45}            & \multicolumn{1}{c|}{80.43}          & \multicolumn{1}{c|}{91.80}          & \multicolumn{1}{c|}{98.26}            & 88 M                                                                                \\ \hline
\multicolumn{1}{|l|}{Virchow \cite{Vorontsov2024FoundationModelCP}}                         & \multicolumn{1}{c|}{92.51}        & \multicolumn{1}{c|}{92.35}        & \multicolumn{1}{c|}{98.63}            & \multicolumn{1}{c|}{79.02}        & \multicolumn{1}{c|}{91.23}        & \multicolumn{1}{c|}{98.28}            & \multicolumn{1}{c|}{82.05}        & \multicolumn{1}{c|}{91.69}        & \multicolumn{1}{c|}{98.45}            & \multicolumn{1}{c|}{79.02}          & \multicolumn{1}{c|}{91.23}          & \multicolumn{1}{c|}{98.28}            & 305 M                                                                               \\ \hline
\multicolumn{1}{|l|}{VIM4PATH \cite{10678515}}                        & \multicolumn{1}{c|}{91.51}        & \multicolumn{1}{c|}{92.66}        & \multicolumn{1}{c|}{98.53}            & \multicolumn{1}{c|}{75.41}        & \multicolumn{1}{c|}{91.50}        & \multicolumn{1}{c|}{98.27}            & \multicolumn{1}{c|}{79.10}        & \multicolumn{1}{c|}{92.01}        & \multicolumn{1}{c|}{98.40}            & \multicolumn{1}{c|}{75.41}          & \multicolumn{1}{c|}{91.50}          & \multicolumn{1}{c|}{98.29}            & 86 M                                                                                \\ \hline
\multicolumn{14}{|c|}{\textbf{Customized models (trained from scratch)}}                                                                                                                                                                                                                                                                                                                                                                                                                                                                                                                                        \\ \hline
\multicolumn{1}{|l|}{CNN}                             & \multicolumn{1}{c|}{83.97}        & \multicolumn{1}{c|}{78.45}        & \multicolumn{1}{c|}{98.10}            & \multicolumn{1}{c|}{64.21}        & \multicolumn{1}{c|}{68.10}        & \multicolumn{1}{c|}{97.81}            & \multicolumn{1}{c|}{68.12}        & \multicolumn{1}{c|}{66.39}        & \multicolumn{1}{c|}{97.93}            & \multicolumn{1}{c|}{64.21}          & \multicolumn{1}{c|}{68.10}          & \multicolumn{1}{c|}{97.81}            & 3.9 M                                                                               \\ \hline
\multicolumn{1}{|l|}{Histo-ViT}                        & \multicolumn{1}{c|}{86.17}        & \multicolumn{1}{c|}{76.45}        & \multicolumn{1}{c|}{96.88}            & \multicolumn{1}{c|}{69.48}        & \multicolumn{1}{c|}{67.16}        & \multicolumn{1}{c|}{96.32}            & \multicolumn{1}{c|}{72.01}        & \multicolumn{1}{c|}{67.77}        & \multicolumn{1}{c|}{96.52}            & \multicolumn{1}{c|}{69.48}          & \multicolumn{1}{c|}{67.16}          & \multicolumn{1}{c|}{96.32}            & 86 M                                                                                \\ \hline
\end{tabular}

}
\end{table} 

\subsection{Multi-Class Tissue Classification} \label{subsec:classification}
All models trained on STARC-9 demonstrated exceptional performance on the STANFORD-CRC-HE-VAL-LARGE validation dataset (Table \ref{table1}). Among the baseline models, EfficientNet-B7 trained on STARC-9 achieved the highest overall accuracy (98.80\%), with a 14.7\% improvement over the best model trained on NCT (84.25\%) and an 8.6\% improvement over the best model trained on HMU (90.29\%). In the SOTA category, Swin Transformer (Swin Trans-base) trained on STARC-9 achieved 98.79\% accuracy, a 16.1\% improvement over ConvNeXT-base trained on NCT (82.82\%) and a 6.9\% improvement over Swin Trans-base trained on HMU (91.88\%). Among the pathology-specific foundation models, CTransPath trained on STARC-9 with 87M parameters achieved 99\% accuracy, significantly outperforming UNI trained on NCT (80.43\%) and HiPT trained on HMU (91.99\%), emphasizing the importance of domain-specific pretraining. Custom models trained from scratch, such as a CNN and Histo-ViT trained on STARC-9, achieved accuracies of 97.81\% and 96.32\%, respectively, highlighting the ability of high-quality, domain-specific training data to enable effective representation learning without the overhead of pre-training and risk of overfitting. Overall, these results emphasize the importance of diverse, high-quality training samples for developing robust tissue classification models. The consistent improvements in precision, recall, and F1-macro scores across all tissue types highlight the advantage of STARC-9’s data diversity, which contributed to the over 97\% accuracy, even for models without extensive pretraining. CTransPath trained on STARC-9 consistently outperformed ViT-base (trained on NCT) and HiPT (trained on HMU) across all evaluation metrics on external validation sets. 

Table \ref{table2} reports the precision, recall, F1‑macro, and accuracy metrics for the top‑performing models (with respect to accuracy in Table \ref{table1}), when validated on STANFORD-CRC-HE-VAL-SMALL, STANFORD-CRC-HE-VAL-LARGE, and CURATED-TCGA-CRC-HE-VAL-20K. These top-performing models were: ViT‑Base trained on NCT, HiPT trained on HMU, and CTransPath trained on STARC‑9. For STANFORD-CRC-HE-VAL-SMALL, the STARC-9-trained CTransPath model achieved 99.75\% precision, 99.73\% recall, 99.74\% F1- macro, and 99.73\% accuracy - significantly higher than the other models, which showed lower recall and F1-macro scores. Similarly, on STARC-9-HE-VAL-LARGE, CTransPath maintained its lead, with 99.34\% precision, 99.00\% recall, 99.16\% F1-macro, and 99.00\% accuracy. Even on the more challenging STANFORD-TCGA-CRC-HE-20K set, CTransPath consistently achieved near-perfect precision (99.05\%), recall (98.88\%), and F1-macro (98.96\%), demonstrating excellent generalization and robustness across diverse tissue types. To further evaluate the generalizability of the model trained on STARC-9, we curated a small test set of the seven common tissue classes taken from 10 WSI from the IMP‑CRS10K biopsy/polypectomy dataset \cite{Neto2024IMPWSI}. In total, 1,093 image tiles were annotated for model performance validation, in which the STARC‑9-trained model achieved a 95.55\% F1‑macro and 96.61\% accuracy, consistently outperforming the HMU and NCT–trained models. It would also be interesting to evaluate the performance of the models trained on STARC-9 on the NCT and HMU datasets. However, as noted in Section 2 (and in reference \cite{Ignatov2025}), the publicly available validation sets from NCT and HMU contain a substantial fraction of artifact‐laden or mislabeled tiles, as well as "ambiguous" tiles with more than one tissue type represented within the same tile, despite only a single tissue-type label being assigned to the tile. In order to utilize these two datasets as reliable validation datasets, pathologist re‑verification and correction/refinement of the tile‐level labels would be necessary, which is labor‐intensive and infeasible, given that the original WSI used to generate these two datasets were not publicly available for verification of label accuracy. 

\begin{table} [t!]  
\large
\centering
\caption{Multi-class tissue classification performance of the best-performing models trained on HMU, NCT, and STARC-9 for seven common tissue types on the validation sets.}
\label{table2}
\resizebox{\textwidth}{!}{%
\begin{tabular}{|l|ccc|ccc|ccc|ccc|}
\hline
\multicolumn{1}{|c|}{}                                                 & \multicolumn{3}{c|}{\textbf{Precision}}                                                                                              & \multicolumn{3}{c|}{\textbf{Recall}}                                                                                                 & \multicolumn{3}{c|}{\textbf{F1-macro}}                                                                                               & \multicolumn{3}{c|}{\textbf{Accuracy}}                                                                                                        \\ \cline{2-13} 
\multicolumn{1}{|c|}{\multirow{-2}{*}{\textbf{Validation dataset}}}    & \multicolumn{1}{c|}{\textbf{NCT}}                 & \multicolumn{1}{c|}{\textbf{HMU}}                 & \textbf{STARC-9}             & \multicolumn{1}{c|}{\textbf{NCT}}                 & \multicolumn{1}{c|}{\textbf{HMU}}                 & \textbf{STARC-9}             & \multicolumn{1}{c|}{\textbf{NCT}}                 & \multicolumn{1}{c|}{\textbf{HMU}}                 & \textbf{STARC-9}             & \multicolumn{1}{c|}{\textbf{NCT}}                 & \multicolumn{1}{c|}{\textbf{HMU}}                 & \textbf{STARC-9}                      \\ \hline
\begin{tabular}[c]{@{}l@{}}STANFORD-CRC-\\ HE-VAL-SMALL\end{tabular}   & \multicolumn{1}{c|}{88.52}                        & \multicolumn{1}{c|}{90.22}                        & 99.75                        & \multicolumn{1}{c|}{76.19}                        & \multicolumn{1}{c|}{88.34}                        & 99.73                        & \multicolumn{1}{c|}{79.34}                        & \multicolumn{1}{c|}{89.16}                        & 99.74                        & \multicolumn{1}{c|}{76.19}                        & \multicolumn{1}{c|}{88.34}                        & \textbf{99.73}                        \\ \hline
\begin{tabular}[c]{@{}l@{}}STANFORD-CRC-\\ HE-VAL-LARGE\end{tabular}   & \multicolumn{1}{c|}{92.71}                        & \multicolumn{1}{c|}{93.21}                        & 99.34                        & \multicolumn{1}{c|}{84.25}                        & \multicolumn{1}{c|}{91.99}                        & 99.00                        & \multicolumn{1}{c|}{87.30}                        & \multicolumn{1}{c|}{92.54}                        & 99.16                        & \multicolumn{1}{c|}{84.25}                        & \multicolumn{1}{c|}{91.99}                        & \textbf{99.00}                        \\ \hline
\begin{tabular}[c]{@{}l@{}}CURATED-TCGA-\\ CRC-HE-VAL-20K\end{tabular} & \multicolumn{1}{c|}{89.69}                        & \multicolumn{1}{c|}{92.21}                        & 99.03                        & \multicolumn{1}{c|}{72.42}                        & \multicolumn{1}{c|}{90.9}                         & 98.85                        & \multicolumn{1}{c|}{76.74}                        & \multicolumn{1}{c|}{91.45}                        & 98.94                        & \multicolumn{1}{c|}{72.42}                        & \multicolumn{1}{c|}{90.9}                         & \textbf{98.85}                        \\ \hline
{\color[HTML]{000000} IMP-CRS10K }                             & \multicolumn{1}{c|}{{\color[HTML]{000000} 63.29}} & \multicolumn{1}{c|}{{\color[HTML]{000000} 65.06}} & {\color[HTML]{000000} 96.70} & \multicolumn{1}{c|}{{\color[HTML]{000000} 42.77}} & \multicolumn{1}{c|}{{\color[HTML]{000000} 61.99}} & {\color[HTML]{000000} 94.88} & \multicolumn{1}{c|}{{\color[HTML]{000000} 45.85}} & \multicolumn{1}{c|}{{\color[HTML]{000000} 62.46}} & {\color[HTML]{000000} 95.55} & \multicolumn{1}{c|}{{\color[HTML]{000000} 69.62}} & \multicolumn{1}{c|}{{\color[HTML]{000000} 76.40}} & {\color[HTML]{000000} \textbf{96.61}} \\ \hline
\end{tabular}
}
\end{table} 

\begin{wrapfigure}{r}{0.6\textwidth} 
    \centering
    \includegraphics[width=0.6\textwidth]{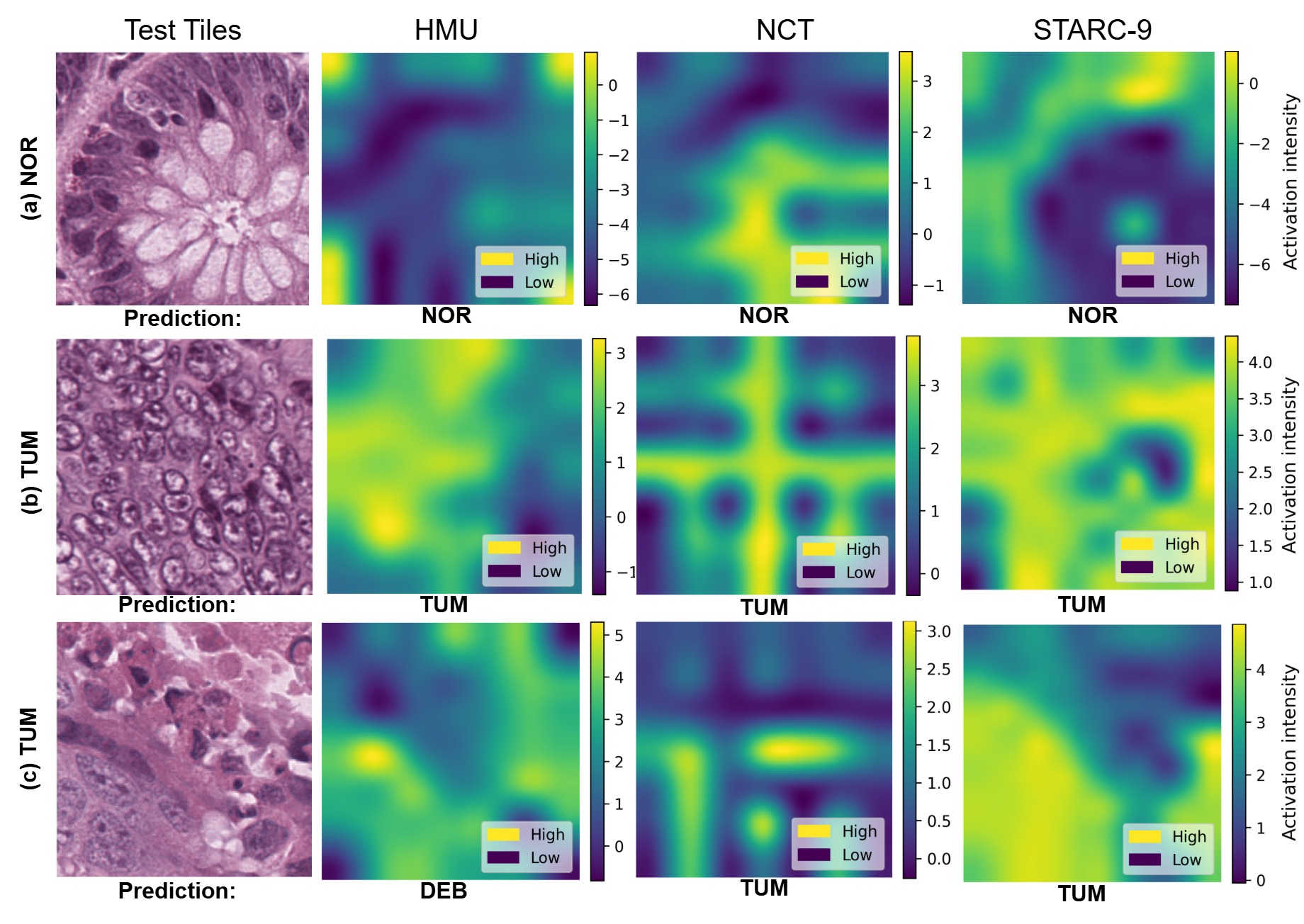}
    \caption{Feature map visualizations for the best models trained on HMU, NCT, and STARC-9. }
    \label{fig:3}
\end{wrapfigure}

\textit{\textbf{Feature map visualization analysis}}: Figure \ref{fig:3} illustrates the significant impact of training data quality on model feature activations for multi-class tissue classification. The figure presents activation maps generated by models trained on HMU, NCT, and STARC-9 for three representative ground truth input tiles (NOR, TUM, and mixed TUM) in panels (a), (b), and (c), respectively. Models trained on STARC-9 consistently focused on diagnostically relevant histologic features, aligning closely with pathologist evaluation patterns, while those trained on NCT and HMU often activated less diagnostically relevant regions. In Figure \ref{fig:3}(a), while all three models correctly predicted the normal (NOR) class, the model trained on HMU activated more dispersed, less relevant regions, reflecting its exposure to less-representative training data. The NCT-trained model captures some vague cellular architecture, but lacks comprehensive coverage of relevant structures. 
In contrast, the STARC-9 model accurately focuses on the regions critical for the diagnosis, demonstrating the impact of well-curated, diverse training samples on models' ability to   capture subtle, diagnostically relevant histologic 
features. In Figure \ref{fig:3}(b), for a tumor (TUM) tile, both HMU and NCT-trained models highlight broad, non-specific regions, missing critical cellular features necessary for precise tumor identification. However, the STARC-9-trained model effectively captures the full structural context of the tumor, aligning closely with the pathologist’s focus on tightly packed, hypercellular regions typical of tumor tissue. In the challenging case of a mixed tissue-type tile containing both tumor (TUM) and necrosis (Figure \ref{fig:3}(c)), the HMU-trained model incorrectly classifies the tile as containing necrosis (NCS) and the NCT-trained model correctly classifies it as tumor (TUM), but with poorly localized feature activations, indicating a less precise spatial understanding. In contrast, the model trained on STARC-9, which contains the complex, mixed tissue-type context often found in real-world WSI, accurately identifies the most clinically significant tumor regions. These feature map visualizations illustrate the high generalization capacity of models trained on STARC-9, further emphasizing the importance of diverse, high-quality training samples for robust, clinically relevant tissue classification. 

\textit{\textbf{Tissue map visualization}}: Figure \ref{fig:8} in Technical Appendices Section \ref{sec:prediction_maps} shows tissue segmentation maps generated by remapping the tile-level classifications from models trained on STARC-9, NCT, and HMU back onto their respective WSI. This approach provides a quick, intuitive overview of WSI-level tissue composition for pathologist verification. For the sample regions highlighted for normal mucosa (NOR), necrosis (NCS), tumor (TUM), muscle (MUS), lymphoid tissue (LYM), mucin (MUC), and adipose tissue (ADI) in Figure \ref{fig:8}(a), the model trained on STARC-9 consistently produced more accurate and contextually relevant predictions (Figure \ref{fig:8}(d)), closely aligning with pathologist assessments. In contrast, models trained on NCT (Figure \ref{fig:8}(c)) and HMU (Figure \ref{fig:8}(b)) exhibited significant misclassification, particularly within challenging regions containing mixed tissue-type tiles. Notably, NCS classification was over 45\% and 90\% more accurate, when compared to the models trained on HMU and NCT, respectively. Additionally, blood-containing (BLD) regions, which were frequently misclassified as NCS by both the NCT and HMU-trained models, were correctly identified by the STARC-9-trained model (Figure \ref{fig:8}(e)). Furthermore, the STARC-9-trained model demonstrated significantly lower confusion (over 80\% error rate reduction) between the NOR, TUM, and MUC classes, compared to the models trained on HMU and NCT. Similarly, TUM regions, often misclassified as MUC by HMU and NCT (over a 30\% error rate), were better delineated by the STARC-9-trained model. While all three models performed consistently across simple tissue types such as LYM and ADI, the STARC-9-trained model achieved over 85\% accuracy on mixed tissue-type tiles, significantly outperforming the models trained on HMU (55\%) and NCT (42\%). 

\begin{figure}[!b]
	\centering
	\includegraphics[width=0.9\textwidth]{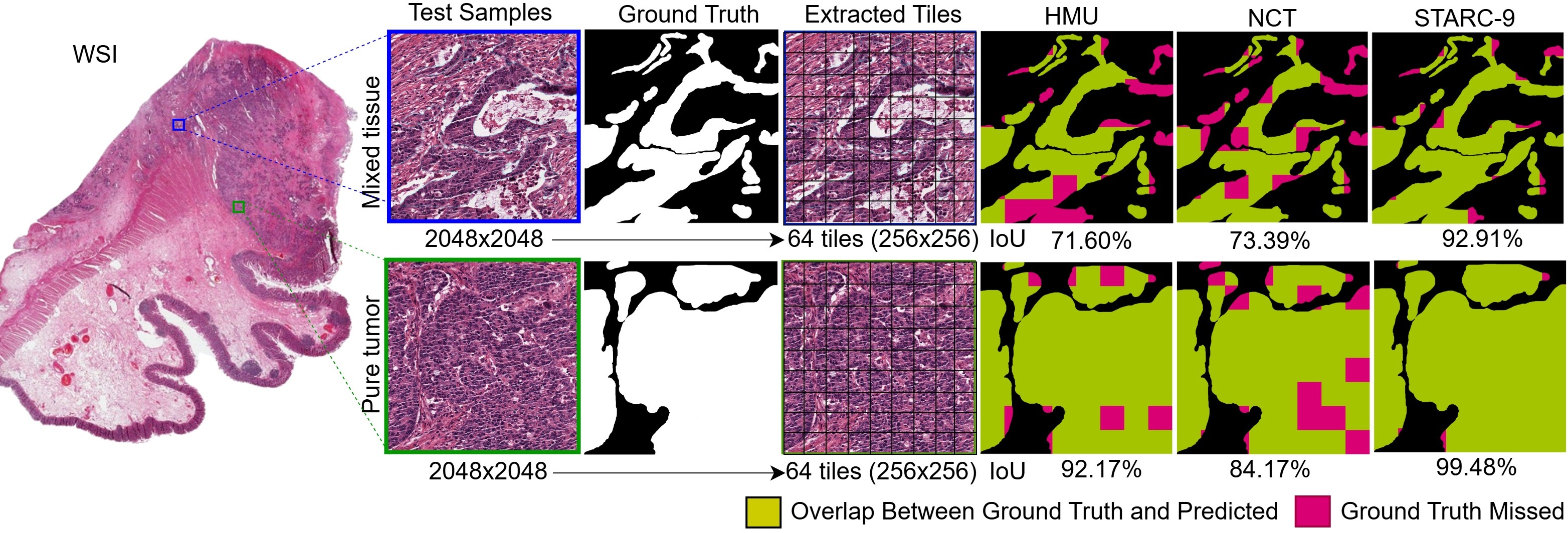}
	\caption{Tumor segmentation within 2048x2048 regions from a WSI from the CURATED-TCGA-CRC-HE-VAL dataset using tile-level classifiers trained on HMU, NCT, and STARC-9. }
	\label{fig:4}
\end{figure}  

\subsection{Tumor Tissue Segmentation}
Among the most common downstream applications for multi-class tissue classification is tissue segmentation, especially tumor region segmentation, which allows for the automated identification and cropping of tumor-containing regions for subsequent annotation, ROI selection, diagnosis, and prognostication. This approach also facilitates downstream applications such as MSI \cite{YAMASHITA2021132} and other biomarker status prediction, and supports survival modeling \cite{Kather2019} for risk stratification and personalized treatment planning. In this context, we conducted experiments to evaluate the effectiveness of models trained on HMU, NCT, and STARC-9 for tumor segmentation, focusing on their ability to accurately identify tumor regions that are important for clinical decision-making. As there were no publicly available CRC WSI repositories with readily usable, high‑quality TUM masks for a larger scale experiment, and existing weakly‑supervised tools did not provide the precision required for generating tissue segmentation masks, we prepared a test set by enlisting pathologists to manually annotate (ground truth) the TUM region in patches of size 2048×2048 pixels using QuPath \cite{Bankhead2017QuPath}. We selected 45 patches (3 per slide) from 15 Stanford WSI and 50 patches (2 per slide) from 25 TCGA-CRC WSI, which were fully independent of our training and validation sets, for a more controlled evaluation of model performance \cite{Bokhorst2023DeepLearning}. Some patches contained mixed tissue types in order to evaluate the effectiveness of the trained models. For example, as shown in Figure \ref{fig:4}, one region contained predominantly tumor, while the other included a mix of tumor and non-tumor tissue (NCS), providing more challenging segmentation.  

For segmentation evaluation, each 2048x2048 pixel region was divided into 64 non-overlapping 256x256 pixel tiles and normalized to match the input requirements of the trained models. Tile-level classification was then performed using the best-performing model trained on each dataset. If a tile was correctly classified as TUM, its location within the ground truth segmentation mask was highlighted in green, while misclassified TUM tiles were marked in red, as shown in Figure \ref{fig:4}. This approach allowed a direct visual comparison of each model’s ability to accurately identify tumor regions. We observed that the model trained on STARC-9 significantly outperformed those trained on NCT and HMU, achieving an Intersection-over-Union (IoU) score of 92.91\% for the mixed tissue-type sample, compared to 73.39\% for NCT and 71.6\% for HMU. This reflects the STARC-9-trained model's exceptional ability to capture fine-grained tissue features and effectively distinguish tumor regions, even within heterogeneous tissue contexts. For the pure tumor sample, the STARC-9-trained model also demonstrated higher performance, with a 99.48\% IoU, significantly surpassing that of the models trained on NCT (84.17\%) and HMU (92.17\%). These results emphasize the critical role of diverse, high-quality training samples in developing robust, clinically relevant tissue classification models, particularly for challenging segmentation tasks.

Table \ref{table3} reports IoU and Dice scores for tumor segmentation on these held-out sets. Models trained on STARC-9 achieved mean Dice scores of 90.47±8.14\% on the Stanford dataset and 89.38±9.14\% on the TCGA-CRC dataset, approximately 14\% and 17\% higher than those trained on NCT and HMU when evaluated on the Stanford dataset, and 35\% and 23\% higher when evaluated on the TCGA-CRC dataset, respectively. Moreover, STARC-9-trained models exhibited substantially narrower standard deviations in both IoU and Dice scores, demonstrating more consistent and robust tumor delineation across diverse samples.   

\begin{table} [t!]  
\tiny
\centering
\caption{Model evaluation for TUM segmentation on the Stanford and TCGA-CRC datasets.}
\label{table3}
\resizebox{\textwidth}{!}{%
\begin{tabular}{|l|ccc|ccc|}
\hline
\multicolumn{1}{|c|}{{\color[HTML]{000000} }}                                   & \multicolumn{3}{c|}{{\color[HTML]{000000} \textbf{IoU (\%)}}}                                                                                                & \multicolumn{3}{c|}{{\color[HTML]{000000} \textbf{Dice score (\%)}}}                                                                                              \\ \cline{2-7} 
\multicolumn{1}{|c|}{\multirow{-2}{*}{{\color[HTML]{000000} \textbf{Dataset}}}} & \multicolumn{1}{c|}{\textbf{NCT}}                         & \multicolumn{1}{c|}{\textbf{HMU}}                         & \textbf{STARC-9}                     & \multicolumn{1}{c|}{\textbf{NCT}}                         & \multicolumn{1}{c|}{\textbf{HMU}}                         & \textbf{STARC-9}                    \\ \hline
{\color[HTML]{000000} Stanford}                                                 & \multicolumn{1}{c|}{{\color[HTML]{000000} 67.19 ± 21.53}} & \multicolumn{1}{c|}{{\color[HTML]{000000} 64.68 ± 24.21}} & {\color[HTML]{000000} 89.33 ± 8.76}  & \multicolumn{1}{c|}{{\color[HTML]{000000} 78.20 ± 17.01}} & \multicolumn{1}{c|}{{\color[HTML]{000000} 75.49 ± 21.01}} & {\color[HTML]{000000} 90.47 ± 8.14} \\ \hline
{\color[HTML]{000000} TCGA-CRC}                                                 & \multicolumn{1}{c|}{{\color[HTML]{000000} 51.94 ± 37.94}} & \multicolumn{1}{c|}{{\color[HTML]{000000} 58.89 ± 29.42}} & {\color[HTML]{000000} 88.81 ± 10.90} & \multicolumn{1}{c|}{{\color[HTML]{000000} 58.90 ± 31.38}} & \multicolumn{1}{c|}{{\color[HTML]{000000} 68.85 ± 22.10}} & {\color[HTML]{000000} 89.38 ± 9.14} \\ \hline
\end{tabular}
}
\end{table} 

\section{Conclusion}
In this work, we introduce STARC-9, a large-scale, high-quality dataset for multi-class tissue classification for CRC histopathology. Comprising 630,000 non-overlapping high-resolution image tiles across nine clinically relevant tissue types, STARC-9 addresses critical limitations in existing datasets, including class imbalance, low tissue diversity, and low-quality tiles. We also present DeepCluster++, a flexible framework that combines unsupervised feature extraction, clustering, and equal-frequency binning to efficiently select diverse representative training examples from each WSI. Extensive benchmarking studies utilizing a wide range of deep learning models, including baseline CNNs, state-of-the-art transformers, pathology-specific foundation models, and custom deep learning models trained from scratch, demonstrate the superior classification performance of models trained on STARC-9 versus the publicly available NCT and HMU datasets, achieving over 98\% accuracy on various independent validation datasets. The STARC-9-trained model also exhibited higher tumor segmentation accuracy, effectively capturing fine-grained tumor features critical for diagnosis and risk stratification, highlighting the importance of high-quality, diverse training data in model development.

\textit{\textbf{Limitations and future scope}}: While STARC-9 contains extensive CRC tissue diversity across 9 tissue types, these may not exhaustively cover all potential tissue types found in CRC resections. Future work might focus on incorporating additional, more granular tissue classes, as well as expanding the dataset for multi-modal applications through the addition of large-scale image-caption pairs. Additionally, as STARC-9 is limited to CRC patients, its relevance for model validation for other cancer types not sharing similar tumor morphologies or background non-tumor tissue classes (for example, central nervous system tumors) remains to be explored. STARC-9 reflects local demographics, with limited Black and Native American representation. While race may not affect tissue morphology, broader inclusion is vital for fair, generalizable models. Lastly, we acknowledge that our dataset originates from a single institution and emphasize the need for future extensions incorporating multi-institutional data to enhance diversity and ensure fairness in downstream biomedical AI models.	

\textit{\textbf{Acknowledgments}}: Funding for this study was provided by the United States National Cancer Institute (NCI), National Institutes of Health (NIH) (R01 CA270437).

\bibliography{neurips_2025}

@article {Morgan338,
	author = {Morgan, Eileen and Arnold, Melina and Gini, A and Lorenzoni, V and Cabasag, C J and Laversanne, Mathieu and Vignat, Jerome and Ferlay, Jacques and Murphy, Neil and Bray, Freddie},
	title = {Global burden of colorectal cancer in 2020 and 2040: incidence and mortality estimates from GLOBOCAN},
	volume = {72},
	number = {2},
	pages = {338--344},
	year = {2023},
	doi = {10.1136/gutjnl-2022-327736},
	publisher = {BMJ Publishing Group}, 
	URL = {https://gut.bmj.com/content/72/2/338}, 
	journal = {Gut}
}

@article{QIMS135171,
	author = {Qi Ke and Wun-She Yap and Yee Kai Tee and Yan Chai Hum and Hua Zheng and Yu-Jian Gan},
	title = {Advanced deep learning for multi-class colorectal cancer histopathology: integrating transfer learning and ensemble methods},
	journal = {Quantitative Imaging in Medicine and Surgery},
	volume = {15},
	number = {3},
	year = {2025}, 
	issn = {2223-4306},	
    url = {https://qims.amegroups.org/article/view/135171}
}

@article{PAN2025106933,
title = {EL-CNN: An enhanced lightweight classification method for colorectal cancer histopathological images},
journal = {Biomedical Signal Processing and Control},
volume = {100},
pages = {106933},
year = {2025},
issn = {1746-8094},
doi = {https://doi.org/10.1016/j.bspc.2024.106933}, 
author = {Xing-Liang Pan and Bo Hua and Ke Tong and Xia Li and Jin-Long Luo and Hua Yang and Ju-Rong Ding} 
}

@article{Bokhorst2023DeepLearning,
  author    = {Bokhorst, J. M. and Nagtegaal, I. D. and Fraggetta, F. and others},
  title     = {Deep learning for multi-class semantic segmentation enables colorectal cancer detection and classification in digital pathology images},
  journal   = {Scientific Reports},
  year      = {2023},
  volume    = {13},
  pages     = {8398},
  doi       = {10.1038/s41598-023-35491-z},
  url       = {https://doi.org/10.1038/s41598-023-35491-z}
}

@article{JIAO2021106047,
title = {Deep learning-based tumor microenvironment analysis in colon adenocarcinoma histopathological whole-slide images},
journal = {Computer Methods and Programs in Biomedicine},
volume = {204},
pages = {106047},
year = {2021},
issn = {0169-2607},
doi = {https://doi.org/10.1016/j.cmpb.2021.106047}, 
author = {Yiping Jiao and Junhong Li and Chenqi Qian and Shumin Fei} 
}

@article{YAMASHITA2021132,
title = {Deep learning model for the prediction of microsatellite instability in colorectal cancer: a diagnostic study},
journal = {The Lancet Oncology},
volume = {22},
number = {1},
pages = {132-141},
year = {2021},
issn = {1470-2045},
doi = {https://doi.org/10.1016/S1470-2045(20)30535-0}, 
author = {Rikiya Yamashita and Jin Long and Teri Longacre and Lan Peng and Gerald Berry and Brock Martin and John Higgins and Daniel L Rubin and Jeanne Shen} 
}

@article{Kather2019,
    doi = {10.1371/journal.pmed.1002730},
    author = {Kather, Jakob Nikolas AND Krisam, Johannes AND Charoentong, Pornpimol AND Luedde, Tom AND Herpel, Esther AND Weis, Cleo-Aron AND Gaiser, Timo AND Marx, Alexander AND Valous, Nektarios A. AND Ferber, Dyke AND Jansen, Lina AND Reyes-Aldasoro, Constantino Carlos AND Zörnig, Inka AND Jäger, Dirk AND Brenner, Hermann AND Chang-Claude, Jenny AND Hoffmeister, Michael AND Halama, Niels},
    journal = {PLOS Medicine},
    publisher = {Public Library of Science},
    title = {Predicting survival from colorectal cancer histology slides using deep learning: A retrospective multicenter study},
    year = {2019},
    month = {01},
    volume = {16}, 
    pages = {1-22},
    number = {1}
}

@article{Lou2025GastricCancer,
  author    = {Lou, S. and Ji, J. and Li, H. and others},
  title     = {A large histological images dataset of gastric cancer with tumour microenvironment annotation for AI},
  journal   = {Scientific Data},
  year      = {2025},
  volume    = {12},
  pages     = {138},
  doi       = {10.1038/s41597-025-04489-9}
}

@article{TCGANetwork2012ColonRectal,
  author    = {{The Cancer Genome Atlas Network}},
  title     = {Comprehensive molecular characterization of human colon and rectal cancer},
  journal   = {Nature},
  year      = {2012},
  volume    = {487},
  pages     = {330--337},
  doi       = {10.1038/nature11252} 
}

@article{Tani2025,
author = {Tani, Ishrat Zahan and Goh, Kah Ong Michael and Islam, Md Nazmul and Aziz, Md Tarek and Mahmud, S. M. Hasan and Nandi, Dip},
title = {Addressing Label Noise in Colorectal Cancer Classification Using Cross-Entropy Loss and pLOF Methods With Stacking-Ensemble Technique},
journal = {Applied Computational Intelligence and Soft Computing},
volume = {2025},
number = {1},
pages = {6552580}, 
doi = {https://doi.org/10.1155/acis/6552580},  
year = {2025}
}

@InProceedings{Ignatov2025,
author="Ignatov, Andrey and Malivenko, Grigory",  
title="NCT-CRC-HE: Not All Histopathological Datasets are Equally Useful",
booktitle="Computer Vision -- ECCV 2024 Workshops",
year="2025",
publisher="Springer Nature Switzerland", 
pages="300--317"  
}

@article{Kheiri2025BiasFactors,
  author    = {Kheiri, F. and Rahnamayan, S. and Makrehchi, M. and Asilian Bidgoli, Azam},
  title     = {Investigation on potential bias factors in histopathology datasets},
  journal   = {Scientific Reports},
  year      = {2025},
  volume    = {15},
  pages     = {11349},
  doi       = {10.1038/s41598-025-89210-x},
  url       = {https://doi.org/10.1038/s41598-025-89210-x}
}

@article{Bankhead2017QuPath,
  author    = {Bankhead, P. and Loughrey, M. B. and Fernández, J. A. and et al.},
  title     = {QuPath: Open source software for digital pathology image analysis},
  journal   = {Scientific Reports},
  year      = {2017},
  volume    = {7},
  pages     = {16878},
  doi       = {10.1038/s41598-017-17204-5},
  url       = {https://doi.org/10.1038/s41598-017-17204-5}
}

@article{Kather2016TextureAnalysis,
  author    = {Kather, Jakob Nikolas and Weis, Cleo-Aron and Bianconi, Francesco and Melchers, Susanne M. and Schad, Lothar R. and Gaiser, Timo and Marx, Alexander and Z{\"o}llner, Frank Gerrit},
  title     = {Multi-class texture analysis in colorectal cancer histology},
  journal   = {Scientific Reports},
  year      = {2016},
  volume    = {6},
  pages     = {27988},
  doi       = {10.1038/srep27988},
  url       = {https://doi.org/10.1038/srep27988}
}

@conference{bioimaging24,
author={Ardhendu Sekhar and Ravi Gupta and Amit Sethi},
title={Few-Shot Histopathology Image Classification: Evaluating State-of-the-Art Methods and Unveiling Performance Insights},
booktitle={Proceedings of the 17th International Joint Conference on Biomedical Engineering Systems and Technologies - BIOIMAGING},
year={2024},
pages={244-253},
publisher={SciTePress},
organization={INSTICC},
doi={10.5220/0012568000003657},
isbn={978-989-758-688-0},
issn={2184-4305}
}

@article{WANG2025107608,
title = {Prediction of colorectal cancer microsatellite instability and tumor mutational burden from histopathological images using multiple instance learning},
journal = {Biomedical Signal Processing and Control},
volume = {104},
pages = {107608},
year = {2025},
issn = {1746-8094},
doi = {https://doi.org/10.1016/j.bspc.2025.107608}, 
author = {Wenyan Wang and Wei Shi and Chuanqi Nie and Weipeng Xing and Hailong Yang and Feng Li and Jinyang Liu and Geng Tian and Bing Wang and Jialiang Yang}
}

@misc{Stettler_HistopathologyDatasets,
  author       = {Stettler, Marie (Duc)},
  title        = {Histopathology Datasets for Machine Learning},
  howpublished = {\url{https://github.com/marieduc/Histopathology-Datasets-for-Machine-Learning}},
  note         = {GitHub repository},
  year         = {n.d.}
}

@INPROCEEDINGS{Ismail2022,
  author={Ismail, Mohamed Tarek and Sharara, Hossam and Madkour, Kareem and Seddik, Karim},
  booktitle={2022 ACM/IEEE 4th Workshop on Machine Learning for CAD (MLCAD)}, 
  title={Autoencoder-Based Data Sampling for Machine Learning-Based Lithography Hotspot Detection}, 
  year={2022},
  volume={},
  number={},
  pages={91-96}, 
  doi={10.1109/MLCAD55463.2022.9900096}
}

@InProceedings{Caron_2018_ECCV,
author = {Caron, Mathilde and Bojanowski, Piotr and Joulin, Armand and Douze, Matthijs},
title = {Deep Clustering for Unsupervised Learning of Visual Features},
booktitle = {Proceedings of the European Conference on Computer Vision (ECCV)},
month = {September},
year = {2018}
}

@article{RIASATIAN2021102032,
title = {Fine-Tuning and training of densenet for histopathology image representation using TCGA diagnostic slides},
journal = {Medical Image Analysis},
volume = {70},
pages = {102032},
year = {2021},
issn = {1361-8415},
doi = {https://doi.org/10.1016/j.media.2021.102032}, 
author = {Abtin Riasatian and Morteza Babaie and Danial Maleki and Shivam Kalra and Mojtaba Valipour and Sobhan Hemati and Manit Zaveri and Amir Safarpoor and Sobhan Shafiei and Mehdi Afshari and Maral Rasoolijaberi and Milad Sikaroudi and Mohd Adnan and Sultaan Shah and Charles Choi and Savvas Damaskinos and Clinton JV Campbell and Phedias Diamandis and Liron Pantanowitz and Hany Kashani and Ali Ghodsi and H.R. Tizhoosh}
}

@INPROCEEDINGS{Macenko2009,
  author={Macenko, Marc and Niethammer, Marc and Marron, J. S. and Borland, David and Woosley, John T. and Xiaojun Guan and Schmitt, Charles and Thomas, Nancy E.},
  booktitle={2009 IEEE International Symposium on Biomedical Imaging: From Nano to Macro}, 
  title={A method for normalizing histology slides for quantitative analysis}, 
  year={2009},
  volume={},
  number={},
  pages={1107-1110}, 
  doi={10.1109/ISBI.2009.5193250}
}

@misc{Stanford_Carina,
  author       = {{Stanford University}},
  title        = {Carina Computing Platform},
  howpublished = {\url{https://carina.stanford.edu}},
  note         = {Accessed: 2025-10-15},
  year         = {n.d.}
}

@misc{kapfer_2025_14751899,
  author       = {Kapfer, Craig and Stine, Kurt and Narasimhan, Balasubramanian and Mentzel, Christopher and Candes, Emmanuel},
  title        = {Marlowe: Stanford's GPU-based Computational Instrument },
  month        = jan,
  year         = 2025,
  publisher    = {Zenodo},
  version      = {0.1},
  doi          = {10.5281/zenodo.14751899}
}

@inproceedings{He2016DeepResidual,
  author    = {He, Kaiming and Zhang, Xiangyu and Ren, Shaoqing and Sun, Jian},
  title     = {Deep Residual Learning for Image Recognition},
  booktitle = {Proceedings of the IEEE Conference on Computer Vision and Pattern Recognition (CVPR)},
  year      = {2016},
  pages     = {770--778},
  doi       = {10.1109/CVPR.2016.90} 
}

@inproceedings{Tan2019EfficientNet,
  author    = {Tan, Mingxing and Le, Quoc V.},
  title     = {EfficientNet: Rethinking Model Scaling for Convolutional Neural Networks},
  booktitle = {Proceedings of the 36th International Conference on Machine Learning (ICML) / PMLR 97},
  year      = {2019},
  pages     = {6105--6114},
  url       = {https://proceedings.mlr.press/v97/tan19a/tan19a.pdf}
}

@inproceedings{Dosovitskiy2021ViT,
  author    = {Dosovitskiy, Alexey and Beyer, Lucas and Kolesnikov, Alexander and Weissenborn, Dirk and Zhai, Xiaohua and Unterthiner, Thomas and Dehghani, Mostafa and Minderer, Matthias and Heigold, Georg and Gelly, Sylvain and Uszkoreit, Jakob and Houlsby, Neil},
  title     = {An Image is Worth 16×16 Words: Transformers for Image Recognition at Scale},
  booktitle = {International Conference on Learning Representations (ICLR)},
  year      = {2021}, 
  doi       = {10.48550/arXiv.2010.11929}, 
  archivePrefix = {arXiv} 
}

@InProceedings{pmlr-v139-touvron21a,
  title = 	 {Training data-efficient image transformers and distillation through attention},
  author =       {Touvron, Hugo and Cord, Matthieu and Douze, Matthijs and Massa, Francisco and Sablayrolles, Alexandre and Jegou, Herve},
  booktitle = 	 {Proceedings of the 38th International Conference on Machine Learning},
  pages = 	 {10347--10357},
  year = 	 {2021}, 
  volume = 	 {139},
  series = 	 {Proceedings of Machine Learning Research},
  month = 	 {18--24 Jul},
  publisher =    {PMLR}
}

@inproceedings{Liu2022SwinTransformer,
  author    = {Liu, Ze and Lin, Yutong and Cao, Yue and Hu, Han and Wei, Yixuan and Zhang, Zheng and Lin, Stephen and Guo, Baining},
  title     = {Swin Transformer: Hierarchical Vision Transformer using Shifted Windows},
  booktitle = {Proceedings of the IEEE/CVF International Conference on Computer Vision (ICCV)},
  year      = {2022},
  pages     = {9992--10002},
  doi       = {10.1109/ICCV48234.2021.00981} 
}

@inproceedings{Liu2022ConvNet2020s,
  author    = {Liu, Zhuang and Mao, Hanzi and Wu, Chao-Yuan and Feichtenhofer, Christoph and Darrell, Trevor and Xie, Saining},
  title     = {A ConvNet for the 2020s},
  booktitle = {Proceedings of the IEEE/CVF Conference on Computer Vision and Pattern Recognition (CVPR)},
  year      = {2022},
  pages     = {11976--11986},
  doi       = {10.1109/CVPR52688.2022.01167} 
}

@article{WANG2022102559,
title = {Transformer-based unsupervised contrastive learning for histopathological image classification},
journal = {Medical Image Analysis},
volume = {81},
pages = {102559},
year = {2022},
issn = {1361-8415},
doi = {https://doi.org/10.1016/j.media.2022.102559}, 
author = {Xiyue Wang and Sen Yang and Jun Zhang and Minghui Wang and Jing Zhang and Wei Yang and Junzhou Huang and Xiao Han}
}

@inproceedings{Chen2022ScalingViT,
  author    = {Chen, Richard J. and Chen, Chengkuan and Li, Yicong and Chen, Tiffany Y. and Trister, Andrew D. and Krishnan, Rahul G. and Mahmood, Faisal},
  title     = {Scaling Vision Transformers to Gigapixel Images via Hierarchical Self-Supervised Learning},
  booktitle = {Proceedings of the IEEE/CVF Conference on Computer Vision and Pattern Recognition (CVPR)},
  year      = {2022},
  month     = {June},
  pages     = {16144--16155},
  doi       = {10.1109/CVPR52688.2022.01590} 
}

@article{Xu2024WholeSlideFoundationModel,
  author    = {Xu, H. and Usuyama, N. and Bagga, J. and et al.},
  title     = {A whole-slide foundation model for digital pathology from real-world data},
  journal   = {Nature},
  year      = {2024},
  volume    = {630},
  pages     = {181--188},
  doi       = {10.1038/s41586-024-07441-w} 
}

@inproceedings{Alfasly2024RotationAgnostic,
  author    = {Alfasly, Saghir and Shafique, Abubakr and Nejat, Peyman and Khan, Jibran and Alsaafin, Areej and Alabtah, Ghazal and Tizhoosh, H.\,R.},
  title     = {Rotation-Agnostic Image Representation Learning for Digital Pathology},
  booktitle = {Proceedings of the IEEE/CVF Conference on Computer Vision and Pattern Recognition (CVPR)},
  year      = {2024},  
  doi       = {10.1109/CVPR60427.2024.00640} 
}

@article{Lu2024VisualLanguageCP,
  author    = {Lu, Ming Y. and Chen, Bowen and Williamson, Drew F. K. and Chen, Richard J. and Liang, Ivy and Ding, Tong and Jaume, Guillaume and Odintsov, Igor and Le, Long Phi and Gerber, Georg and Parwani, Anil V. and Zhang, Andrew and Mahmood, Faisal},
  title     = {A visual-language foundation model for computational pathology},
  journal   = {Nature Medicine},
  year      = {2024},
  volume    = {30},
  pages     = {863--874},
  doi       = {10.1038/s41591-024-02856-4} 
}

@article{Chen2024GeneralPurposeCP,
  author    = {Chen, Richard J. and Ding, Tong and Lu, Ming Y. and Williamson, Drew F. K. and Jaume, Guillaume and Song, Andrew H. and Chen, Bowen and Zhang, Andrew and Shao, Daniel and Shaban, Muhammad and Williams, Mane and Oldenburg, Lukas and Weishaupt, Luca L. and Wang, Judy J. and Vaidya, Anurag and Le, Long Phi and Gerber, Georg and Sahai, Sharifa and Williams, Walt and Mahmood, Faisal},
  title     = {Towards a general-purpose foundation model for computational pathology},
  journal   = {Nature Medicine},
  year      = {2024},
  volume    = {30},
  pages     = {850--862},
  doi       = {10.1038/s41591-024-02857-3} 
}

@article{Vorontsov2024FoundationModelCP,
  author    = {Vorontsov, Eugene and Bozkurt, Alican and Casson, Adam and Shaikovski, George and Zelechowski, Michal and Severson, Kristen and Zimmermann, Eric and Hall, James and Tenenholtz, Neil and Fusi, Nicolo and Yang, Ellen and Mathieu, Philippe and van Eck, Alexander and Lee, Donghun and Viret, Julian and Robert, Eric and Wang, Yi Kan and Kunz, Jeremy D. and Lee, Matthew C. H. and Bernhard, Jan H. and Godrich, Ran A. and Oakley, Gerard and Millar, Ewan and Hanna, Matthew and Wen, Hannah and Retamero, Juan A. and Moye, William A. and Yousfi, Razik and Kanan, Christopher and Klimstra, David S. and Rothrock, Brandon and Liu, Siqi and Fuchs, Thomas J.},
  title     = {A foundation model for clinical-grade computational pathology and rare cancers detection},
  journal   = {Nature Medicine},
  year      = {2024},
  volume    = {30},
  pages     = {2924--2935},
  doi       = {10.1038/s41591-024-03141-0} 
}

@INPROCEEDINGS {10678515,
author = { Nasiri-Sarvi, Ali and Trinh, Vincent Quoc-Huy and Rivaz, Hassan and Hosseini, Mahdi S. },
booktitle = { 2024 IEEE/CVF Conference on Computer Vision and Pattern Recognition Workshops (CVPRW) },
title = { Vim4Path: Self-Supervised Vision Mamba for Histopathology Images },
year = {2024}, 
pages = {6894-6903}, 
doi = {10.1109/CVPRW63382.2024.00683}, 
publisher = {IEEE Computer Society},
address = {Los Alamitos, CA, USA},
month =Jun}

@article{FU2024105826,
title = {Whole slide images classification model based on self-learning sampling},
journal = {Biomedical Signal Processing and Control},
volume = {90},
pages = {105826},
year = {2024},
issn = {1746-8094},
doi = {https://doi.org/10.1016/j.bspc.2023.105826}, 
author = {Zhibing Fu and Qingkui Chen and Mingming Wang and Chen Huang} 
}

@article{Pataki2022HunCRC,
  author    = {Pataki, B{\'a}lint {\'A}rmin and Olar, Alex and Ribli, Dezs{\H{o}} and Pesti, Adri{\'a}n and Kontsek, Endre and Gy{\"o}ngy{\"o}si, Benedek and Bilecz, {\'A}gnes and Kov{\'a}cs, Tekla and Kov{\'a}cs, Krist{\'o}f Attila and Kramer, Zs{\'o}fia and Kiss, Andr{\'a}s and Sz{\'o}cska, Mikl{\'o}s and Pollner, P{\'e}ter and Csabai, Istv{\'a}n},
  title     = {HunCRC: annotated pathological slides to enhance deep learning applications in colorectal cancer screening},
  journal   = {Scientific Data},
  year      = {2022},
  volume    = {9},
  pages     = {370},
  doi       = {10.1038/s41597-022-01450-y} 
}

@article{Koziarski2024DiagSet,
  author    = {Koziarski, Micha{\l} and Cyganek, Bogus{\l}aw and Niedziela, Przemys{\l}aw and Olborski, Bogus{\l}aw and Antosz, Zbigniew and {\.Z}ydak, Marcin and Kwolek, Bogdan and W{\k{a}}sowicz, Pawe{\l} and Buka{\l}a, Andrzej and Swad{\'z}ba, Jakub and Sitkowski, Piotr},
  title     = {DiagSet: a dataset for prostate cancer histopathological image classification},
  journal   = {Scientific Reports},
  year      = {2024},
  volume    = {14},
  pages     = {6780},
  doi       = {10.1038/s41598-024-52183-4} 
}

@article{Li2024ActiveLearning,
  author    = {Li, Xiongquan and Wang, Xukang and Chen, Xuhesheng and Lu, Yao and Fu, Hongpeng and Wu, Ying Cheng},
  title     = {Unlabeled data selection for active learning in image classification},
  journal   = {Scientific Reports},
  year      = {2024},
  volume    = {14},
  pages     = {424},
  doi       = {10.1038/s41598-023-50598-z} 
}

@dataset{Neto2024IMPWSI,
  author    = {Neto, P. C. and Montezuma, D. and Oliveira, S. P. and Oliveira, D. and Fraga, J. and Monteiro, A. and Ribeiro, L. and Gon{\c{c}}alves, S. and Reinhard, S. and Zlobec, I. and Pinto, I. M. and Cardoso, J. S.},
  title     = {IMP Whole-Slide Images of Colorectal Samples 2024},
  year      = {2024},
  publisher = {INESC TEC},
  doi       = {10.25747/FB1Q-J507},
  note      = {Data set}
}

@InProceedings{shi2023transnext,
  author    = {Dai Shi},
  title     = {TransNeXt: Robust Foveal Visual Perception for Vision Transformers},
  booktitle = {Proceedings of the IEEE/CVF Conference on Computer Vision and Pattern Recognition (CVPR)},
  month     = {June},
  year      = {2024},
  pages     = {17773-17783}
}

@INPROCEEDINGS{Lou11092403,
  author={Lou, Meng and Yu, Yizhou},
  booktitle={2025 IEEE/CVF Conference on Computer Vision and Pattern Recognition (CVPR)}, 
  title={OverLoCK: An Overview-first-Look-Closely-next ConvNet with Context-Mixing Dynamic Kernels}, 
  year={2025},
  volume={},
  number={},
  pages={128-138}, 
  doi={10.1109/CVPR52734.2025.00021}
}

@inproceedings{beit2022,
title={{BEiT}: {BERT} Pre-Training of Image Transformers},
author={Hangbo Bao and Li Dong and Songhao Piao and Furu Wei},
booktitle={International Conference on Learning Representations},
year={2022},
url={https://openreview.net/forum?id=p-BhZSz59o4}
}

@misc{sun2024pathgen16m16millionpathology,
      title={PathGen-1.6M: 1.6 Million Pathology Image-text Pairs Generation through Multi-agent Collaboration}, 
      author={Yuxuan Sun and Yunlong Zhang and Yixuan Si and Chenglu Zhu and Zhongyi Shui and Kai Zhang and Jingxiong Li and Xingheng Lyu and Tao Lin and Lin Yang},
      year={2024}, 
      archivePrefix={arXiv},
      primaryClass={cs.CV},
      url={https://arxiv.org/abs/2407.00203}, 
}

% --- Print the appendices table of contents AFTER references ---

\newpage

\listofappendices

\newpage
\appendix
\beginappendixtoc     % redirect section entries to .atoc

\section{Patient demographic details}\label{sec:demographic_details}
\begin{figure}[h!]
	\centering
	\includegraphics[width=1\textwidth]{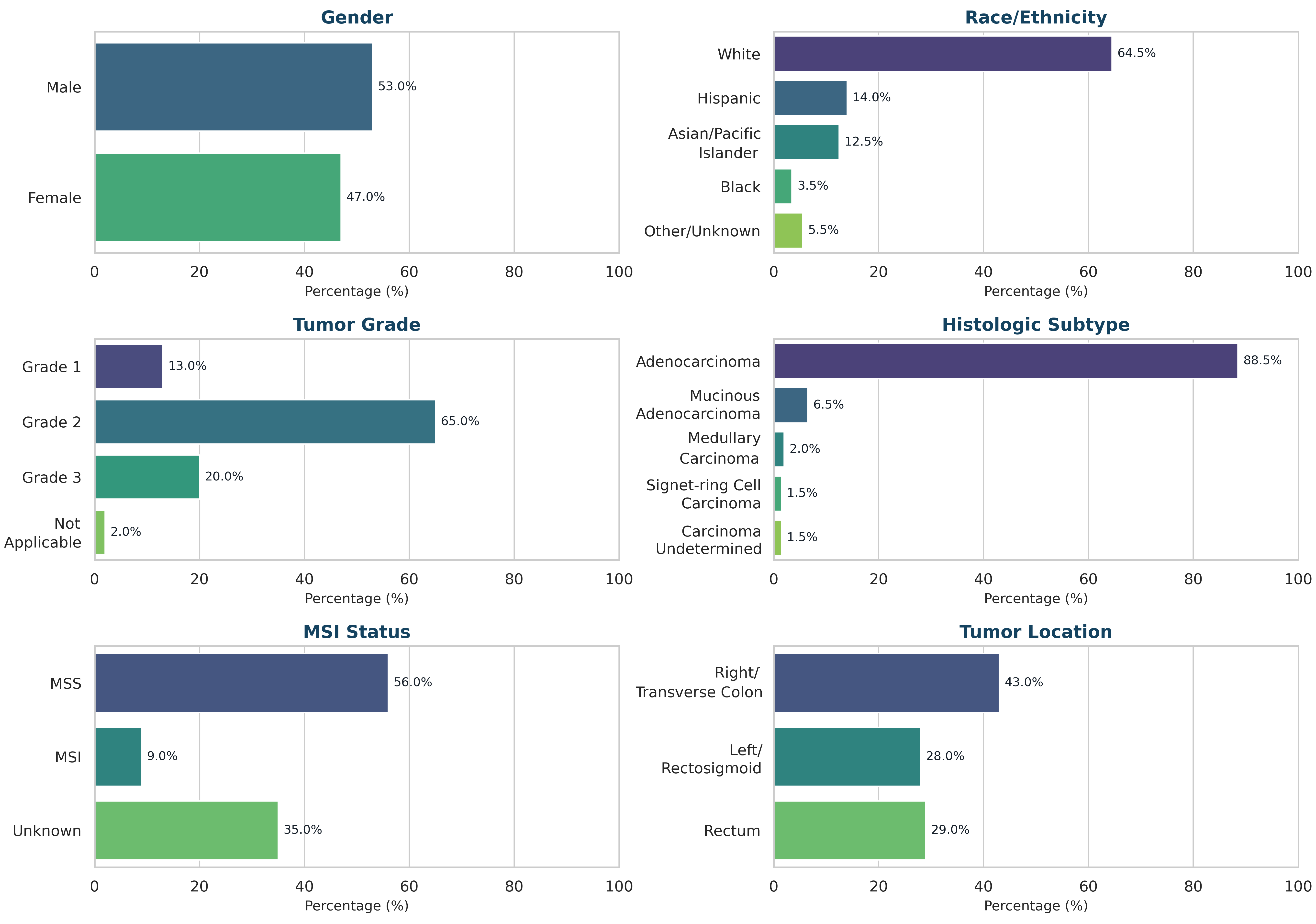}
	\caption{STARC-9 patient demographic details.}
	\label{fig:5}
\end{figure}

For the STARC-9 dataset, as shown in Figure \ref{fig:5}, 53\% of patients were male and 47\% female, with racial/ethnic distribution 64.5\% White, 14\% Hispanic, 12.5\% Asian/Pacific Islander, 3.5\% Black, and 5.5\% Other/Unknown. Age range: 23-97 yrs (mean 62.9 yrs, standard deviation 16 yrs). Tumor grade distribution: 13\% Grade 1 (n=26), 65\% Grade 2 (n=130), 20\% Grade 3 (n=4), and 2\% Grade Not Applicable (n=4, all medullary carcinomas). Histologic subtypes: 88.5\% (n=177) Adenocarcinoma, 6.5\% (n=13) Mucinous Adenocarcinoma, 2\% (n=4) Medullary Carcinoma, 1.5\% (n=3) Signet-ring Cell Carcinoma, and 1.5\% (n=3) Carcinoma, Type Undetermined. Regarding microsatellite instability status: 56\% (n=112) microsatellite stable (MSS), 9\% (n=18) microsatellite unstable (MSI), 35\% MSI status unknown. Tumor location: 43\% (n=86) Right/ transverse colon, 28\% (n=56) Left colon/splenic flexure/rectosigmoid, and 29\% (n=58) Rectum.

As WSI naturally vary in the amount and composition of tissue present on a slide, the per‑patient and even per‑tissue class tile counts were inherently imbalanced (for example, WSI with a high tumor volume tended to contain more NCS and MUC tiles). As this imbalanced tile distribution reflects the naturally diverse/heterogeneous tissue-type class distribution, it would be infeasible to balance the number of tiles of each tissue class per patient without running into the issues of (1) having an insufficient number of tiles from some patients and (2) needing to discard valid/informative tiles from some patients. Therefore, we did not seek to balance the number of tiles per patient.

\section{Advantages of using an AutoEncoder for feature extraction}\label{sec:advantagesof_auto_encoder}
In constructing a high-quality, diverse, and representative histopathology dataset, the choice of a feature extractor is critical. We chose a custom‑trained autoencoder (AE\_CRC) over off‑the‑shelf pathology foundation models for three main reasons:  

\textit{\textbf{Task-specific features}}: To understand the role of encoders in sample selection, we analyzed 9,000 samples across nine tissue types from STARC-9 (1,000 per class). After feature extraction, embeddings were reduced to 256 dimensions and clustered using K-means (400 samples per cluster). As shown in Figure \ref{ab_fig}, supervised encoders such as, ResNet50 (trained on natural images) (Figure \ref{ab_fig}(a)) and KimiaNet (trained on images from pathology WSI) (Figure \ref{ab_fig} (b)) exhibited scattered and 

\begin{figure}[htbp]
  \centering
  \begin{subfigure}[b]{1\textwidth}
    \centering
    \includegraphics[width=\linewidth]{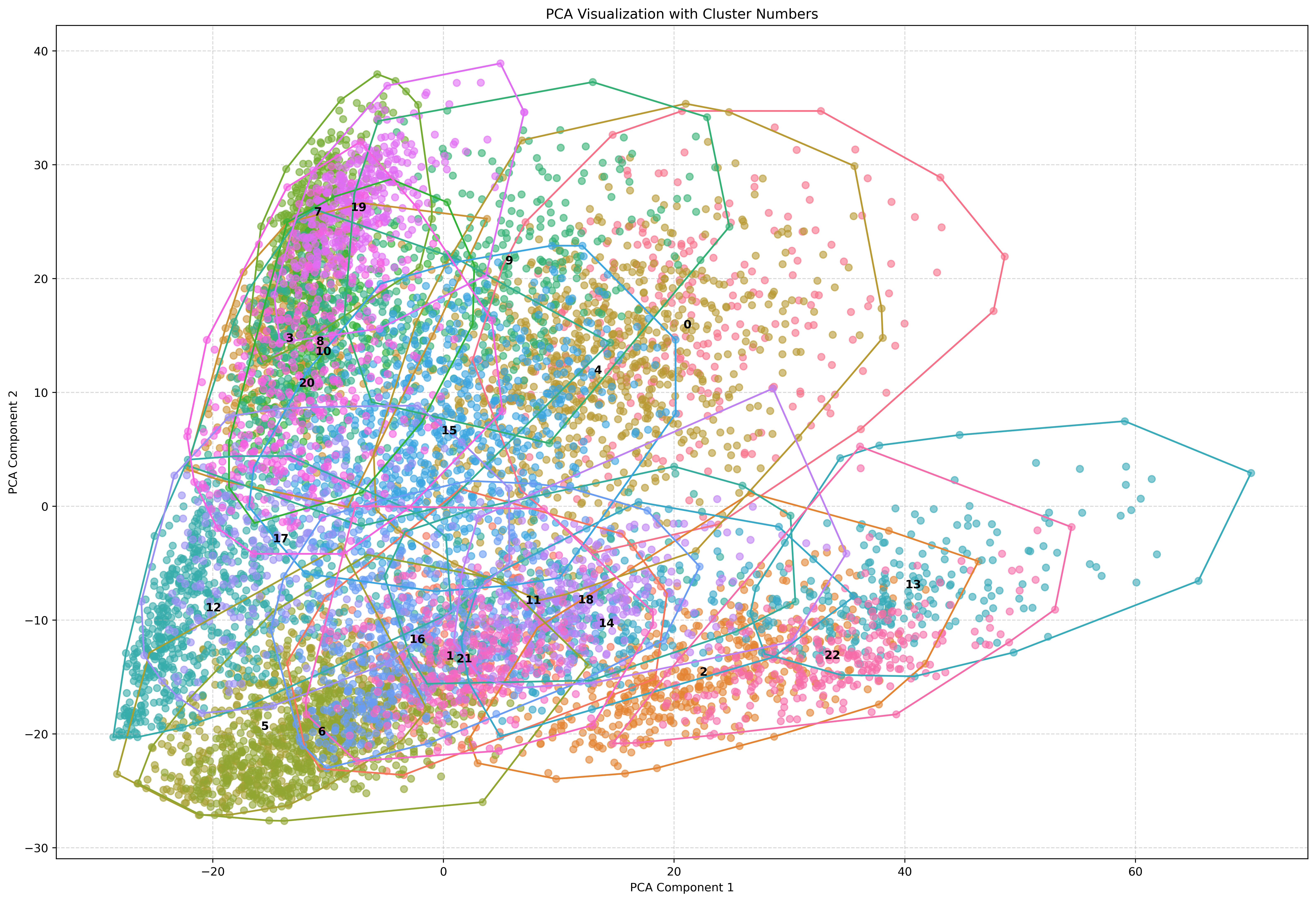}
    \caption{ResNet50}
    \label{ab_fig1}
  \end{subfigure}
  \hfill
  \begin{subfigure}[b]{1\textwidth}
    \centering
    \includegraphics[width=\linewidth]{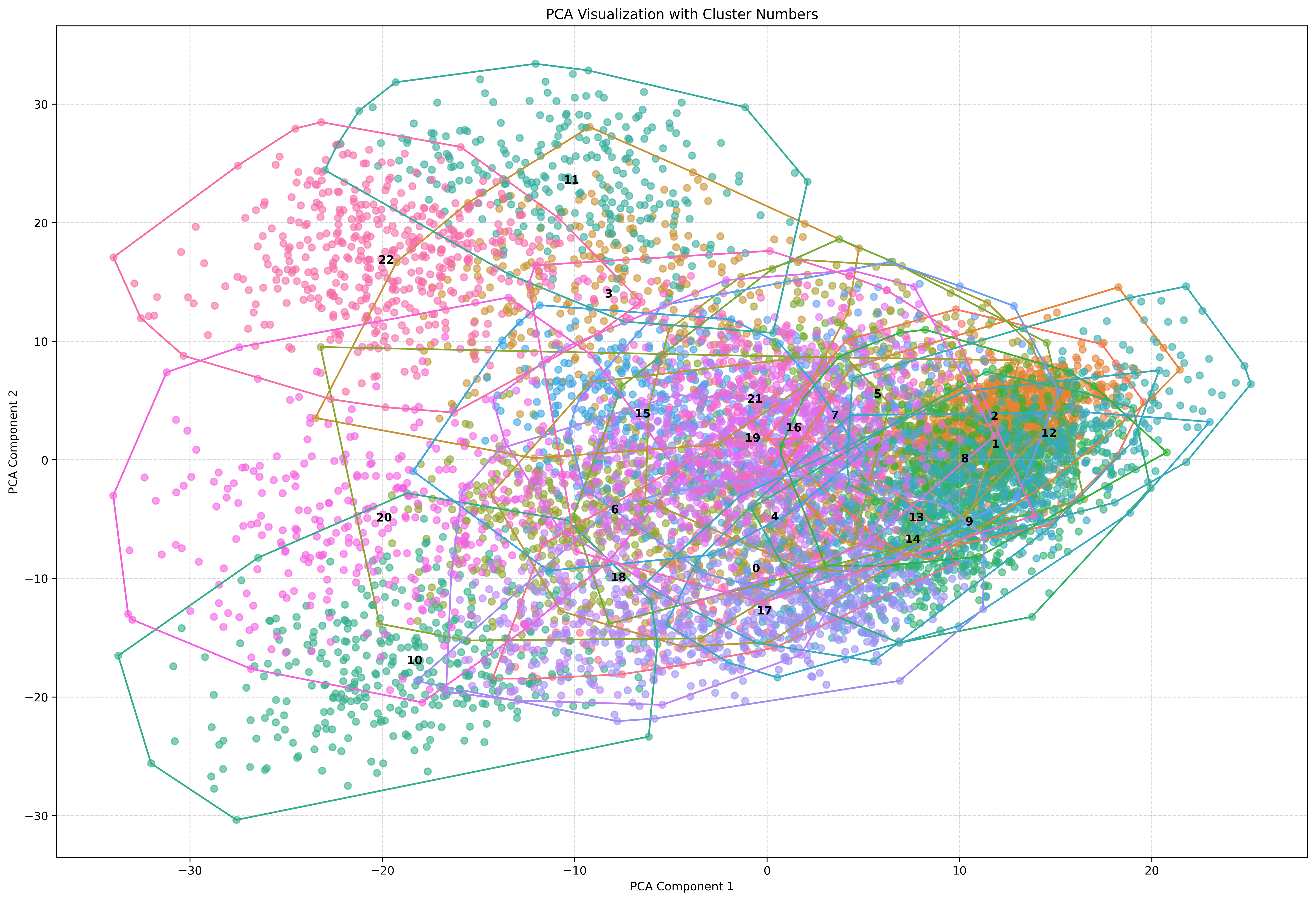}
    \caption{KimiaNet}
    \label{ab_fig2}
  \end{subfigure}
\end{figure} 
% Continued float: third subfigure on next page
\begin{figure}[htbp]\ContinuedFloat
  \centering
  \begin{subfigure}[b]{1\textwidth}
    \centering
    \includegraphics[width=\linewidth]{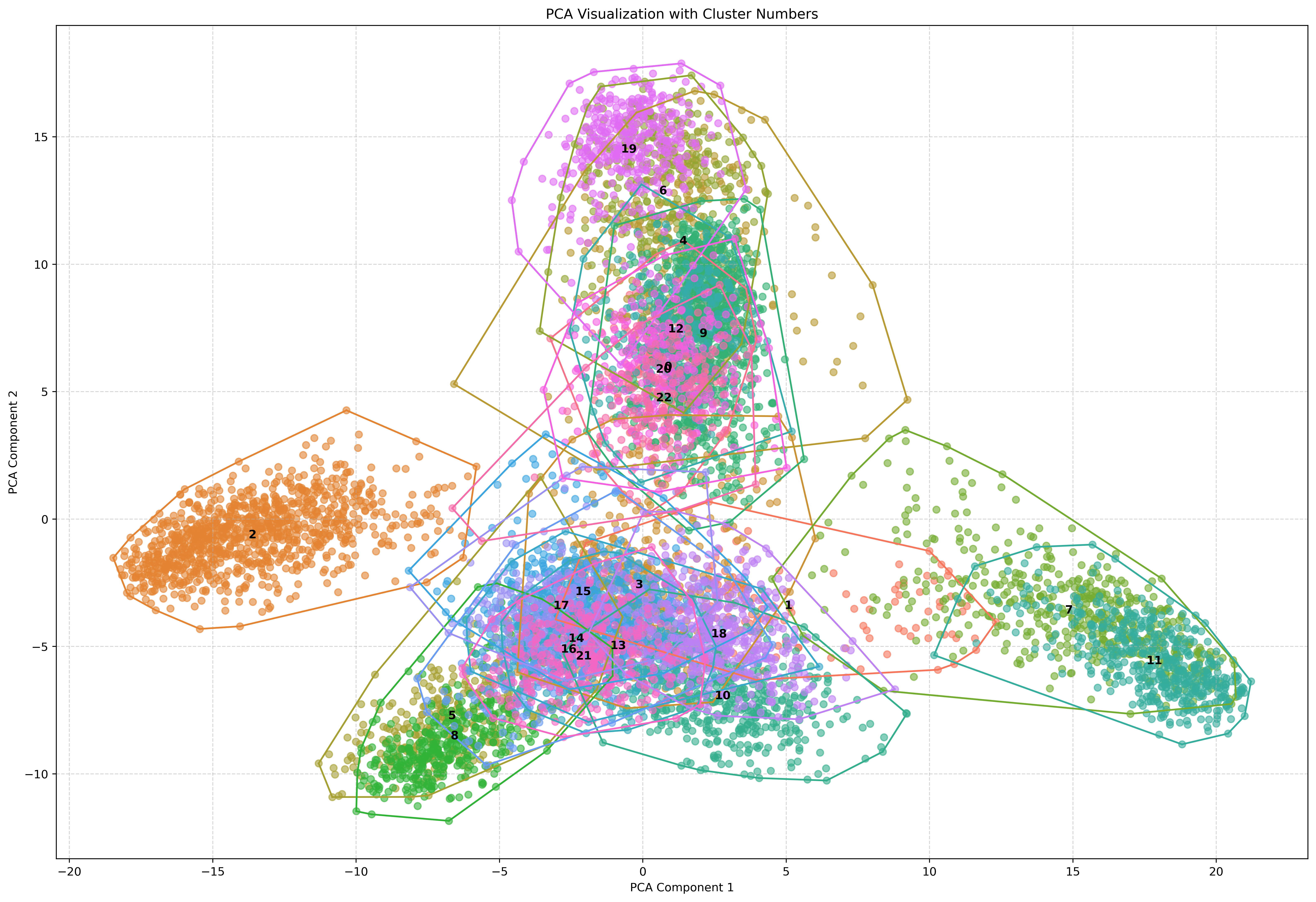}
    \caption{UNI}
    \label{ab_fig3}
  \end{subfigure}
    \hfill
  \begin{subfigure}[b]{1\textwidth}
    \centering
    \includegraphics[width=\linewidth]{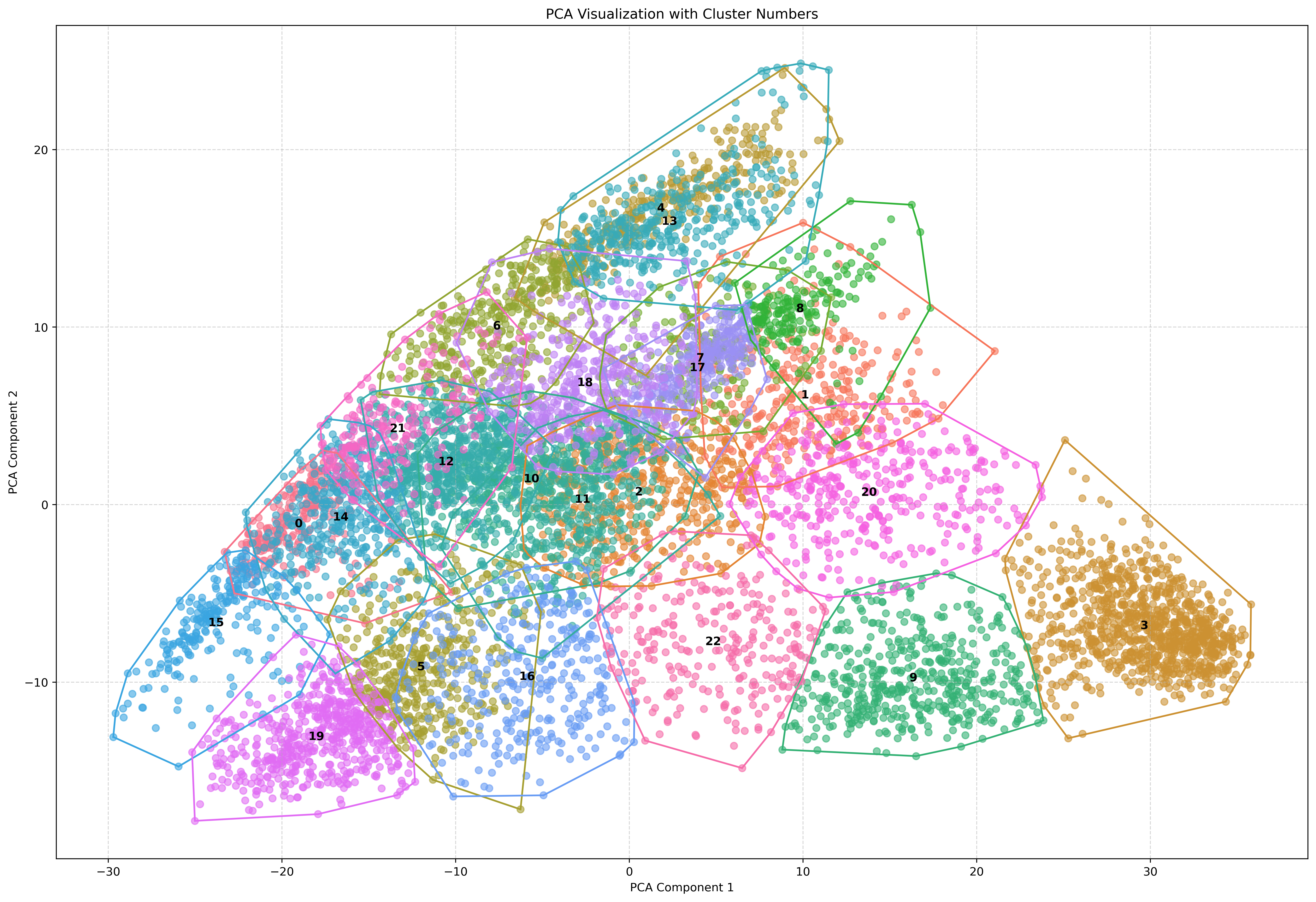}
    \caption{AutoEncoder}
    \label{ab_fig4}
  \end{subfigure}
      \caption{Feature visualization of different encoders (a) ResNet50, (b) KimiaNet, (c) UNI, and (d) encoder from custom trained AutoEncoder, for 9000 samples of nine different tissue types.}
      \label{ab_fig}
\end{figure}

overlapping representations among tissue types. Pathology foundation models are typically trained with classification or contrastive losses to distinguish major diagnostic categories. However, these discriminative objectives often fail to capture subtle intra-class variations essential for representing the full morphologic spectrum of tissue types. For example, despite UNI (Figure \ref{ab_fig}(c)) achieves strong self-supervised representations, they tend to over-separate biologically related tissues, reducing intra-class coherence.

These perform well on label-based separation but overlook fine-grained structural differences critical for interpretability and intra-class diversity. In contrast, our autoencoder-based encoder (Figure \ref{ab_fig4}(d)) learns structure-preserving, domain-specific representations by optimizing a reconstruction loss using the Structural Similarity Index (SSIM). This aligns clusters according to true morphologic proximity, preserving biologically meaningful relationships. This structure-preserving representation enables balanced sampling, helps exploring the surrounding clusters for similar tissue types, and supports the construction of morphologically diverse, clinically meaningful datasets such as STARC-9, where representational fidelity is critical for robust downstream model development.

\textit{\textbf{Domain‑specific sensitivity}}: By training exclusively on over 100,000 CRC tiles, AE\_CRC becomes finely attuned to colorectal (and other tubular gastrointestinal tract) histopathology. It learns to distinguish the fine-grained features present within tissue types such as NOR, MUC, TUM, and NCS. This is in contrast to existing pathology foundation models, which are trained on a wide variety of organs and tasks, causing these models to overlook the specific fine-grained features necessary for accurate representation within the latent space.

\textit{\textbf{Efficiency and scalability}}: Large foundation models (e.g., CTransPath, UNI, CONCH) require substantial GPU resources and slower embedding times. AE\_CRC, in contrast, is lightweight and fast to implement on standard hardware, making it practical for clustering 630,000 tiles in DeepCluster++ without incurring prohibitive compute costs.

In our experiments, clustering with AE\_CRC embeddings produced more coherent and morphology-driven groups compared to models trained on natural images (e.g., ImageNet) or contrastive learning-based pathology encoders. This allowed DeepCluster++ to effectively sample both prototypical and edge-case tiles, ensuring comprehensive coverage of histologic diversity. The reconstruction-based learning objective thus aligns well with the goal of building a large-scale, diverse histopathology benchmark dataset. However, when applying this framework to a different dataset, it may be necessary to retrain the AE\_CRC model on the target data before integration into DeepCluster++.

\section{SSIM vs MSE loss functions in AutoEncoder}\label{sec:msevsssim}
To ensure the AutoEncoder learned higher-level histologic and morphologic structures rather than low-level pixel statistics, we trained it using a structural similarity loss (SSIM). Compared to a model trained with mean squared error (MSE) loss, the SSIM-based AutoEncoder achieved significantly better reconstruction quality on the validation set STANFORD-CRC-HE-VAL-SMALL, showing lower pixel error (0.0012 vs 0.0015), higher SSIM (0.9262 vs 0.8863), and greater Peak Signal-to-Noise Ratio (PSNR) (32.48 dB vs 28.53 dB) on average. These metrics confirm high-fidelity reconstruction of complex tissue types such as necrosis (NCS) and tumor (TUM), as illustrated in Figure \ref{fig:2}(a). By explicitly optimizing for texture, contrast, and spatial structure, SSIM encourages the encoder to capture perceptually meaningful features rather than minimizing pixel-level intensity differences. This results in a latent space rich in fine-grained morphologic representations, leading to distinct and coherent clusters in DeepCluster++. In contrast, MSE-based models often blur subtle variations and reduce cluster purity by focusing on pixel accuracy rather than structural integrity. The observed reconstruction quality demonstrates that the AutoEncoder-based feature extractor surpasses task-specific and contrastive encoders in capturing diverse and biologically relevant tissue morphologies, thereby offering a more reliable foundation for clustering and representation learning.

\section{Additional experiments with recent models}\label{sec:additional_experiments}
We trained and evaluated recent models such as TransNeXt \cite{shi2023transnext}, OverLoCK \cite{Lou11092403}, and Beit-base \cite{beit2022} with masked image modeling on the NCT, HMU, and STARC-9 datasets. The comprehensive results are presented in Table \ref{table5}. While these recent models demonstrate competitive performance, they did not surpass our best-performing combinations: CTransPath on STARC-9, HiPT on HMU, and ViT-base on NCT (as reported in Table \ref{table1}. Notably, models trained on STARC-9 consistently achieve superior performance across all architectures, validating our dataset's quality and diversity.  

\begin{table} [t!]  
\small
\centering
\caption{Evaluation of additional models on the validation datasets.}
\label{table5} 
\begin{tabular}{|lccccccc|}
\hline
\multicolumn{1}{|l|}{{\color[HTML]{000000} }}                                 & \multicolumn{3}{c|}{{\color[HTML]{000000} \textbf{F1-Macro (\%)}}}                                                                                                                 & \multicolumn{3}{c|}{{\color[HTML]{000000} \textbf{Accuracy (\%)}}}                                                                                                                 &                                                                                  \\ \cline{2-7}
\multicolumn{1}{|l|}{\multirow{-2}{*}{{\color[HTML]{000000} \textbf{Model}}}} & \multicolumn{1}{c|}{{\color[HTML]{000000} \textbf{NCT}}} & \multicolumn{1}{c|}{{\color[HTML]{000000} \textbf{HMU}}} & \multicolumn{1}{c|}{{\color[HTML]{000000} \textbf{STARC-9}}} & \multicolumn{1}{c|}{{\color[HTML]{000000} \textbf{NCT}}} & \multicolumn{1}{c|}{{\color[HTML]{000000} \textbf{HMU}}} & \multicolumn{1}{c|}{{\color[HTML]{000000} \textbf{STARC-9}}} & \multirow{-2}{*}{\textbf{\begin{tabular}[c]{@{}c@{}}Model \\ size\end{tabular}}} \\ \hline
\multicolumn{8}{|c|}{{\color[HTML]{000000} \textbf{STANFORD-CRC-HE-VAL-SMALL}}}                                                                                                                                                                                                                                                                                                                                                                                                                                                            \\ \hline
\multicolumn{1}{|l|}{{\color[HTML]{000000} TransNeXt \cite{shi2023transnext}}}                        & \multicolumn{1}{c|}{{\color[HTML]{000000} 59.68}}        & \multicolumn{1}{c|}{{\color[HTML]{000000} 56.57}}        & \multicolumn{1}{c|}{{\color[HTML]{000000} 98.60}}            & \multicolumn{1}{c|}{{\color[HTML]{000000} 73.89}}        & \multicolumn{1}{c|}{{\color[HTML]{000000} 64.22}}        & \multicolumn{1}{c|}{{\color[HTML]{000000} 98.61}}            & {\color[HTML]{000000} 110M}                                                      \\ \hline
\multicolumn{1}{|l|}{{\color[HTML]{000000} OverLoCK \cite{Lou11092403}}}                         & \multicolumn{1}{c|}{{\color[HTML]{000000} 63.75}}        & \multicolumn{1}{c|}{{\color[HTML]{000000} 62.09}}        & \multicolumn{1}{c|}{{\color[HTML]{000000} 97.20}}            & \multicolumn{1}{c|}{{\color[HTML]{000000} 79.00}}        & \multicolumn{1}{c|}{{\color[HTML]{000000} 68.98}}        & \multicolumn{1}{c|}{{\color[HTML]{000000} 97.22}}            & {\color[HTML]{000000} 24.3M}                                                     \\ \hline
\multicolumn{1}{|l|}{{\color[HTML]{000000} Beit-base \cite{beit2022}}}                        & \multicolumn{1}{c|}{{\color[HTML]{000000} 59.95}}        & \multicolumn{1}{c|}{{\color[HTML]{000000} 76.48}}        & \multicolumn{1}{c|}{{\color[HTML]{000000} 98.17}}            & \multicolumn{1}{c|}{{\color[HTML]{000000} 74.08}}        & \multicolumn{1}{c|}{{\color[HTML]{000000} 86.17}}        & \multicolumn{1}{c|}{{\color[HTML]{000000} 98.20}}            & {\color[HTML]{000000} 86.5M}                                                     \\ \hline
\multicolumn{8}{|c|}{{\color[HTML]{000000} \textbf{STANFORD-CRC-HE-VAL-LARGE}}}                                                                                                                                                                                                                                                                                                                                                                                                                                                            \\ \hline
\multicolumn{1}{|l|}{{\color[HTML]{000000} TransNeXt \cite{shi2023transnext}}}                        & \multicolumn{1}{c|}{{\color[HTML]{000000} 59.62}}        & \multicolumn{1}{c|}{{\color[HTML]{000000} 55.05}}        & \multicolumn{1}{c|}{{\color[HTML]{000000} 98.58}}            & \multicolumn{1}{c|}{{\color[HTML]{000000} 73.55}}        & \multicolumn{1}{c|}{{\color[HTML]{000000} 62.78}}        & \multicolumn{1}{c|}{{\color[HTML]{000000} 95.68}}            & {\color[HTML]{000000} 110M}                                                      \\ \hline
\multicolumn{1}{|l|}{{\color[HTML]{000000} OverLoCK \cite{Lou11092403}}}                         & \multicolumn{1}{c|}{{\color[HTML]{000000} 63.69}}        & \multicolumn{1}{c|}{{\color[HTML]{000000} 55.56}}        & \multicolumn{1}{c|}{{\color[HTML]{000000} 98.85}}            & \multicolumn{1}{c|}{{\color[HTML]{000000} 79.29}}        & \multicolumn{1}{c|}{{\color[HTML]{000000} 64.09}}        & \multicolumn{1}{c|}{{\color[HTML]{000000} 98.87}}            & {\color[HTML]{000000} 24.3M}                                                     \\ \hline
\multicolumn{1}{|l|}{{\color[HTML]{000000} Beit-base \cite{beit2022}}}                        & \multicolumn{1}{c|}{{\color[HTML]{000000} 62.01}}        & \multicolumn{1}{c|}{{\color[HTML]{000000} 78.11}}        & \multicolumn{1}{c|}{{\color[HTML]{000000} 98.61}}            & \multicolumn{1}{c|}{{\color[HTML]{000000} 77.03}}        & \multicolumn{1}{c|}{{\color[HTML]{000000} 88.40}}        & \multicolumn{1}{c|}{{\color[HTML]{000000} 98.68}}            & {\color[HTML]{000000} 86.5M}                                                     \\ \hline
\multicolumn{8}{|c|}{{\color[HTML]{000000} \textbf{CURATED-TCGA-CRC-HE-VAL-20K}}}                                                                                                                                                                                                                                                                                                                                                                                                                                                          \\ \hline
\multicolumn{1}{|l|}{{\color[HTML]{000000} TransNeXt \cite{shi2023transnext}}}                        & \multicolumn{1}{c|}{{\color[HTML]{000000} 58.95}}        & \multicolumn{1}{c|}{{\color[HTML]{000000} 51.92}}        & \multicolumn{1}{c|}{{\color[HTML]{000000} 95.57}}            & \multicolumn{1}{c|}{{\color[HTML]{000000} 72.65}}        & \multicolumn{1}{c|}{{\color[HTML]{000000} 61.70}}        & \multicolumn{1}{c|}{{\color[HTML]{000000} 96.07}}            & {\color[HTML]{000000} 110M}                                                      \\ \hline
\multicolumn{1}{|l|}{{\color[HTML]{000000} OverLoCK \cite{Lou11092403}}}                         & \multicolumn{1}{c|}{{\color[HTML]{000000} 64.47}}        & \multicolumn{1}{c|}{{\color[HTML]{000000} 53.93}}        & \multicolumn{1}{c|}{{\color[HTML]{000000} 95.38}}            & \multicolumn{1}{c|}{{\color[HTML]{000000} 80.25}}        & \multicolumn{1}{c|}{{\color[HTML]{000000} 62.98}}        & \multicolumn{1}{c|}{{\color[HTML]{000000} 95.56}}            & {\color[HTML]{000000} 24.3M}                                                     \\ \hline
\multicolumn{1}{|l|}{{\color[HTML]{000000} Beit-base \cite{beit2022}}}                        & \multicolumn{1}{c|}{{\color[HTML]{000000} 57.45}}        & \multicolumn{1}{c|}{{\color[HTML]{000000} 72.57}}        & \multicolumn{1}{c|}{{\color[HTML]{000000} 97.93}}            & \multicolumn{1}{c|}{{\color[HTML]{000000} 71.75}}        & \multicolumn{1}{c|}{{\color[HTML]{000000} 82.11}}        & \multicolumn{1}{c|}{{\color[HTML]{000000} 97.87}}            & {\color[HTML]{000000} 86.5M}                                                     \\ \hline
\end{tabular}
\end{table} 

Regarding contrastive learning, we have extensively evaluated several state-of-the-art pathology-specific foundation models (CTransPath, UNI, CONCH, Virchow, etc.) that employ contrastive learning and are pre-trained on histopathologic images, as comprehensively discussed in Section \ref{subsec:classification}. These experiments demonstrate that our STARC-9 dataset enables competitive performance even with the latest architectural advances.   

\begin{figure}[b!]
	\centering
	\includegraphics[width=0.75\textwidth]{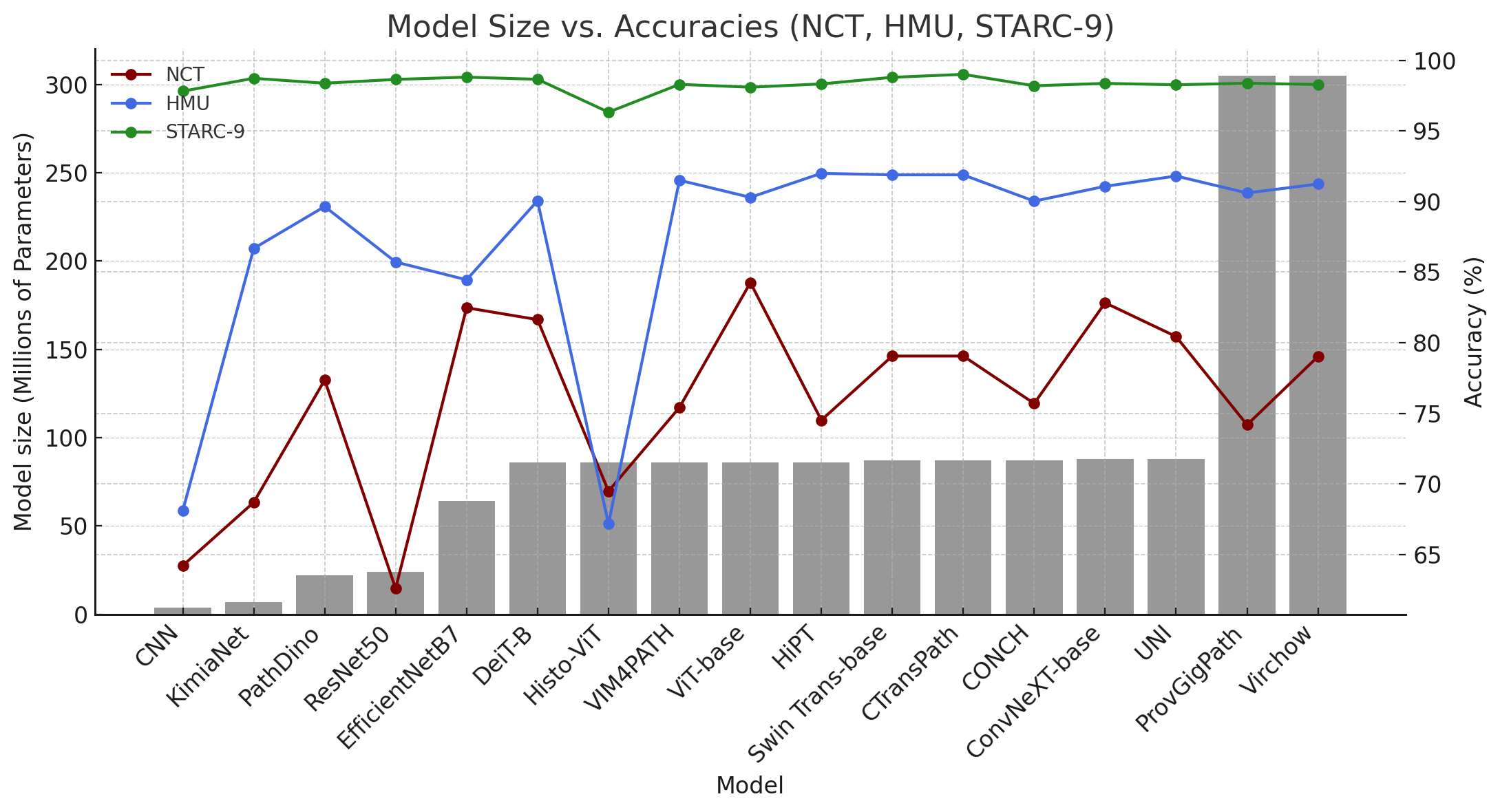}
	\caption{Relationship between model size and dataset-specific performance. }
	\label{fig:6}
\end{figure}  

\section{Relationship between model size and dataset-specific performance}\label{sec:model_size_vs_performance}
Across the NCT, HMU, and STARC-9 datasets, the relationship between model size and classification accuracy did not follow a simple linear trend, as shown in Figure \ref{fig:6}. Instead, model performance appeared to depend far more on architectural design and domain alignment than on the number of parameters. For instance, when trained on the STARC-9 dataset, CTransPath achieved the highest accuracy (99\%), despite having only about 87 million parameters, which is considerably smaller than models such as ProvGigPath or Virchow (both 305 million parameters), as shown in Table \ref{table6}. This result suggests that histopathology-specific pretraining and architectural efficiency enable CTransPath to capture subtle morphologic cues better than very large, general-purpose models that risk overfitting given the moderate dataset size of STARC-9.

When trained on the HMU dataset, HiPT outperformed all other architectures, with an accuracy of 91.99\%. Like CTransPath, HiPT belongs to the class of medium-sized transformer models (around 86 million parameters). Its hierarchical patch-embedding design effectively integrates local and global contextual features, which seems particularly beneficial for tissue patterns requiring multiscale spatial reasoning. Larger models such as ProvGigPath and Virchow again offered no significant performance improvement, implying diminishing returns once models exceed roughly 100 million parameters. The HMU dataset thus appears best served by architectures that balance representational power with generalization capacity rather than raw scale.

For the NCT dataset, ViT-base achieved the best performance (84.25\%) among all evaluated models. Although it shares a similar parameter range with HiPT and CTransPath, ViT’s pure attention mechanism captures patch-level variations and color normalization differences characteristic of the NCT slides. In contrast, smaller CNN-based models (e.g., a 3.9-million-parameter CNN) underperformed due to limited capacity for modeling long-range dependencies, while much larger networks did not yield further gains. This reinforces that the optimal model capacity for histopathology datasets often lies within a moderate range where sufficient complexity is achieved without overfitting risk.

\begin{table} [t!]  
\small
\centering
\caption{Model size and performance (bold-face denotes highest performance on each dataset).}
\label{table6} 
\begin{tabular}{|l|c|c|c|c|}
\hline
\multicolumn{1}{|c|}{{\color[HTML]{000000} \textbf{Model}}} & {\color[HTML]{000000} \textbf{No. of params.}} & {\color[HTML]{000000} \textbf{NCT}} & {\color[HTML]{000000} \textbf{HMU}} & {\color[HTML]{000000} \textbf{STARC-9}} \\ \hline
{\color[HTML]{000000} CNN}                                  & {\color[HTML]{000000} 3.9 M}                   & {\color[HTML]{000000} 64.21}        & {\color[HTML]{000000} 68.1}         & {\color[HTML]{000000} 97.81}            \\ \hline
{\color[HTML]{000000} KimiaNet}                    & {\color[HTML]{000000} 7M}                      & {\color[HTML]{000000} 68.69}        & {\color[HTML]{000000} 86.67}        & {\color[HTML]{000000} 98.72}            \\ \hline
{\color[HTML]{000000} PathDino }                    & {\color[HTML]{000000} 22 M}                    & {\color[HTML]{000000} 77.35}        & {\color[HTML]{000000} 89.64}        & {\color[HTML]{000000} 98.37}            \\ \hline
{\color[HTML]{000000} ResNet50}                    & {\color[HTML]{000000} 24 M}                    & {\color[HTML]{000000} 62.59}        & {\color[HTML]{000000} 85.71}        & {\color[HTML]{000000} 98.64}            \\ \hline
{\color[HTML]{000000} EfficientNet-B7}              & {\color[HTML]{000000} 64 M}                    & {\color[HTML]{000000} 82.47}        & {\color[HTML]{000000} 84.45}        & {\color[HTML]{000000} 98.8}             \\ \hline
{\color[HTML]{000000} ViT-base}                    & {\color[HTML]{000000} 86 M}                    & {\color[HTML]{000000} \textbf{84.25}}        & {\color[HTML]{000000} 90.29}        & {\color[HTML]{000000} 98.09}            \\ \hline
{\color[HTML]{000000} HiPT}                        & {\color[HTML]{000000} 86 M}                    & {\color[HTML]{000000} 74.51}        & {\color[HTML]{000000} \textbf{91.99}}        & {\color[HTML]{000000} 98.32}            \\ \hline
{\color[HTML]{000000} Histo-ViT}                            & {\color[HTML]{000000} 86 M}                    & {\color[HTML]{000000} 69.48}        & {\color[HTML]{000000} 67.16}        & {\color[HTML]{000000} 96.32}            \\ \hline
{\color[HTML]{000000} VIM4PATH}                    & {\color[HTML]{000000} 86 M}                    & {\color[HTML]{000000} 75.41}        & {\color[HTML]{000000} 91.5}         & {\color[HTML]{000000} 98.29}            \\ \hline
{\color[HTML]{000000} DeiT-B}                      & {\color[HTML]{000000} 86 M}                    & {\color[HTML]{000000} 81.63}        & {\color[HTML]{000000} 90.05}        & {\color[HTML]{000000} 98.65}            \\ \hline
{\color[HTML]{000000} Swin Trans-base}             & {\color[HTML]{000000} 87 M}                    & {\color[HTML]{000000} 79.05}        & {\color[HTML]{000000} 91.88}        & {\color[HTML]{000000} 98.79}            \\ \hline
{\color[HTML]{000000} CTransPath}                  & {\color[HTML]{000000} 87 M}                    & {\color[HTML]{000000} 79.05}        & {\color[HTML]{000000} 91.88}        & {\color[HTML]{000000} \textbf{99}}               \\ \hline
{\color[HTML]{000000} CONCH}                       & {\color[HTML]{000000} 87 M}                    & {\color[HTML]{000000} 75.69}        & {\color[HTML]{000000} 90.02}        & {\color[HTML]{000000} 98.19}            \\ \hline
{\color[HTML]{000000} ConvNeXT-base}               & {\color[HTML]{000000} 88 M}                    & {\color[HTML]{000000} 82.82}        & {\color[HTML]{000000} 91.07}        & {\color[HTML]{000000} 98.36}            \\ \hline
{\color[HTML]{000000} UNI}                         & {\color[HTML]{000000} 88 M}                    & {\color[HTML]{000000} 80.43}        & {\color[HTML]{000000} 91.8}         & {\color[HTML]{000000} 98.26}            \\ \hline
{\color[HTML]{000000} ProvGigPath}                 & {\color[HTML]{000000} 305 M}                   & {\color[HTML]{000000} 74.18}        & {\color[HTML]{000000} 90.6}         & {\color[HTML]{000000} 98.37}            \\ \hline
{\color[HTML]{000000} Virchow}                     & {\color[HTML]{000000} 305 M}                   & {\color[HTML]{000000} 79.02}        & {\color[HTML]{000000} 91.23}        & {\color[HTML]{000000} 98.28}            \\ \hline
\end{tabular}
\end{table} 

When comparing the training results across all three datasets, a trend is consistent: medium-sized transformer architectures, typically between 80 and 90 million parameters, deliver the most reliable and generalizable performance. Larger models do not necessarily outperform smaller ones, as the marginal benefit of additional parameters diminishes once the representational capacity surpasses the diversity of the dataset. These findings emphasize that, for computational pathology, model design and domain pretraining (resulting in effective representation learning tailored to tissue morphology and staining variability) are far more important than model size.

\section{Advantages of DeepCluster++ for computational pathology}\label{sec:advantages_of_deepcluster}
The proposed approach significantly reduces the manual burden of annotation and tile selection, compared to the conventional approach to constructing tissue-type classification datasets (which involves manual pathologist delineation of ROI within a WSI, followed by extraction of tiles from these ROIs). In contrast, with our automated DeepCluster++ framework, once the tiles have been collected for each tissue class within a WSI (which would normally require a significant amount of human time and effort using the conventional manual approach), a pathologist can simply use QuPath software to re-map the collected tiles back onto the original WSI from which they were taken, then confirm through a quick WSI-level visual inspection that the tiles for each tissue class were correctly assigned by DeepCluster++. This quality‐control pass takes less than five minutes per slide, which is significantly less time than would be required to perform manual ROI annotation for each tissue class (following the conventional annotation and tile selection approach). By restricting the pathologist's workload to this final WSI-level verification step, our DeepCluster++ allows for the collection of tissue-type specific datasets with high‐confidence labels and significantly reduced manual effort.  

\section{Ablation study}\label{sec:ablation_study}
For the DeepCluster++ framework, performing a comprehensive ablation study by varying multiple configuration parameters and extracting large-scale datasets from 200 WSI is time-intensive, as it requires tile verification prior to downstream task evaluation. Therefore, we carried out a targeted ablation on 10 randomly selected WSI (independent of training/validation) for the TUM and NCS classes:

\textit{\textbf{Number of samples per cluster ($m$)}}: Determining the number of clusters ($K$) and $m$ is very challenging, and especially with histopathology-based tiles, it is difficult to set one value for the number of clusters and samples. We tested $m=100$ to $800$, finding that a small $(m \approx 100)$ yielded many tiny, redundant clusters, while a large $(m>800)$ resulted in mixing of tissue types. We found that $m \approx 400$ balanced intra-cluster purity and inter-cluster diversity. Table \ref{table7} shows the trade-off between the size of the cluster and the purity (inverse of the Shannon entropy) based on the fixed number of significant variations in TUM and NCS tissue morphology. As $m$ increases, purity decreases (entropy rises). We selected $m=400$, as it yielded substantial morphologic variation while maintaining low tissue-type admixture, representing the best balance for our goal. If deciding $m$ is complex for a dataset, it is recommended that the number of clusters and samples per cluster be set using $K=m=\sqrt{T}$ \cite{sun2024pathgen16m16millionpathology}, where $T$ is the number of tiles from a WSI.  

\begin{table} [h!]  
\small
\centering
\caption{Cluster quality analysis.}
\label{table7} 
\begin{tabular}{|c|c|l|}
\hline
{\color[HTML]{000000} \textbf{$m$}} & {\color[HTML]{000000} \textbf{Average Entropy}} & \multicolumn{1}{c|}{{\color[HTML]{000000} \textbf{Morphologic variation}}} \\ \hline
{\color[HTML]{000000} 200}        & {\color[HTML]{000000} 0.12}                     & {\color[HTML]{000000} Low}                                                 \\ \hline
{\color[HTML]{000000} 400}        & {\color[HTML]{000000} 0.41}                     & {\color[HTML]{000000} High}                                                \\ \hline
{\color[HTML]{000000} 600}        & {\color[HTML]{000000} 1.73}                     & {\color[HTML]{000000} High (but mixed tissue types)}                       \\ \hline
{\color[HTML]{000000} 800}        & {\color[HTML]{000000} 2.17}                     & {\color[HTML]{000000} High (but highly mixed tissue   types)}              \\ \hline
\end{tabular}
\end{table} 

 \begin{wrapfigure}{r}{0.5\textwidth} 
    \centering
    \includegraphics[width=0.4\textwidth]{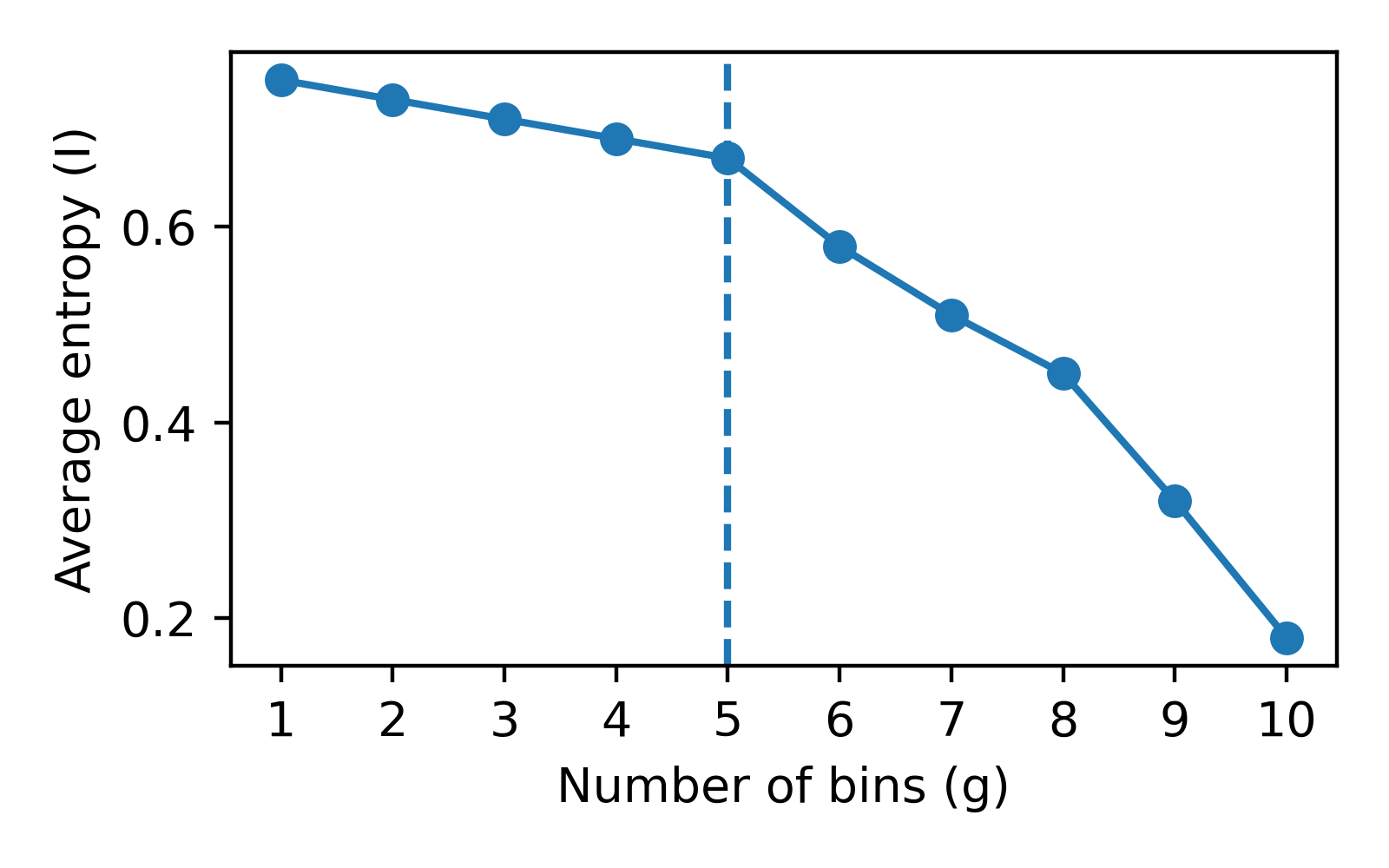}
    \caption{Ablation on the number of bins ($g$) using average inverse Shannon entropy. }
    \label{fig:7}
\end{wrapfigure}

\textit{\textbf{Number of bins ($g$)}}: Fixing the number of tiles per cluster at $m=400$ for a single tissue type, we performed an ablation on the number of equal-frequency distance bins, $ g \in \{1,2,...,10\}$. Each cluster contained approximately $4$ to $6$ distinguishable morphologic variants (e.g., structural subtypes within TUM). To assess within-bin homogeneity, we computed the average inverse normalized Shannon entropy, I(g), across bins, capturing the consistency of morphologic patterns within each bin. As illustrated in Figure \ref{fig:7}, $I(1)$ was the highest because a single coarse bin merges all five morphologic variants, causing admixture of heterogeneous tiles and redundancy in sampled images. In contrast, $I(10)$ was the lowest, as the data became excessively fragmented, with similar patterns being split across different bins, with each bin containing few tiles with nearly identical appearances, reducing the overall morphologic diversity.

Interestingly, the range $g=4$ to $6$ provided a balanced configuration: bins exhibited sufficient internal similarity while maintaining broad coverage across the morphologic continuum, from prototypical (near-centroid) to atypical (edge-of-cluster) tiles. In practice, the number of significant variants can differ across tissue types, making it impractical to fix g purely by empirical morphologic counts. Therefore, to mitigate excessive variability at smaller bin counts ($g=4$) and prevent over-fragmentation at larger counts ($g>5$), we identified $g=5$ as the optimal trade-off between within-bin similarity and across-bin diversity. This configuration consistently preserved meaningful morphologic transitions while maintaining stable sampling performance across tissue types. 

Sampling percentage (20\%): Following the selection of g=5, we sampled an equal 20\% of tiles from each bin to ensure a uniform and unbiased representation across the morphologic spectrum. This strategy guaranteed that all morphologic variants, from highly prototypical to rare atypical appearances, were proportionally included in the dataset. The 20\% sampling rate offered a practical balance between computational efficiency and morphologic coverage. Depending on the specific requirements of downstream tasks (e.g., classification, segmentation, or survival modeling), the sampling ratio can be scaled up or down to expand or contract the dataset size while maintaining representational consistency.  

\newpage

\section{Tile-level prediction map overlaid on the WSI.}\label{sec:prediction_maps}
\textit{\textbf{NCT}}: adipose (ADI), lymphocytes (LYM), smooth muscle (MUS), mucus (MUC), debris (DEB), colorectal adenocarcinoma epithelium (TUM), normal colon mucosa (NORM), cancer-associated stroma (STR), background (BACK)  \\[0.5em]
\textit{\textbf{HMU}}: adipose tissue (ADI), lymphocyte aggregates (LYM), muscle (MUS), mucus (MUC), debris (DEB), tumor epithelium (TUM), normal mucosa (NORM), stroma (STR)  \\[0.5em]
\textit{\textbf{STARC-9}}: adipose tissue (ADI), lymphoid tissue (LYM), muscle (MUS), fibroconnective tissue (FCT), mucin (MUC), necrosis (NCS), blood (BLD), tumor (TUM), and normal mucosa (NOR)  \\[0.5em]
\textit{\textbf{Names for the same tissue type in different datasets}}: DEB in NCT/HMU corresponds to NCS in STARC-9 and NORM in NCT/HMU corresponds to NOR in STARC-9.\\[0.5em]
\begin{figure}[!h]
	\centering
	\includegraphics[width=0.75\textwidth]{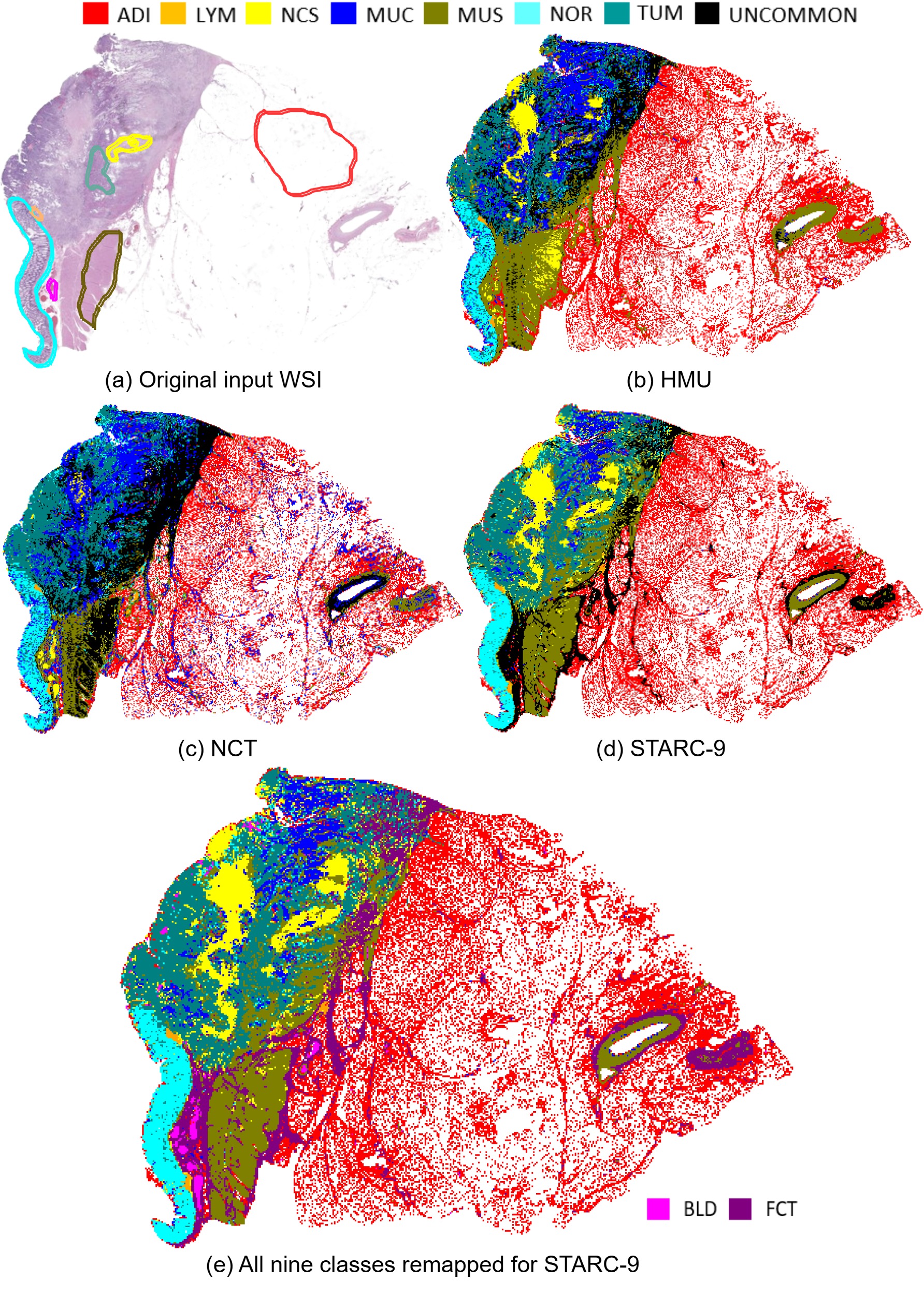}
	\caption{Tile-level prediction maps overlaid on a given input WSI (a) using the best-performing models trained on (b) HMU, (c) NCT, and (d) STARC-9 for the seven common tissue classes (ADI, LYM, MUS, MUC, NCS, TUM, NOR). Tiles assigned to classes outside these seven (e.g., stroma-STR in HMU/NCT and BLD and FCT in STARC-9) are shown in black (Uncommon tissue classes). Panel (e) shows all nine classes (included in STARC-9) mapped back to the input WSI. }
	\label{fig:8}
\end{figure}  

\newpage
\section{Confusion matrices for the best-performing models (trained on NCT, HMU, and STARC-9) on STANFORD-CRC-HE-VAL-LARGE for seven common tissue types. } 
\begin{figure}[htbp]
  \centering
  \begin{subfigure}[b]{0.71\textwidth}
    \centering
    \includegraphics[width=\linewidth]{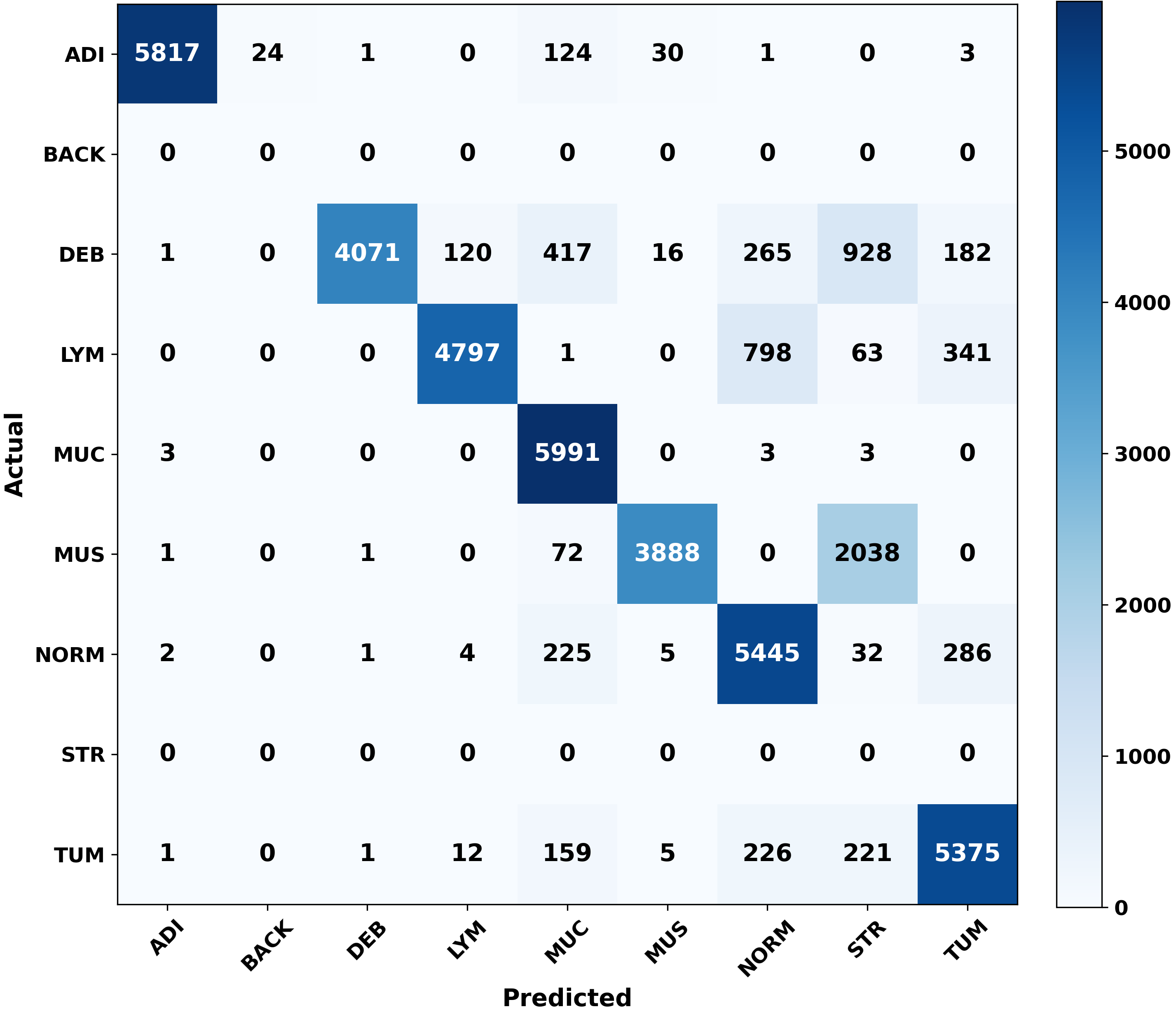}
    \caption{NCT}
    \label{fig:9.1}
  \end{subfigure}
  \hfill
  \begin{subfigure}[b]{0.71\textwidth}
    \centering
    \includegraphics[width=\linewidth]{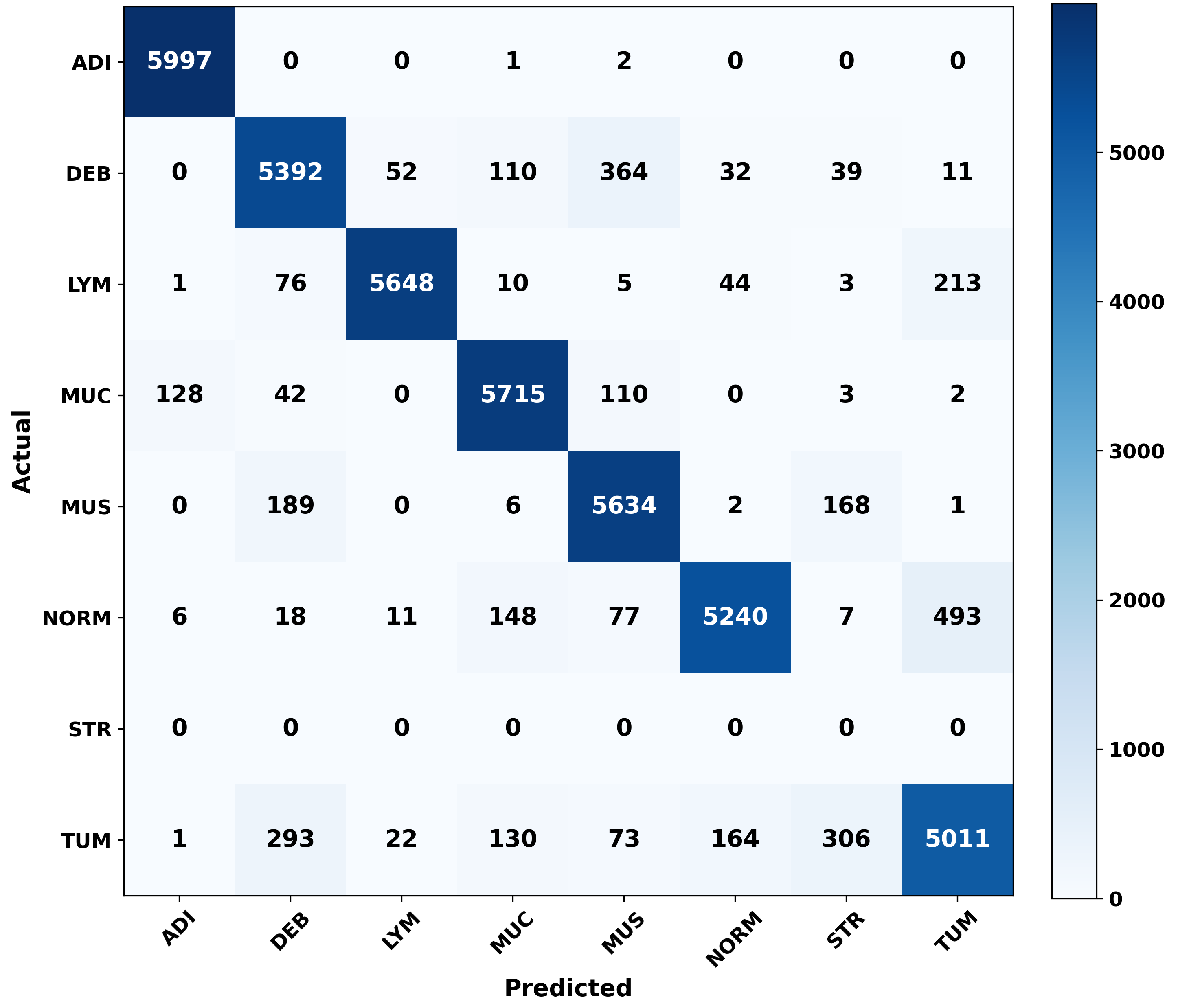}
    \caption{HMU}
    \label{fig:9.2}
  \end{subfigure}
\end{figure}
\clearpage
% Continued float: third subfigure on next page
\begin{figure}[htbp]\ContinuedFloat
  \centering
  \begin{subfigure}[b]{0.71\textwidth}
    \centering
    \includegraphics[width=\linewidth]{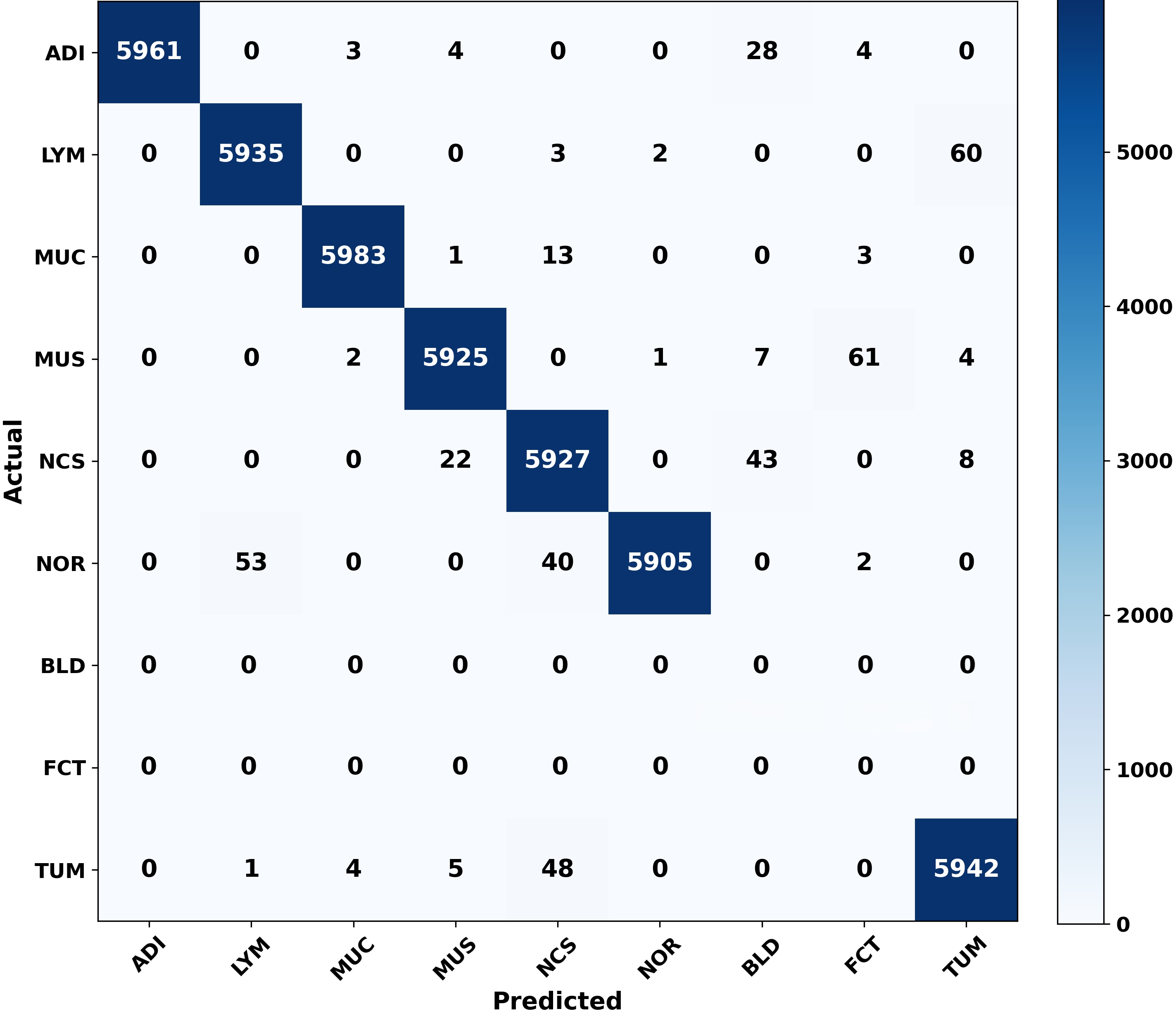}
    \caption{STARC-9}
    \label{fig:9.3}
  \end{subfigure}
    \caption{Confusion matrices for the best-performing models on STANFORD-CRC-HE-VAL-LARGE  for seven common tissue types.}
      \label{fig:9}
\end{figure}

\section{Confusion matrices for the best-performing models (trained on NCT, HMU, and STARC-9) on CURATED-TCGA-CRC-HE-VAL-20K for seven common tissue types. }
\begin{figure}[htbp]
  \centering
  \begin{subfigure}[b]{0.71\textwidth}
    \centering
    \includegraphics[width=\linewidth]{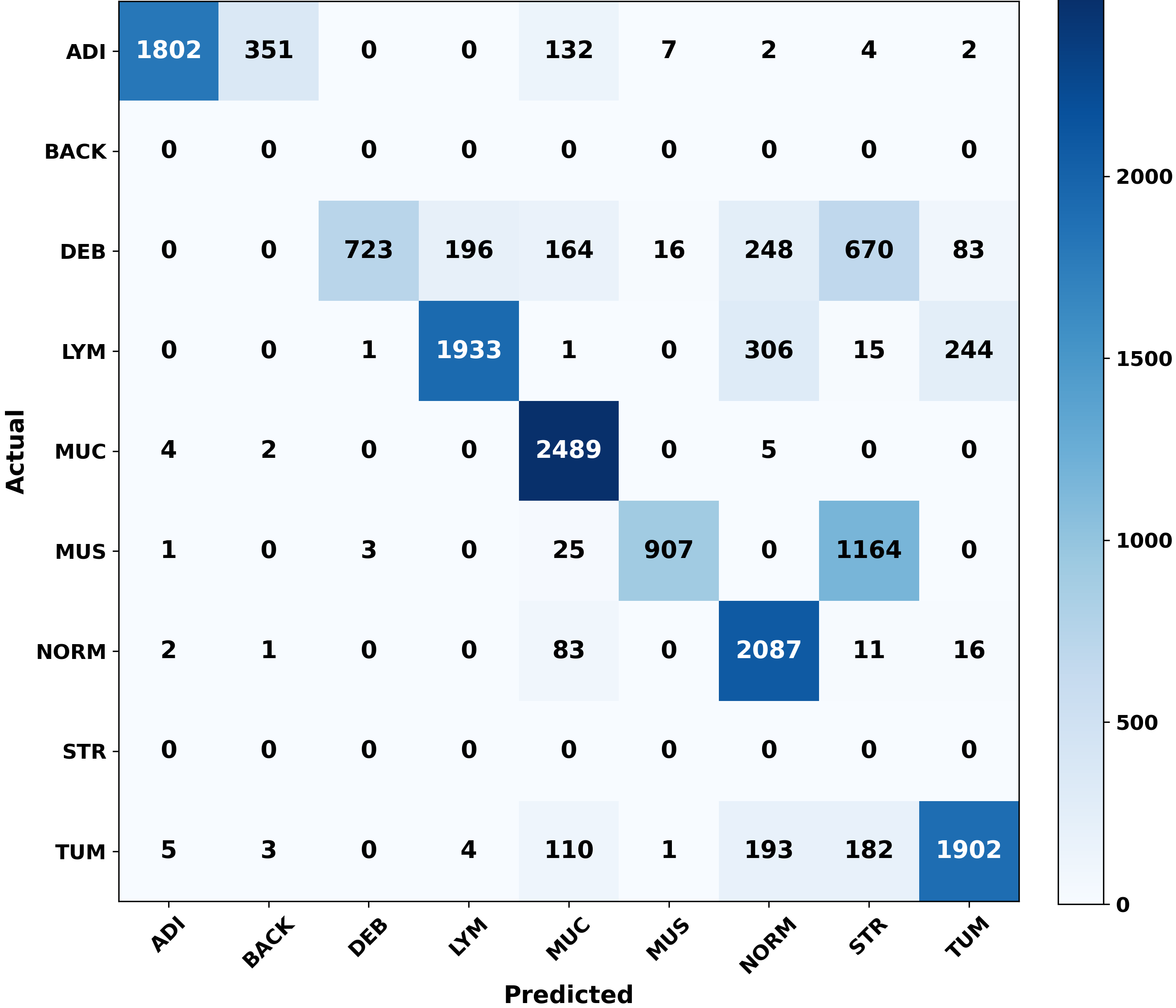}
    \caption{NCT}
    \label{fig:10.1}
  \end{subfigure}
\end{figure}
\clearpage
% Continued float: third subfigure on next page
\begin{figure}[htbp]\ContinuedFloat
  \centering
  \begin{subfigure}[b]{0.71\textwidth}
    \centering
    \includegraphics[width=\linewidth]{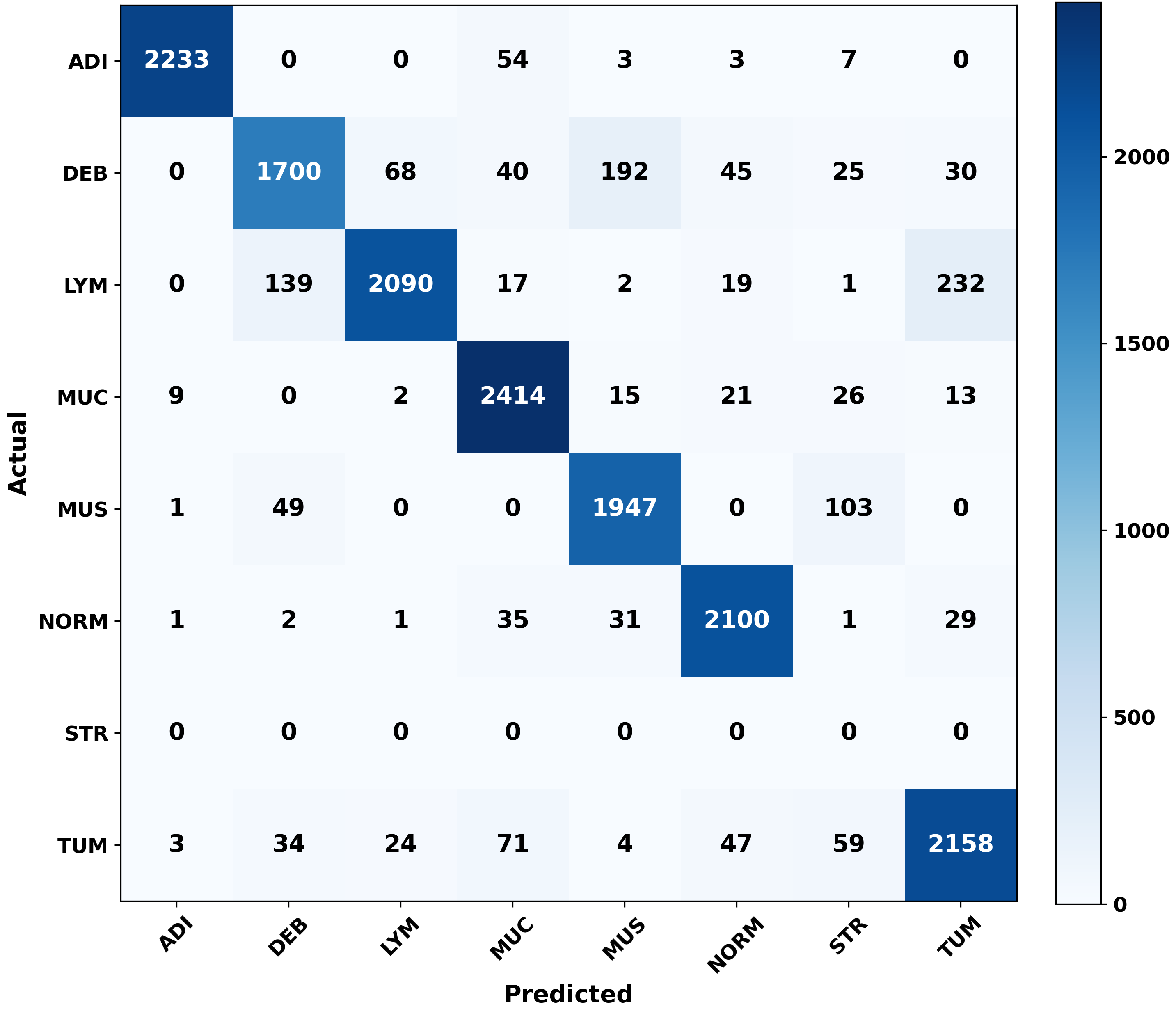}
    \caption{HMU}
    \label{fig:10.2}
  \end{subfigure} 
  \hfill  
  \begin{subfigure}[b]{0.71\textwidth}
    \centering
    \includegraphics[width=\linewidth]{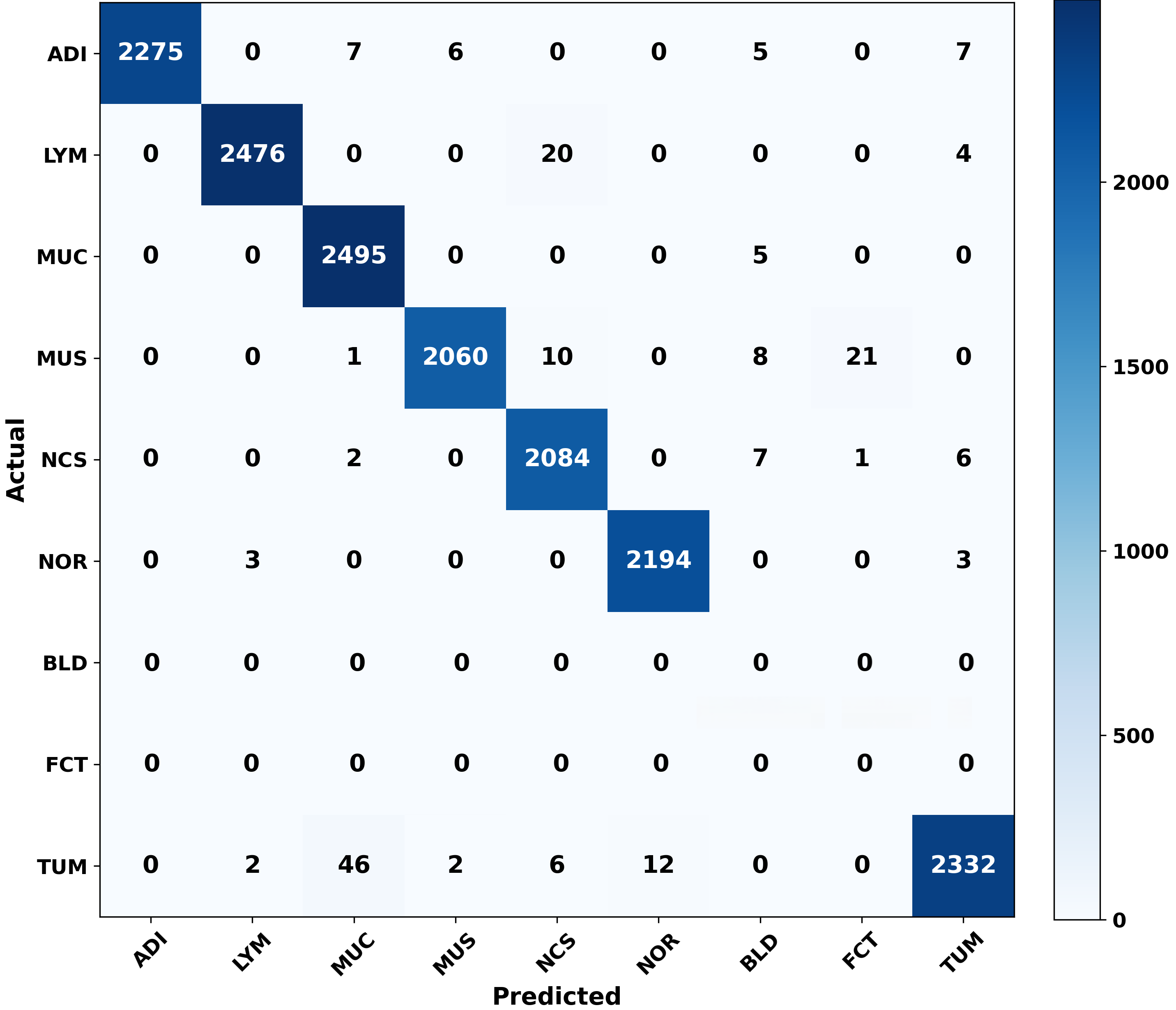}
    \caption{STARC-9}
    \label{fig:10.3}
  \end{subfigure}
    \caption{Confusion matrices for the best-performing models on CURATED-TCGA-CRC-HE-VAL-20K for seven common tissue types.}
    \label{fig:10}
\end{figure}

\newpage
\section{Confusion matrices for the best-performing models (trained on NCT, HMU, and STARC-9) on STANFORD-CRC-HE-VAL-SMALL for seven common tissue types.  } 
\begin{figure}[htbp]
  \centering
  \begin{subfigure}[b]{0.71\textwidth}
    \centering
    \includegraphics[width=\linewidth]{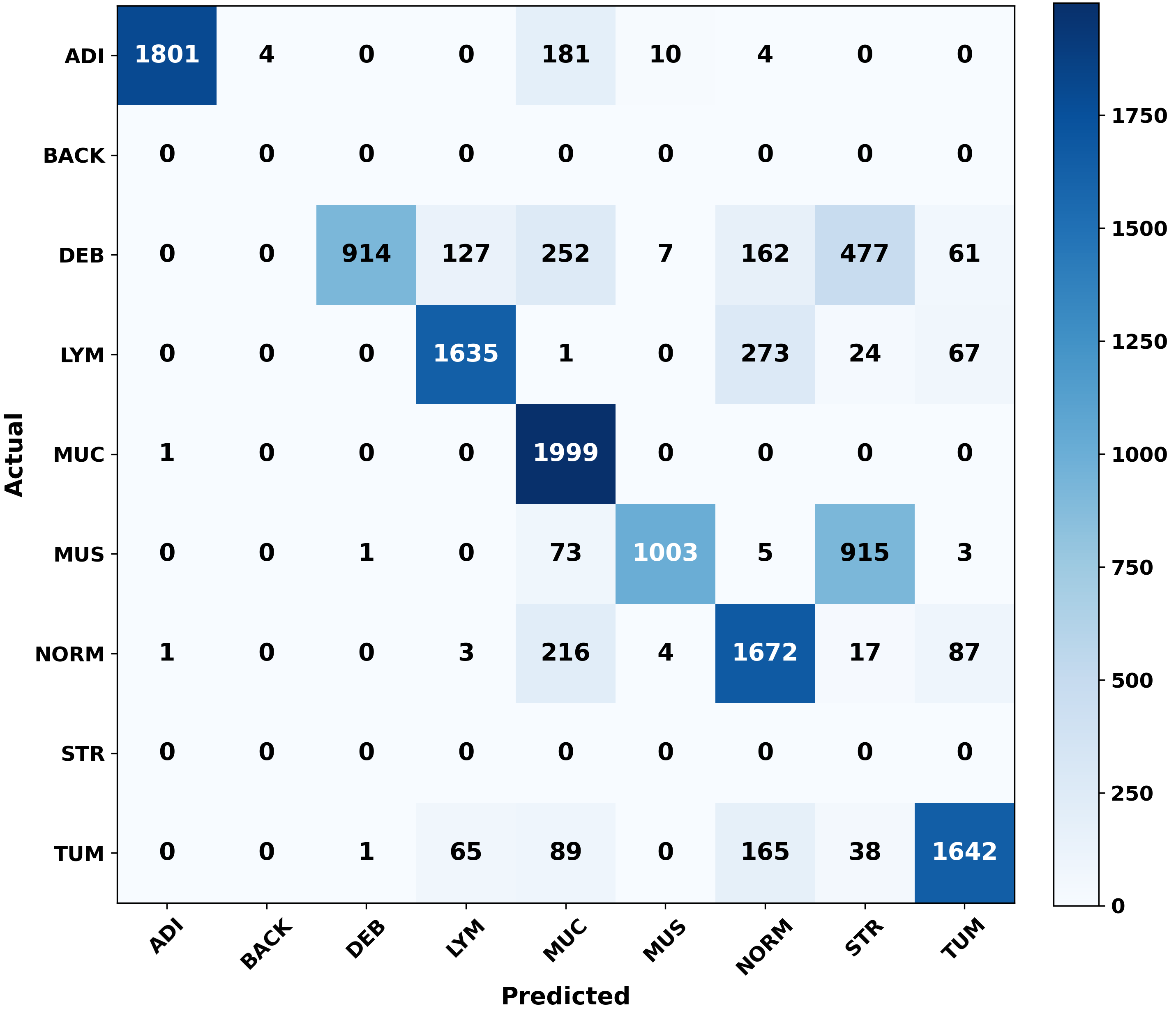}
    \caption{NCT}
    \label{fig:11.1}
  \end{subfigure}
  \hfill
  \begin{subfigure}[b]{0.71\textwidth}
    \centering
    \includegraphics[width=\linewidth]{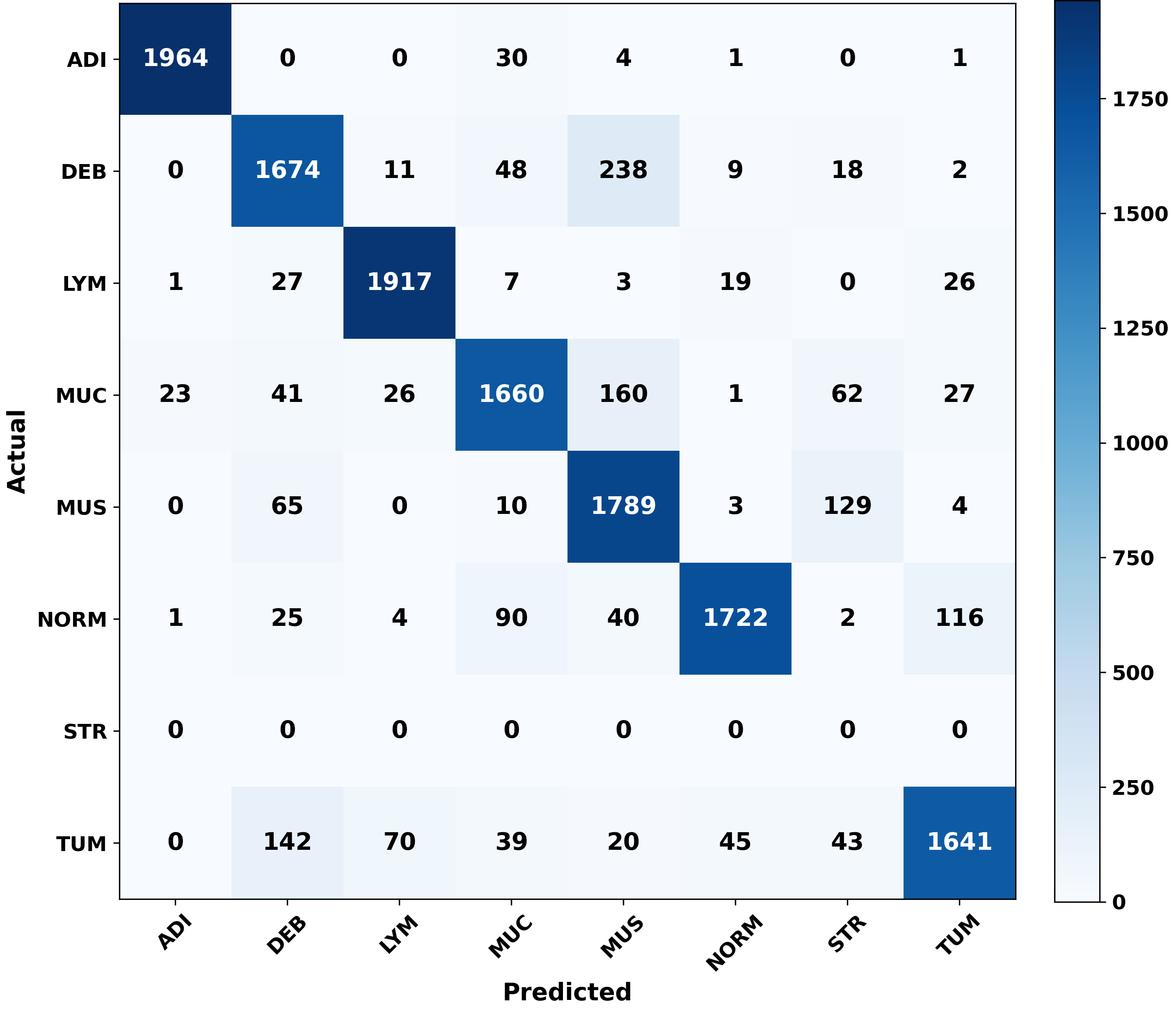}
    \caption{HMU}
    \label{fig:11.2}
  \end{subfigure}
\end{figure}
\clearpage
% Continued float: third subfigure on next page
\begin{figure}[htbp]\ContinuedFloat
  \centering
  \begin{subfigure}[b]{0.71\textwidth}
    \centering
    \includegraphics[width=\linewidth]{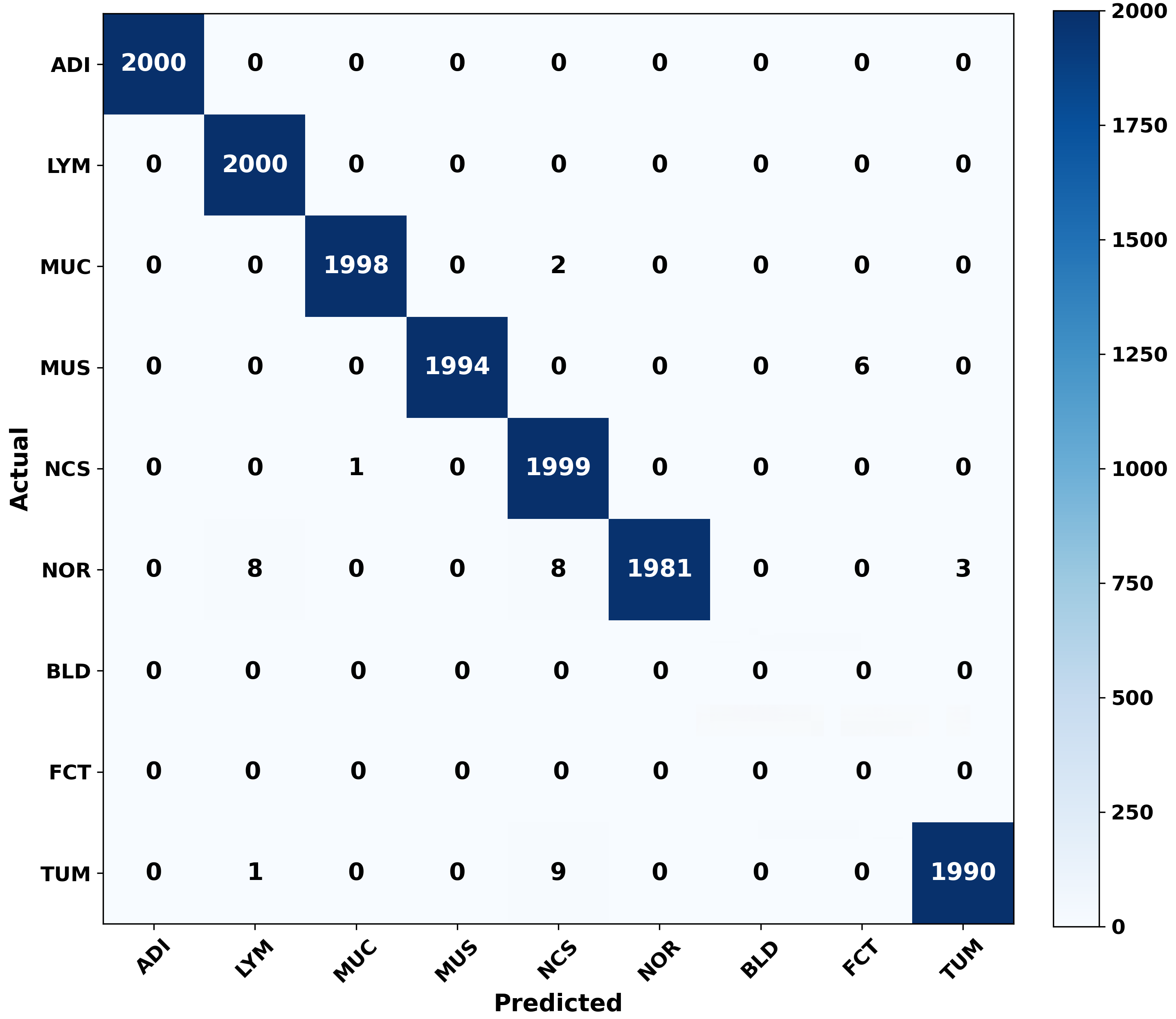}
    \caption{STARC-9}
    \label{fig:11.3}
  \end{subfigure}
    \caption{Confusion matrices for the best-performing models on STANFORD-CRC-HE-VAL-SMALL  for seven common tissue types.}
      \label{fig:11}
\end{figure}

\newpage
\section{ROC curves for the best-performing models (trained on NCT, HMU, and STARC-9) on STANFORD-CRC-HE-VAL-LARGE for seven common tissue types. } 
\begin{figure}[htbp]
  \centering
  \begin{subfigure}[b]{0.65\textwidth}
    \centering
    \includegraphics[width=\linewidth]{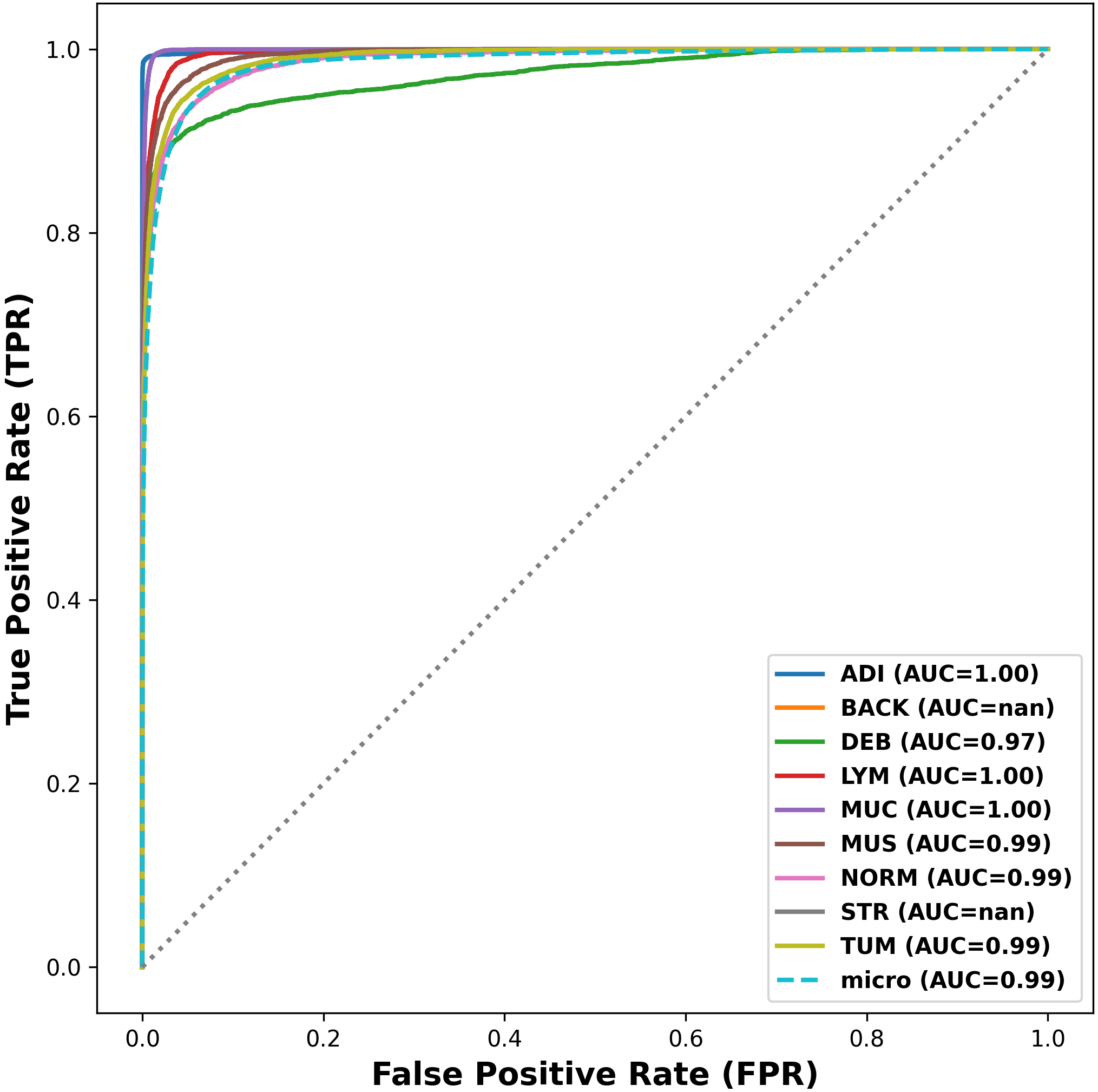}
    \caption{NCT}
    \label{fig:12.1}
  \end{subfigure}
  \hfill
  \begin{subfigure}[b]{0.65\textwidth}
    \centering
    \includegraphics[width=\linewidth]{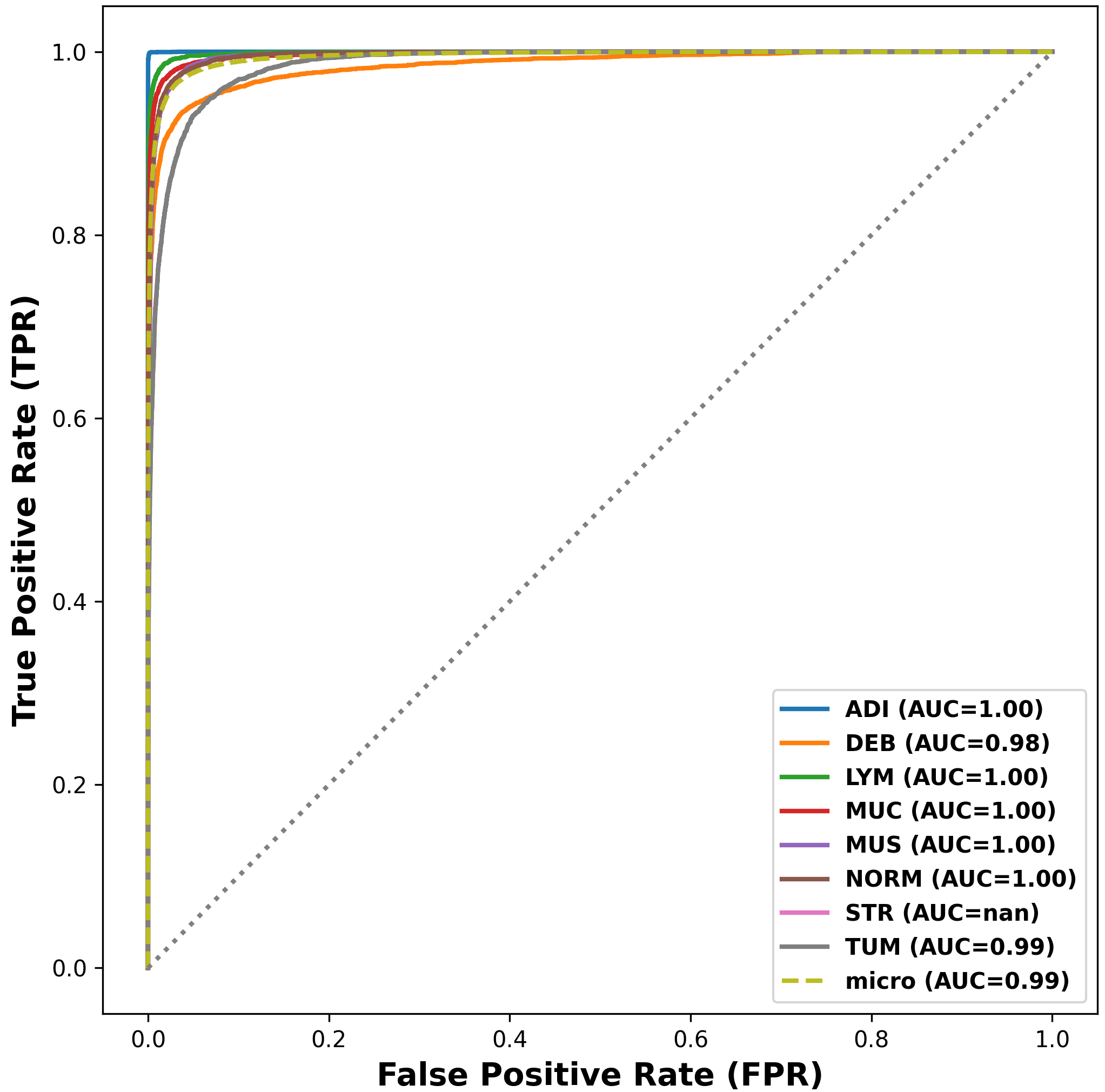}
    \caption{HMU}
    \label{fig:12.2}
  \end{subfigure}
\end{figure}
\clearpage
% Continued float: third subfigure on next page
\begin{figure}[htbp]\ContinuedFloat
  \centering
  \begin{subfigure}[b]{0.6\textwidth}
    \centering
    \includegraphics[width=\linewidth]{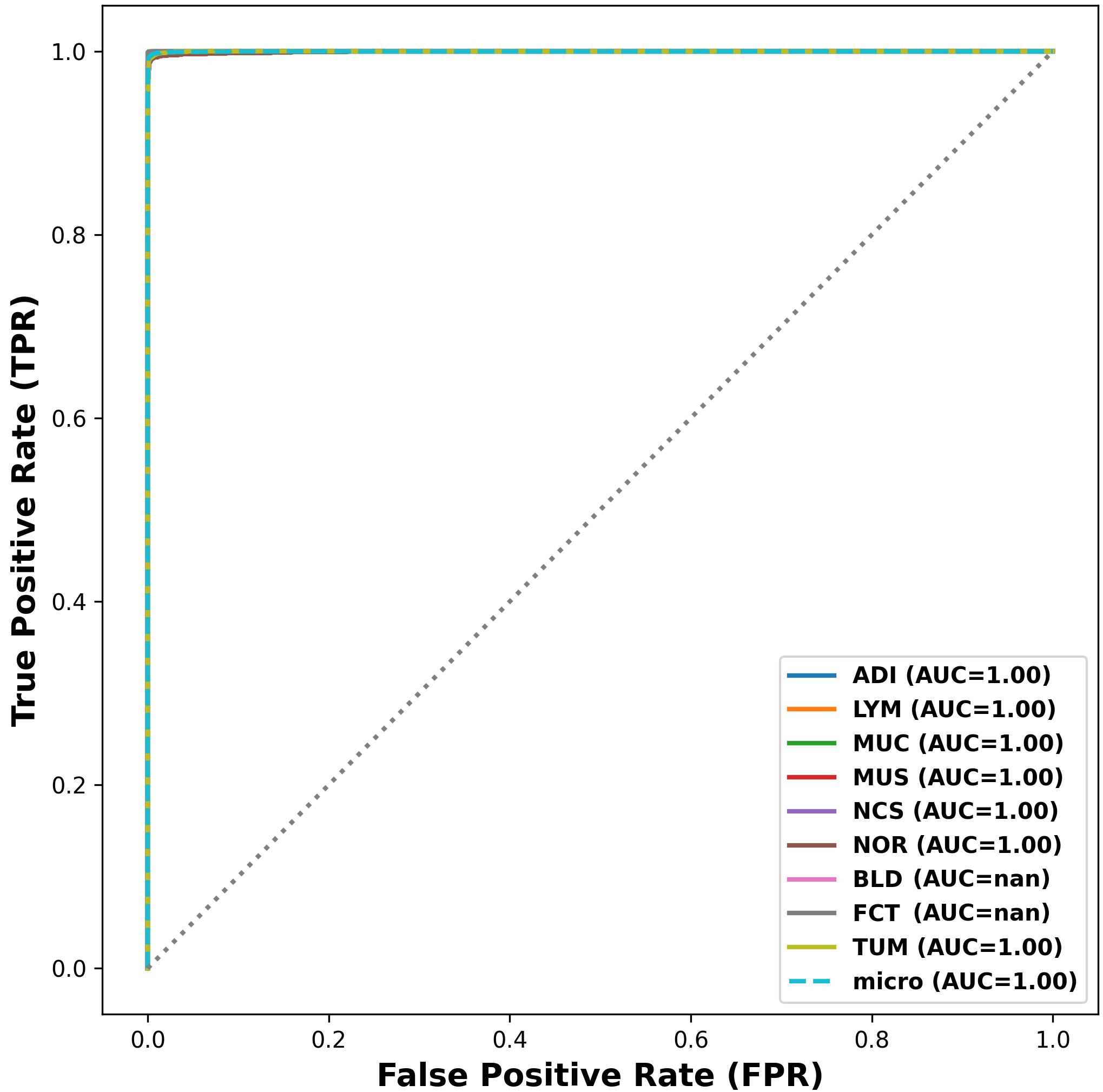}
    \caption{STARC-9}
    \label{fig:12.3}
  \end{subfigure}
    \caption{ROC curves for the best-performing models on STANFORD-CRC-HE-VAL-LARGE  for seven common tissue types.}
      \label{fig:12}
\end{figure}

\section{ROC curves for the best-performing models (trained on NCT, HMU, and STARC-9) on CURATED-TCGA-CRC-HE-VAL-20K for seven common tissue types. }
\begin{figure}[htbp]
  \centering
  \begin{subfigure}[b]{0.6\textwidth}
    \centering
    \includegraphics[width=\linewidth]{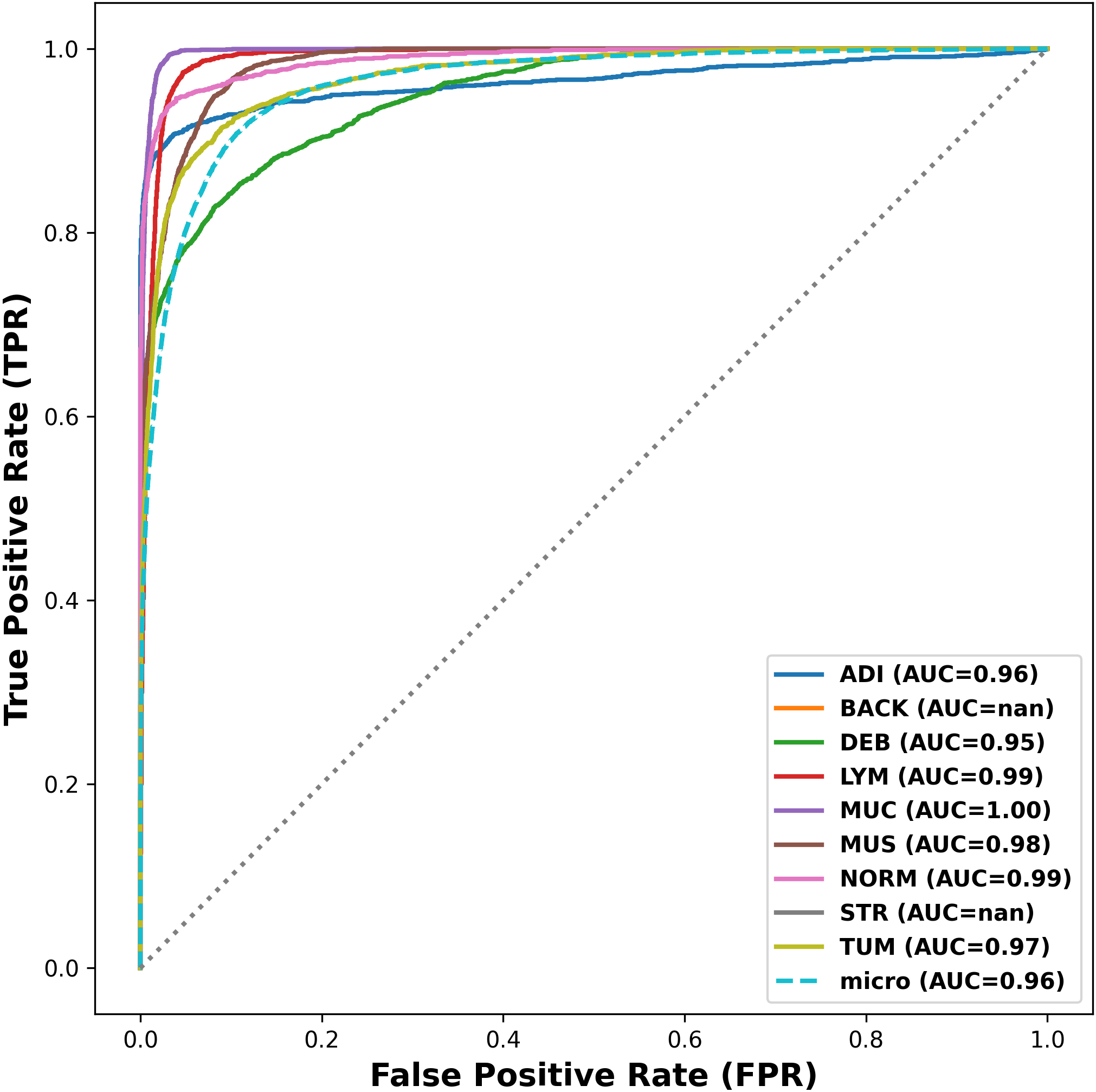}
    \caption{NCT}
    \label{fig:13.1}
  \end{subfigure}
\end{figure}
\clearpage
% Continued float: third subfigure on next page
\begin{figure}[htbp]\ContinuedFloat
  \centering
  \begin{subfigure}[b]{0.62\textwidth}
    \centering
    \includegraphics[width=\linewidth]{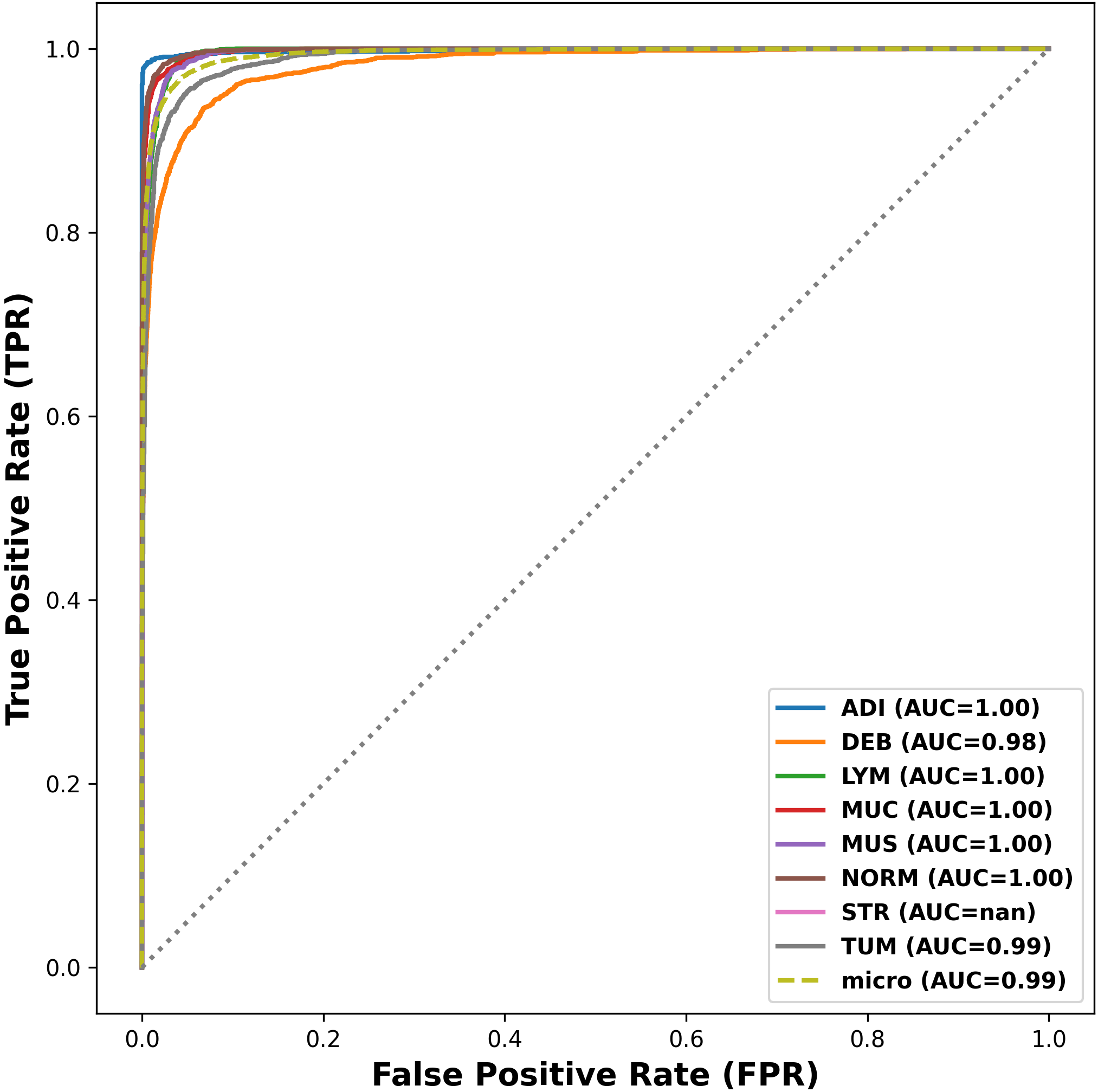}
    \caption{HMU}
    \label{fig:13.2}
  \end{subfigure} 
  \hfill  
  \begin{subfigure}[b]{0.62\textwidth}
    \centering
    \includegraphics[width=\linewidth]{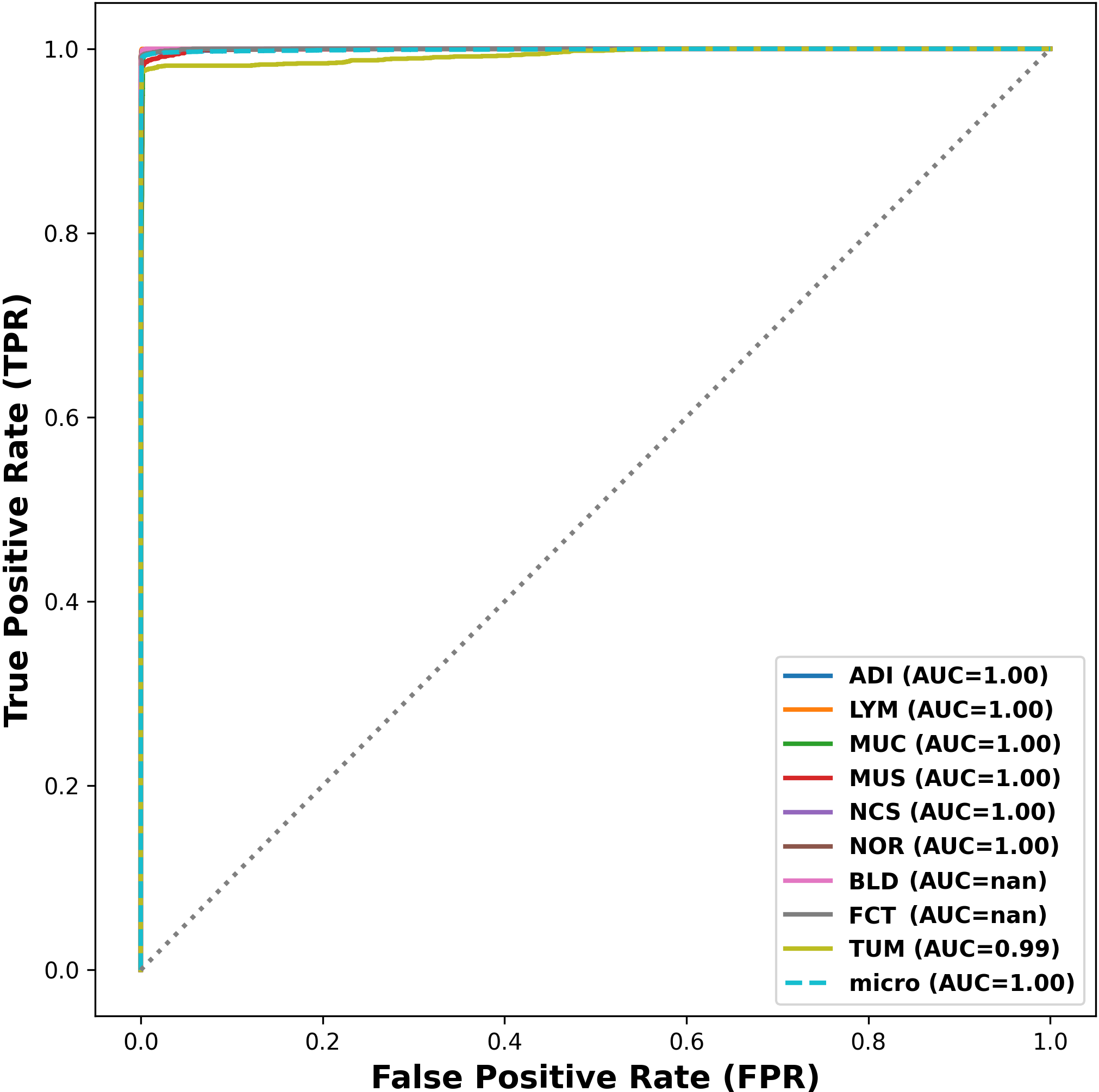}
    \caption{STARC-9}
    \label{fig:13.3}
  \end{subfigure}
    \caption{ROC curves for the best-performing models on CURATED-TCGA-CRC-HE-VAL-20K  for seven common tissue types.}
    \label{fig:13}
\end{figure}

\newpage
\section{ROC curves for the best-performing models (trained on NCT, HMU, and STARC-9) on STANFORD-CRC-HE-VAL-SMALL for seven common tissue types. } 
\begin{figure}[htbp]
  \centering
  \begin{subfigure}[b]{0.68\textwidth}
    \centering
    \includegraphics[width=\linewidth]{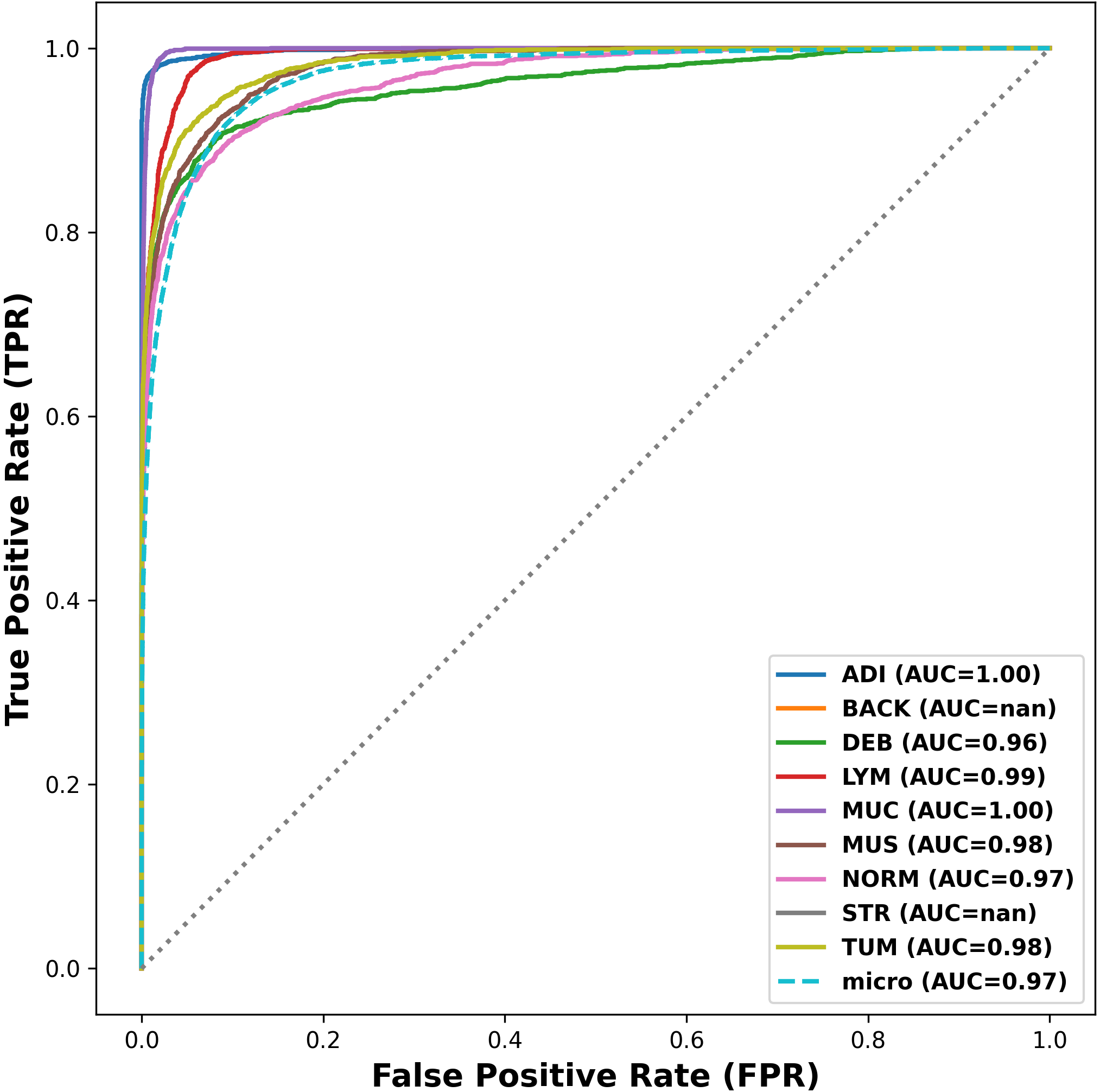}
    \caption{NCT}
    \label{fig:14.1}
  \end{subfigure}
  \hfill
  \begin{subfigure}[b]{0.68\textwidth}
    \centering
    \includegraphics[width=\linewidth]{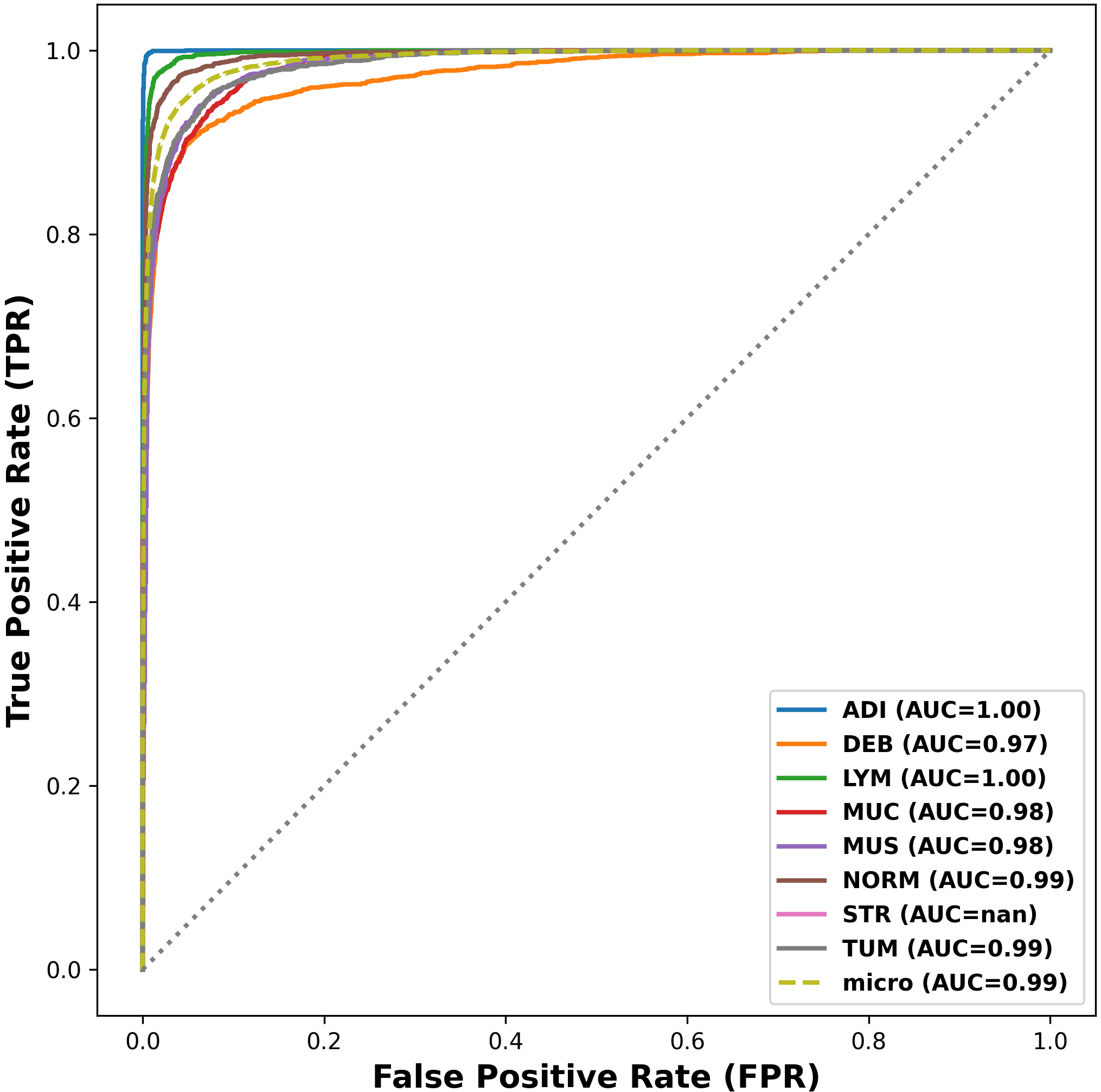}
    \caption{HMU}
    \label{fig:14.2}
  \end{subfigure}
\end{figure}
\clearpage
% Continued float: third subfigure on next page
\begin{figure}[htbp]\ContinuedFloat
  \centering
  \begin{subfigure}[b]{0.68\textwidth}
    \centering
    \includegraphics[width=\linewidth]{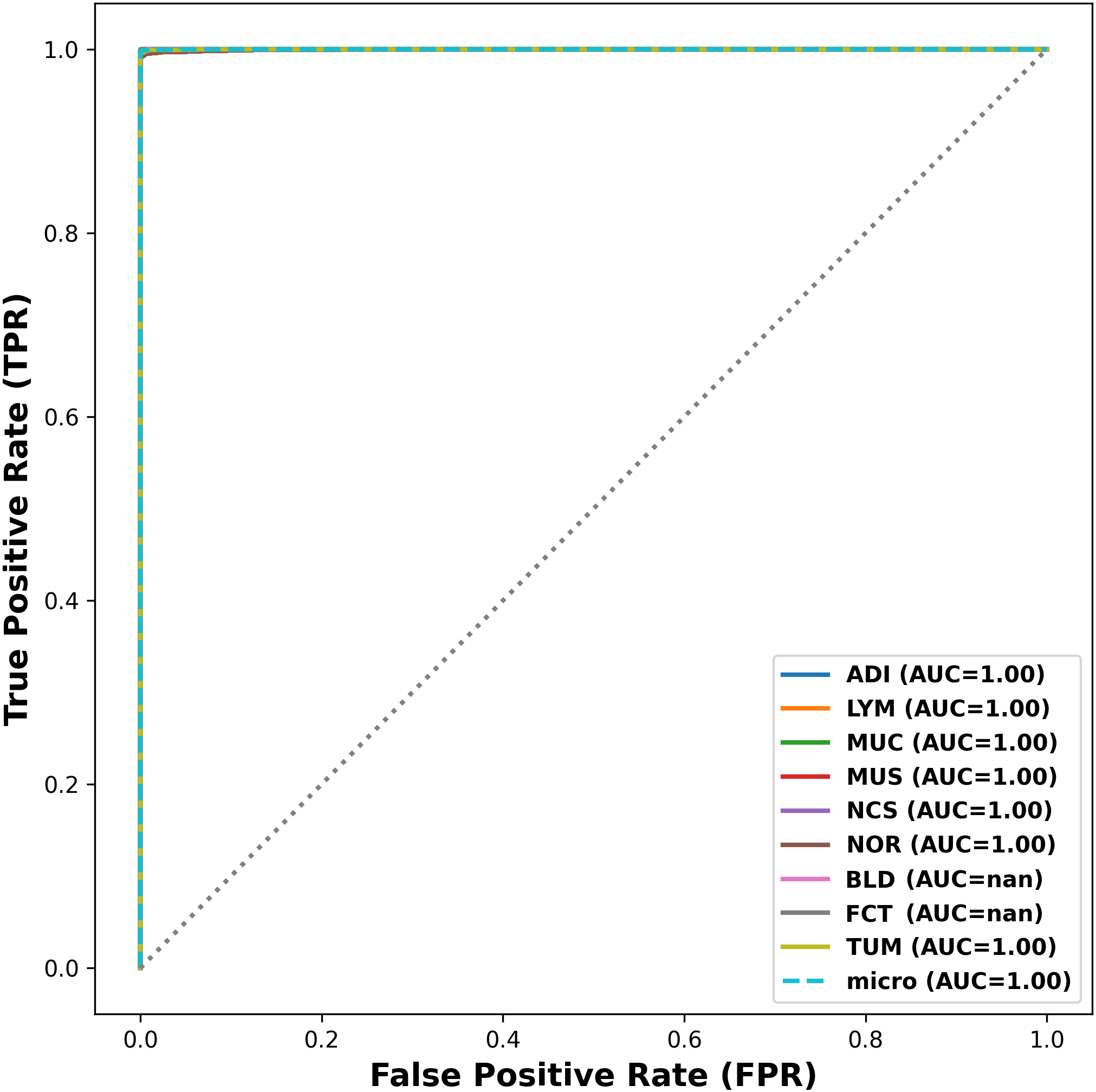}
    \caption{STARC-9}
    \label{fig:14.3}
  \end{subfigure}
    \caption{ROC curves for the best-performing models on STANFORD-CRC-HE-VAL-SMALL  for seven common tissue types.}
      \label{fig:14}
\end{figure}

\section{Tumor segmentation within 2048x2048 regions from a WSI from the STANFORD-CRC-HE-VAL-LARGE dataset. } 
\begin{figure}[!h]
	\centering
	\includegraphics[width=1\textwidth]{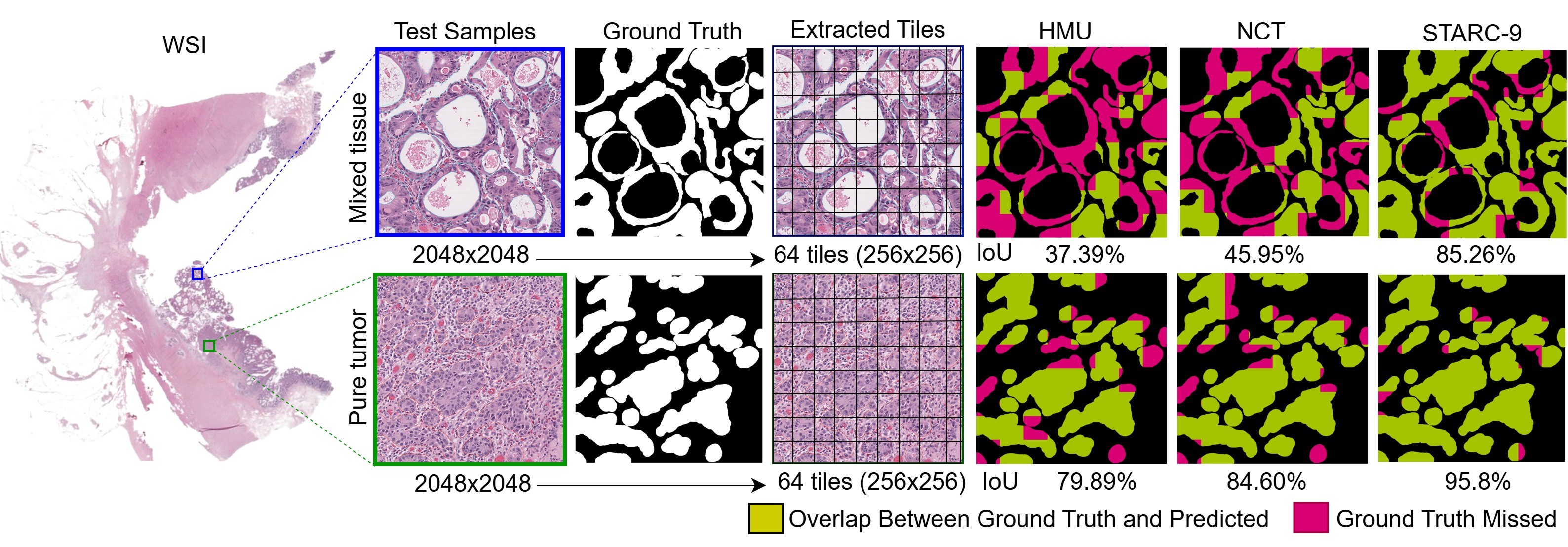}
	\caption{Tumor segmentation within 2048x2048 regions from a WSI from the STANFORD-CRC-HE-VAL-LARGE dataset using tile-level classifiers trained on HMU, NCT, and STARC-9. }
	\label{fig:15}
\end{figure}  

\newpage
\section{Confusion matrices for the best-performing model trained on STARC-9 and run on the validation datasets for all nine tissue types. } 
\begin{figure}[htbp]
  \centering
  \begin{subfigure}[b]{0.71\textwidth}
    \centering
    \includegraphics[width=\linewidth]{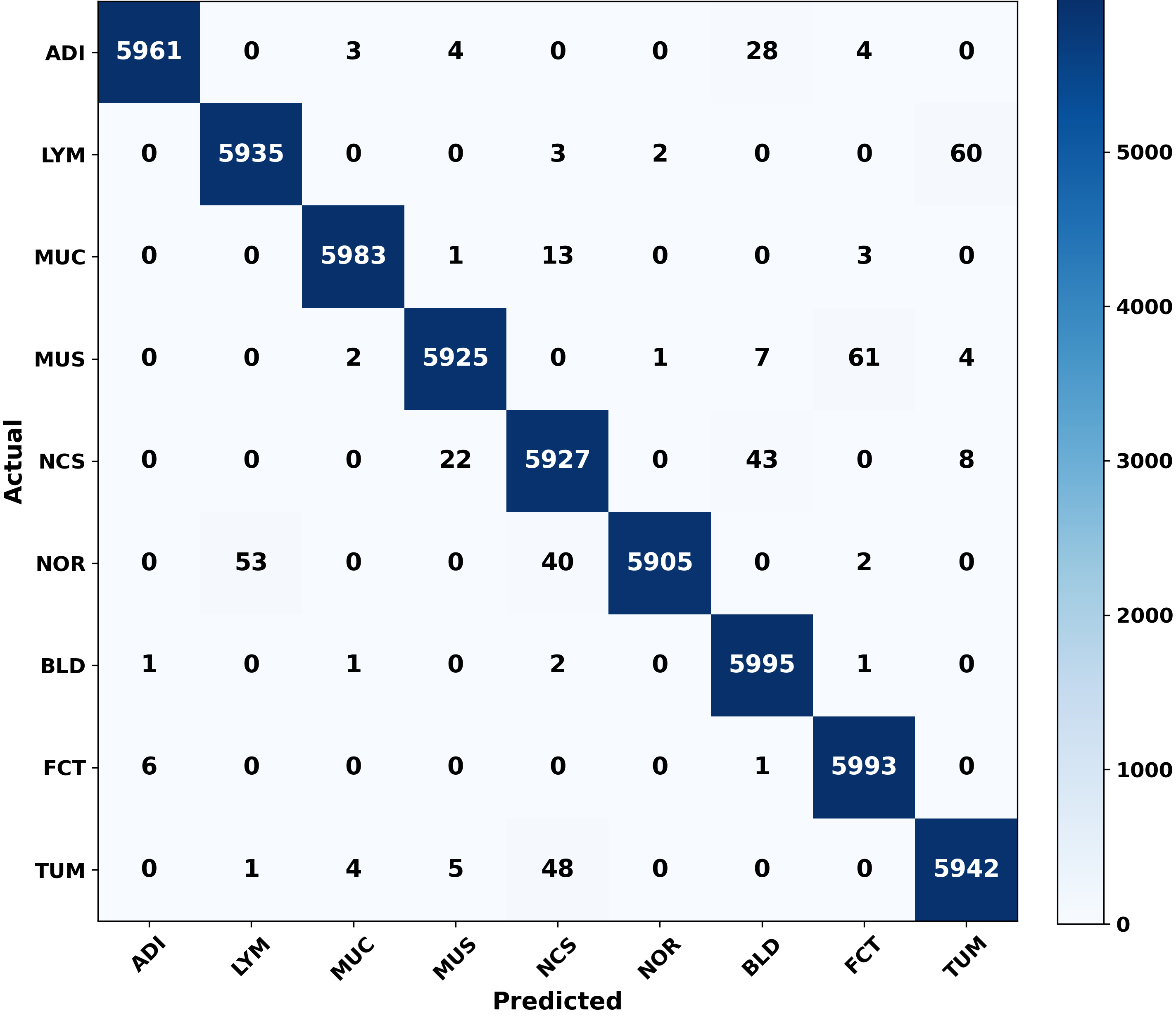}
    \caption{STANFORD-CRC-HE-VAL-LARGE}
    \label{fig:16.1}
  \end{subfigure}
  \hfill
  \begin{subfigure}[b]{0.71\textwidth}
    \centering
    \includegraphics[width=\linewidth]{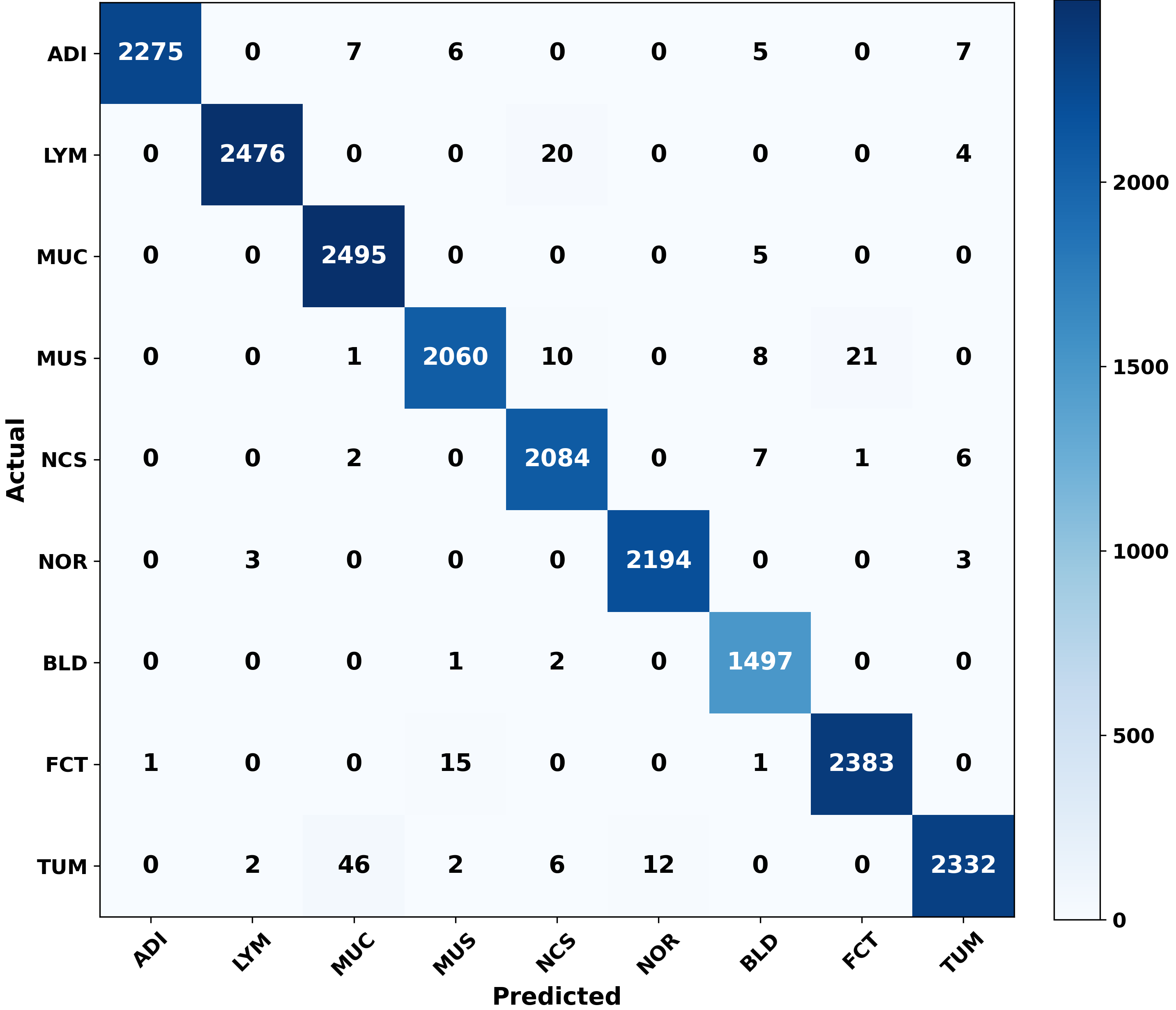}
    \caption{CURATED-TCGA-CRC-HE-VAL-20K}
    \label{fig:16.2}
  \end{subfigure}
\end{figure}
\clearpage
% Continued float: third subfigure on next page
\begin{figure}[htbp]\ContinuedFloat
  \centering
  \begin{subfigure}[b]{0.71\textwidth}
    \centering
    \includegraphics[width=\linewidth]{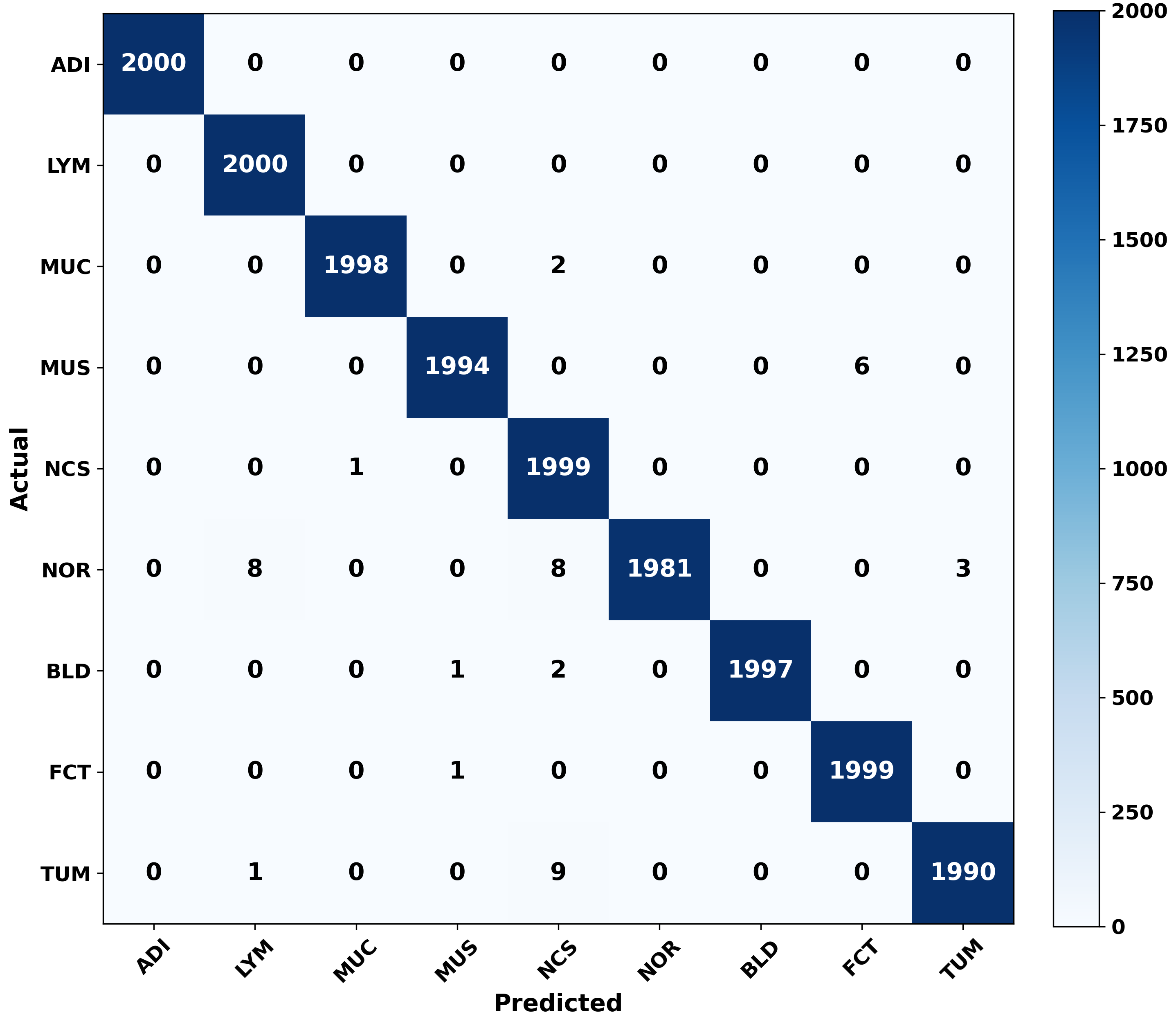}
    \caption{STANFORD-CRC-HE-VAL-SMALL}
    \label{fig:16.3}
  \end{subfigure}
    \caption{Confusion matrices for the best-performing model (trained on STARC-9) on (a) STANFORD-CRC-HE-VAL-LARGE, (b) CURATED-TCGA-CRC-HE-VAL-20K, and (c) STANFORD-CRC-HE-VAL-SMALL.}
      \label{fig:16}
\end{figure}

\newpage
\section{ROC curves for the best-performing model trained on STARC-9 and run on the validation datasets for all nine tissue types. } 
\begin{figure}[htbp]
  \centering
  \begin{subfigure}[b]{0.67\textwidth}
    \centering
    \includegraphics[width=\linewidth]{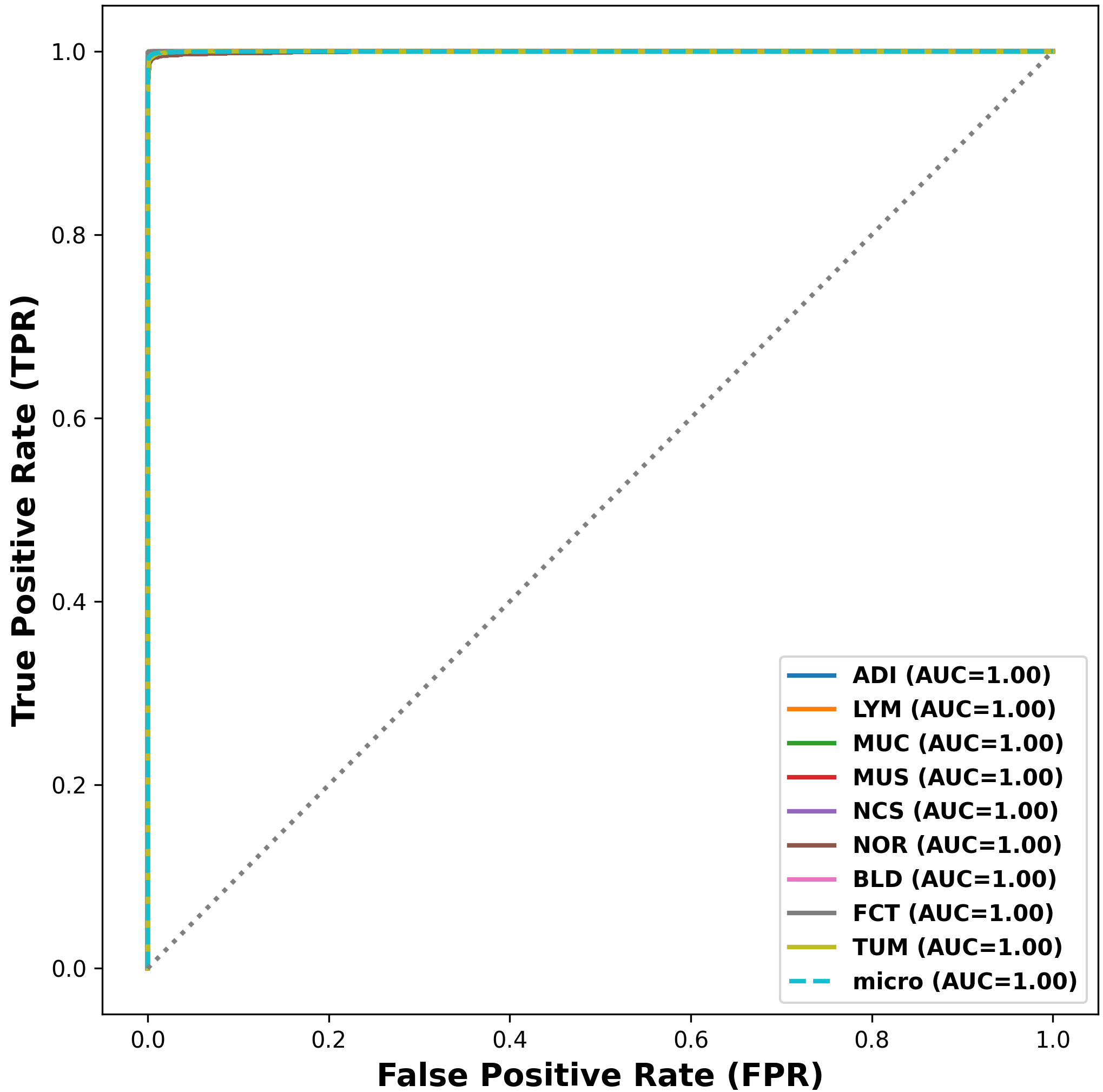}
    \caption{STANFORD-CRC-HE-VAL-LARGE}
    \label{fig:17.1}
  \end{subfigure}
  \hfill
  \begin{subfigure}[b]{0.67\textwidth}
    \centering
    \includegraphics[width=\linewidth]{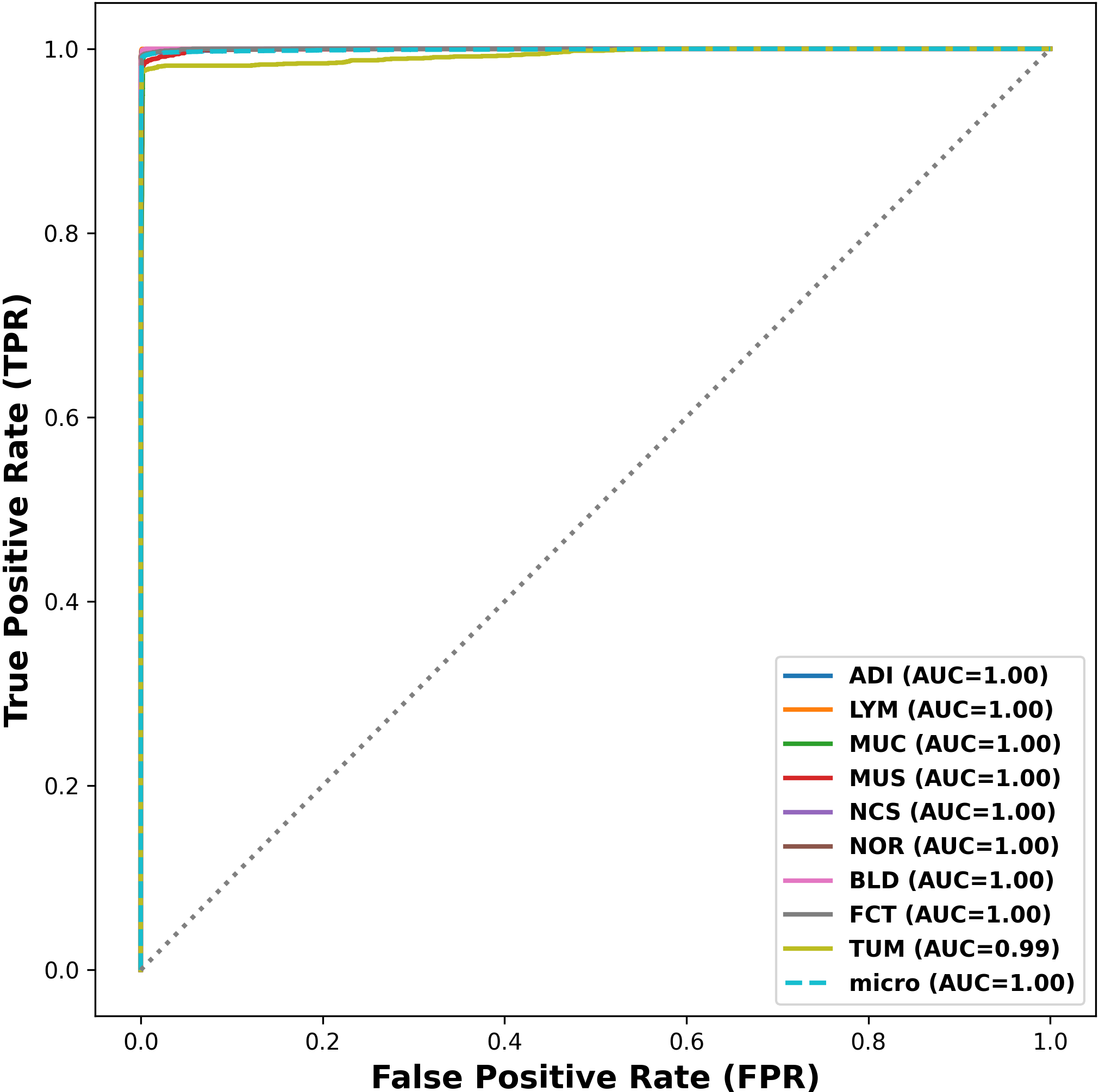}
    \caption{CURATED-TCGA-CRC-HE-VAL-20K}
    \label{fig:17.2}
  \end{subfigure}
\end{figure}
\clearpage
% Continued float: third subfigure on next page
\begin{figure}[htbp]\ContinuedFloat
  \centering
  \begin{subfigure}[b]{0.67\textwidth}
    \centering
    \includegraphics[width=\linewidth]{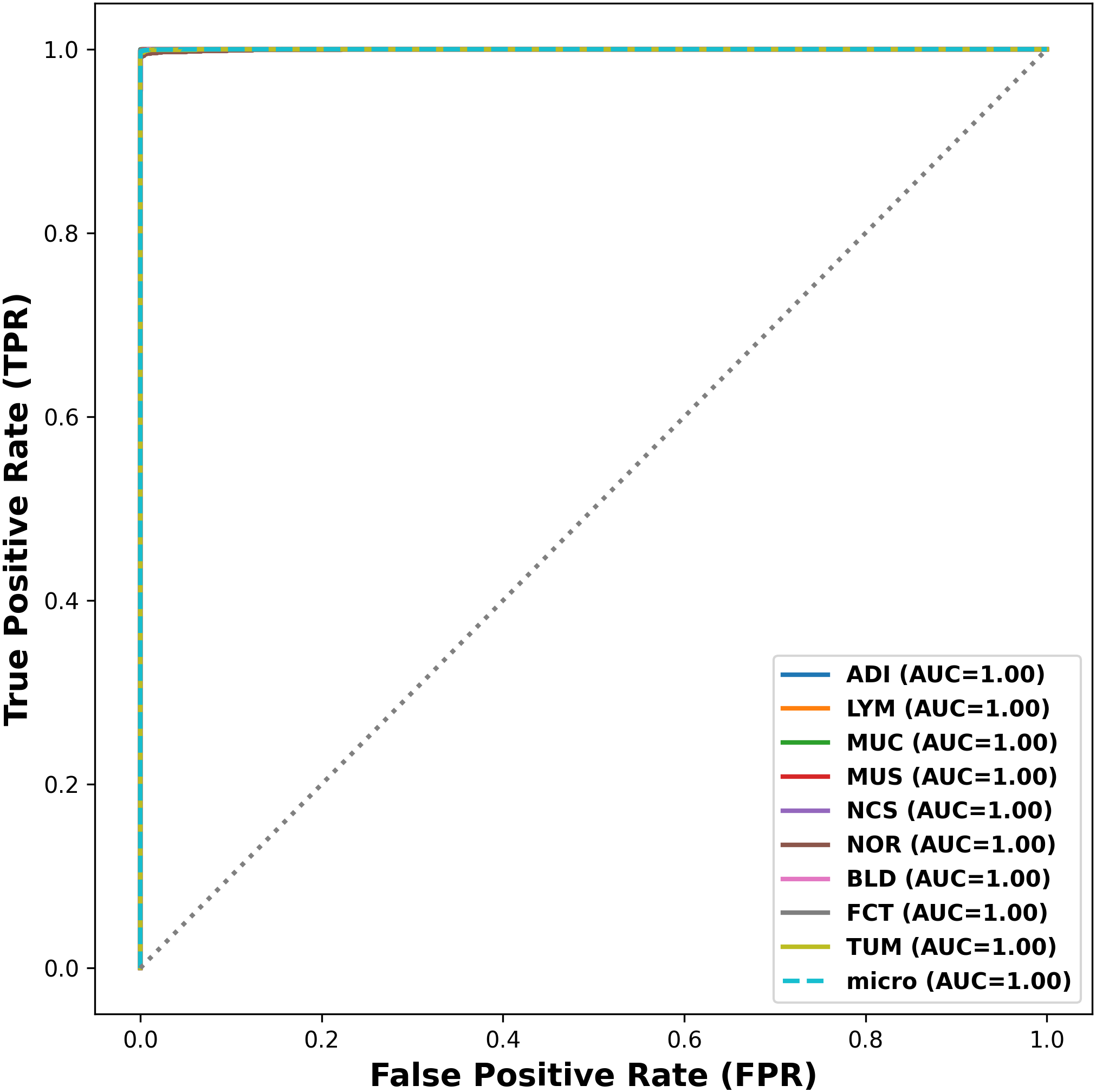}
    \caption{STANFORD-CRC-HE-VAL-SMALL}
    \label{fig:17.3}
  \end{subfigure}
    \caption{Confusion matrices for the best-performing model (trained on STARC-9) on (a) STANFORD-CRC-HE-VAL-LARGE, (b) CURATED-TCGA-CRC-HE-VAL-20K, and (c) STANFORD-CRC-HE-VAL-SMALL.}
      \label{fig:17}
\end{figure}

%%%%%%%%%%%%%%%%%%%%%%%%%%%%%%%%%%%%%%%%%%%%%%%%%%%%%%%%%%%%

\end{document}